\documentclass[12pt]{article}
\usepackage{amsmath}
\usepackage{epsf}
\usepackage{epsfig}
\usepackage{here}
\usepackage{amssymb}
\usepackage{citesort}
\usepackage{graphicx}
\usepackage{latexsym}
\newcommand{\be}{\begin{equation}}
\newcommand{\ee}{\end{equation}}
\newcommand{\bear}{\begin{eqnarray}}
\newcommand{\ear}{\end{eqnarray}}

\newsavebox{\LSIM}
\sbox{\LSIM}{\raisebox{-1ex}{$\ \stackrel{\textstyle<\sim}\ $}}
\newcommand{\lsim}{\usebox{\LSIM}}
\newsavebox{\GSIM}
\sbox{\GSIM}{\raisebox{-1ex}{$\ \stackrel{\textstyle>}{\sim}\ $}}
\newcommand{\gsim}{\usebox{\GSIM}}

\begin{document}
\begin{titlepage}
\begin{flushright}
HD-THEP-00-15\\
hep-ph/0003122
\end{flushright}
$\mbox{ }$
\vspace{.1cm}
\begin{center}
\vspace{.5cm}
{\bf\LARGE Electroweak Baryogenesis: }\\
\vspace{.3cm}
{\bf\LARGE  Concrete in a SUSY Model with a}\\
\vspace{.3cm}
{\bf\LARGE Gauge Singlet}\\
\vspace{1cm}
S.J. Huber$^{a,b,}$\footnote{shuber@udel.edu}
and
M.G. Schmidt$^{a,}$\footnote{m.g.schmidt@thphys.uni-heidelberg.de} \\

\vspace{1cm} {\em
$^a$Institut f\"ur Theoretische Physik, Philosophenweg 16,
D-69120 Heidelberg, Germany\\[.1cm]
$^b$Bartol Research Institute, University of Delaware, Newark, DE 19716, USA
}
\end{center}
\bigskip\noindent
\vspace{1.cm}
\begin{abstract}
SUSY models with a gauge singlet easily allow for a strong first order
electroweak phase transition (EWPT) if the vevs of the singlet and Higgs fields
are of comparable size.
We discuss the profile of the stationary expanding bubble wall and
CP-violation in the effective potential, in particular transitional CP-violation inside
the bubble wall during the EWPT. The dispersion relations for charginos
contain CP-violating terms in the WKB approximation. These
enter as source terms in the Boltzmann equations for the
(particle--antiparticle) chemical potentials and fuel the creation of a
baryon asymmetry through the weak sphaleron in the hot phase. This is
worked out for concrete parameters.
\end{abstract}
\end{titlepage}
\section{Introduction}
Different from models starting at the Grand Unified scale, the
ingredients for a creation of the baryon asymmetry of the universe
in a first order electroweak phase transition (PT) have a fair chance
to be tested experimentally in the near future. Besides non-equilibrium
the other two necessary criteria of Sakharov for baryogenesis, CP and
baryon number violation are naturally fulfilled by the CP violating
bubble wall of a first order PT. The bubble wall separates 
the hot symmetric phase with rapid  sphaleron transitions from 
the Higgs phase, where sphaleron transitions are suppressed by $\exp(v(t)/T)$.
However it turned out that the electroweak
Standard Model (SM) does not have a strong first order PT, indeed
there is a crossover behavior and no PT at all for Higgs masses
beyond the present experimental bound \cite{KLRS96}. Furthermore,
CP-violation by the CKM matrix is very small. In spite of
tremendous successes of the SM  it is commonly believed that it
has to be embedded / enlarged to a more fundamental theory.
Supersymmetry is supposed to be an important facet of such a theory
but it is still not backed by experiments. In concrete SUSY models, in order
to have a strong first order PT one has to strengthen the (one loop)
``$\varphi^3$'' term of the effective potential or to have such a term
already on the tree level. In the minimal SUSY extension of the SM,
the MSSM, the superpartner of the right-handed top - with a mass
below that of the top - gives such a strong $\varphi^3$ loop correction.
With a negative SUSY breaking scalar mass it is almost massless
in the hot phase and has a strong Yukawa coupling to the Higgs fields.
This is confirmed by detailed analytical \cite{BJLS97} and lattice  
calculations \cite{LR98}, showing
a strong first order PT even for Higgs masses as large as 110 GeV
just beyond the experimental bound for MSSM Higgses.
There are also possible CP-violating phases in the Higgs Lagrangian
if 1-loop effects are taken into account \cite{wagnerpil99}. They
are restricted much by the measurements of the neutron electric
dipole moment unless one invokes special conditions. A recent
investigation of a possible spontaneous CP-violation just in the
bubble wall at the  temperatures of the PT gave negative results \cite{HJLS99}.

In this paper\footnote{whose content was presented at the
TMR network meeting {\em Finite Temperature Phase Transitions in Particle Physics}
in Korfu, September 1999.}
we want to discuss baryogenesis in a strong
first order electroweak PT (EWPT) in much detail starting from a model
where all the three criteria given above can be fulfilled without much
problem. This is a supersymmetric model with an additional
gauge singlet superfield. In its original form (NMSSM) is was
designed to substitute the problematic $\mu H_1H_2$ term
of the MSSM superpotential by a coupling $SH_1H_2$ \cite{NMSSM_mu}
\begin{equation}  \label{i1}
W_{Z_3}=\lambda SH_1H_2+\frac{k}{3}S^3 \nonumber
\end{equation}
with a $Z_3$-symmetry whose spontaneous breaking causes dangerous
domain walls in the early cosmos \cite{asw}. The
soft SUSY breaking potential contains a term $\lambda A_{\lambda}SH_1H_2$
which can act as a $\varphi^3$ term if both the vevs of the
Higgses and of the singlet are of the same order of magnitude \cite{pietroni}.
But in a $Z_3$-symmetric model with universal SUSY breaking this turns
out not to be possible if one wants to have a reasonable
spectrum of particles.

In agreement with ref.~\cite{DFM96} we thus add further terms to
the superpotential which now obtains the form
\begin{equation}  \label{i2}
W=\lambda SH_1H_2+\frac{k}{3}S^3 + \mu H_1H_2 + rS
\end{equation}
and is not $Z_3$-symmetric anymore. The soft SUSY breaking
Lagrangian contains scalar masses, gaugino masses
and $A$-terms
\begin{equation}  \label{i3}
{\cal L}^{\rm soft}_A=\lambda A_{\lambda}SH_1H_2+\frac{k}{3}A_kS^3+
Y_eA_e\tilde e^c\tilde lH_1+Y_dA_d\tilde d^c\tilde qH_1
+Y_uA_u\tilde u^c\tilde qH_2+\mbox{h.c.}
\end{equation}
Thus we have reintroduced a $\mu$-term and we have the associated
fine-tuning problem known from the MSSM \cite{polonsky_mu}. Additionally, there is
the danger of quadratically divergent singlet tadpoles  
\cite{ellwrand,abel_div}. Such tadpole
diagrams require three ingredients: (i) a singlet field (ii) non-renormalizable 
interactions and (iii) soft SUSY breaking terms. (i) Giving the ``singlet'' a
charge under some discrete symmetry one can remove the tadpoles
from the very beginning. But the discrete symmetry cannot be exact in order
to avoid unacceptable domain walls at the EWPT \cite{panatam99}. (ii)
In refs. \cite{abel,abel_div} models with
gauged R-symmetry or duality symmetry, both broken at some superheavy scale
have been proposed to forbid dangerous non-renormalizable operators.
On the renormalizable level these
models typically have no discrete symmetries, and therefore
are not plagued by the domain wall problem. While such  models
do not solve the $\mu$-problem, they are still very interesting with
respect to Higgs phenomenology and the EWPT. For that reason
we will study this type of singlet model in the following.
(iii) Another way to evade the tadpole divergences restricts the
soft SUSY breaking terms. ``Gauge mediated SUSY breaking'' (GMSB)
in the context of singlet models does not have domain wall
problems. A $\mu$-parameter is generated by radiative corrections
and the singlet vev \cite{GMSB_NMSSM}. However one of the special properties of 
GMSB models seems to be the strong suppression of $A$-terms.
These however also contain the $\varphi$-type term responsible for
a strong first order PT.

In order not to have too many
parameters we use universality of SUSY breaking at the
GUT scale and run renormalization group equations to get down
to the electroweak scale.
With such a Lagrangian and adding the temperature corrections we
can demonstrate \cite{ich2.3} that one can easily get a strong first order
PT for Higgs masses as high as 115 GeV and superpartner masses
beyond present experimental limits. The effective potential in
$H_1$, $H_2$ and $S$ can be  used to derive differential equations
for the bubble wall profile and for varying CP-violating phases in the
bubble wall arising from constant explicit phases in the theory or,
more interesting, from spontaneous CP-violation. This is described in
sections 4 and 5. As worked out in section 6, in the WKB approximation
the CP-violating phases
in the bubble wall create terms in the dispersion relations which differ
between particles and anti-particles. This gives rise to a driving term
in the Boltzmann equation for the difference of particle and
anti-particle chemical potentials, in particular for left-handed
quarks and their superpartners. This difference is converted into a
baryon asymmetry by the hot phase sphalerons in front of the
proceeding bubbles.

\section{The model}
Weak scale supersymmetric models have many unknown parameters
related to supersymmetry breaking. A considerable part of this parameter
space is excluded because of FCNC or the appearance of charge and color
breaking vacua.  These problems are partially evaded by imposing
universality of the soft terms at the GUT scale, $M_{GUT}\sim 2.6 \cdot 10^{16}$
GeV, where the SM gauge couplings unify. At the GUT scale there is
a common gaugino mass $M_0$, a universal scalar mass squared $m_0^2$
and a universal trilinear coupling $A_0$. Thus, universal SUSY breaking
drastically reduces the parameters of the model to 
\begin{equation} \label{ph1}
y_{t0},\lambda_0,k_0,M_0,A_0,m_0^2,\mu_0,r_0,B_0.
\end{equation}
The subscript ``0'' indicates that the masses and couplings are
evaluated at the GUT scale. $y_t$ denotes the top Yukawa coupling\footnote{
We neglect the Yukawa couplings of the leptons and light quarks.
This approximation is well justified in the range of small and mediate
$\tan\beta$ which we consider in the following.},
and $\mu BH_1H_2$ is the soft Higgs mass term corresponding to the
$\mu$-term in the superpotential. Since the masses of the
top quark and the $Z$-boson mass are known, ``only'' seven parameters of
(\ref{ph1}) are independent.

To evolve the parameters from the GUT scale to the weak scale we use
the renormalization group equations (RGEs) in the 1-loop approximation.
For the $Z_3$-symmetric case the relevant RGEs have already been
given in ref.~\cite{dersav84}. Defining $t=\ln Q^2$ the $Z_3$-breaking
terms $\mu$, $r$ and $B$ are easily included with help of ref.~\cite{mvaughn94}
\begin{eqnarray}\label{RGE}
\frac{d}{dt}\mu&=&\frac{1}{32\pi^2}\mu(3y_t^2+2\lambda^2+2k^2
-3g_2^2-g_1^2)
\nonumber\\
\frac{d}{dt}r&=&\frac{1}{16\pi^2}r\left(\lambda^2+k^2\right)
\nonumber\\
\frac{d}{dt}B&=&\frac{1}{16\pi^2}(2\lambda^2 B+3y_t^2A_t+2\lambda^2 A_{\lambda}
                       +3g_2^2M_2+g_1^2M_1).
\end{eqnarray}

\begin{figure}[t]
\begin{picture}(200,150)
\put(-15,-440){\epsfxsize18cm \epsffile{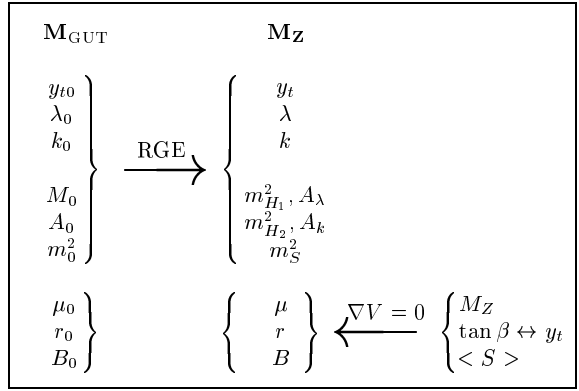}}
\end{picture}
\caption{Sketch of our procedure to fix the weak
scale parameters using RGEs and saddle point conditions.}
\label{f_elim}
\end{figure}

Even with the simple universal pattern of soft breaking terms
we are left with a 9-dimensional parameter
space. Additionally, we must satisfy two constraints coming
from the Z-boson and top quark masses. In the literature
(see e.g.~ref.~\cite{ERS}),
usually ``random shooting'' is applied to deal with such a situation.
This means randomly chosen sets of GUT scale parameters are
evolved to the electroweak scale, where their phenomenological
implications are investigated. Although the physical Z-boson mass
can easily be reproduced by an appropriate rescaling of the dimensionful
parameters, one typically is still plagued by some light unobserved
SUSY particles or an unphysical top quark mass. Therefore this
procedure is rather inefficient. Furthermore, random shooting only
provides statistical averages and correlations. To avoid these shortcomings
we use the
more systematic approach which we introduced in ref.~\cite{ich2.3}.
It allows the elimination of $\mu$, $r$ and $B$ by the Higgs
and singlet vevs, while maintaining universal SUSY breaking.
It relies on the observation that the $Z_3$-breaking terms $\mu$, $r$ and $B$
do not enter the RGEs for the remaining parameters. They can therefore
be calculated without specifying the former ones.
Our procedure consists of the following steps and is also sketched
in fig.~\ref{f_elim}:
\begin{itemize}
\item
Fix the values of $x\equiv\langle S\rangle$, $\tan\beta=v_2/v_1$, $\lambda$  
and $k$ at
the weak scale. The choice of $\tan\beta$  automatically fixes $y_t$ by the relation
$y_t v\sin\beta=m_{top}(=175$ GeV).
\item
Evolve  $y_t$, $\lambda$ and $k$ to the GUT scale.
\item
Choose a set of the parameters $M_0$, $A_0$ and $m_0^2$
at the GUT scale.
\item
Run all parameters down to the weak scale, with exception of $\mu$,
$r$ and $B$ whose initial conditions have not yet been specified at this point.
\item
At the weak scale calculate $\mu$, $r$ and $B$ by using the
saddle point conditions for the 1-loop Higgs potential
$\partial_I V(H_1,H_2,S)=0$ $(I=H_1,H_2,S)$ at $H_1^-=H_2^+=0$,  
$H^0_{1,2}=v_{1,2}$, $S=x$.
\end{itemize}
Finally one can use the corresponding RGEs
to determine the values of $\mu$, $r$ and $B$ at
the GUT scale. Appropriate values of $\mu_0$, $r_0$ and $B_0$
can always be found because of the effectively linear structure of eqs.~(\ref{RGE}). 
Our procedure has the additional benefit that the
constraints coming from the top quark and Z-boson masses,
$M_Z^2=(g_1^2+g_2^2)(v_1^2+v_2^2)/2$,
are automatically built in.
This reduces the number of free parameters by two so we are
finally left with the 7-dimensional parameter space
\begin{equation}\label{rge7}
\tan\beta,x,\lambda,k,M_0,A_0,m_0^2.
\end{equation}

Of course, not every parameter set leads to a viable phenomenology.
The constraints on the model parameters are
related to the particle spectrum and the vacuum structure.
In the elimination procedure discussed before
we assumed that the extremum parametrized
by $v$, $\tan\beta$ and $x$ is indeed the global minimum
of the scalar potential. Thus we must (numerically) verify that
there are no deeper minima in $V(H_1,H_2,S)$.
If there appears a deeper
minimum the parameter set has to be discarded.

Additional constraints arise from the required absence of
charge and color breaking minima (CCB minima) deeper
than the standard minimum, i.e.~from the absence  of squark and
slepton vevs.
We check slepton and squark vevs induced
by large trilinear couplings  \cite{alpolwi83,EH_CCB99}.
Furthermore, there are CCB minima which come from
negative scalar mass squares, typically $m^2_{H_2}$.
These dangerous directions involve Higgs, squark and
slepton fields (``UFB'' directions). The condition for such
a minimum not to be deeper than the standard minimum
implies a lower limit on the ratio $m_0^2/M_0^2$
of ${\cal O}(1)$ \cite{alsav98}. However, the decay
rate of the standard vacuum into a minimum in the
UFB direction is usually negligible compared to the
age of the universe. We therefore allow for a
meta-stable standard vacuum with respect to the UFB
directions and disregard the corresponding constraints.

Up to now, no superpartners of the SM particles have been
detected. From the experimental lower limits on the SUSY
particle masses \cite{pdg} various constraints on the
parameter space can be derived. The experimental limit
on the chargino mass $M_{\tilde \chi_1^{\pm}}>90$ GeV
\cite{ALEPH_char99}
translates into bounds on the universal gaugino
mass and the $\mu$ parameter
\begin{equation} \label{cnstr3}
|M_0| \gtrsim 100 \mbox{ GeV}, \quad |\mu+\lambda x|
\gtrsim 80 \mbox{ GeV}.
\end{equation}
If the bound on $M_0$ is satisfied the gluino mass $M_3$
is automatically above its experimental limit \cite{pdg}. In the case of
$m_0>100$ GeV also the squark and slepton masses are
compatible with the experimental data.

Of particular importance are the properties of the lightest neutral CP-even
'Higgs' mass eigenstate $h$, which is a mixture of the Higgses and the
singlet. If $h$ has a large singlet content its coupling to the Z-boson
and thus its production cross section at LEP is significantly reduced \cite{EGHRZ}. 
Parameter sets (\ref{rge7}) cannot be ruled out by simply calculating the
lightest Higgs mass $M_h$.  For example, if the Higgs production cross section
is reduced by a factor of 10 compared to the SM, Higgs masses down to
about 60 GeV are still in agreement with the data \cite{opal98}.

\begin{figure}[t]
\begin{picture}(200,150)
\put(-80,-223){\epsfxsize12.5cm \epsffile{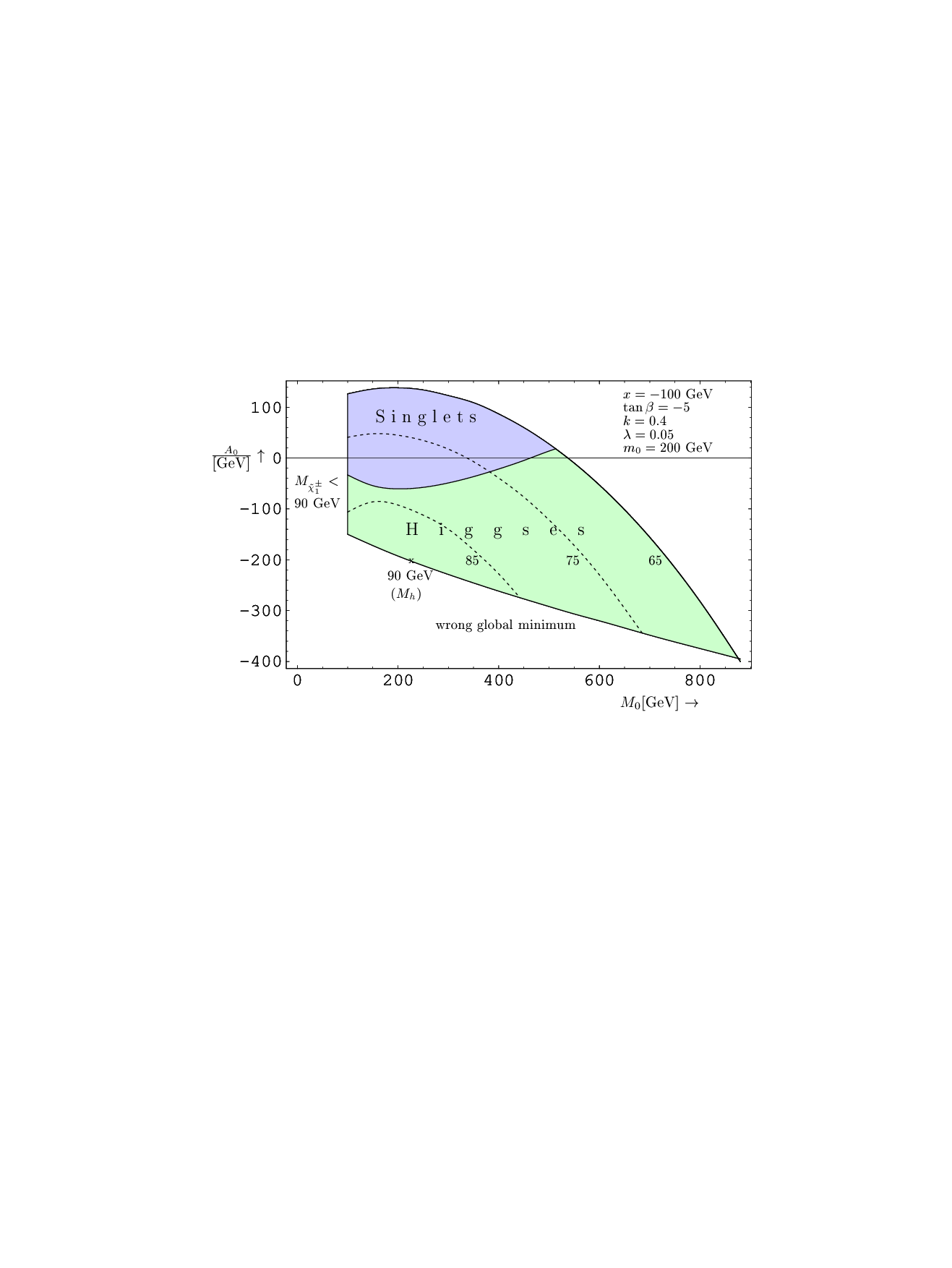}}
\put(180,-240){\epsfxsize11cm \epsffile{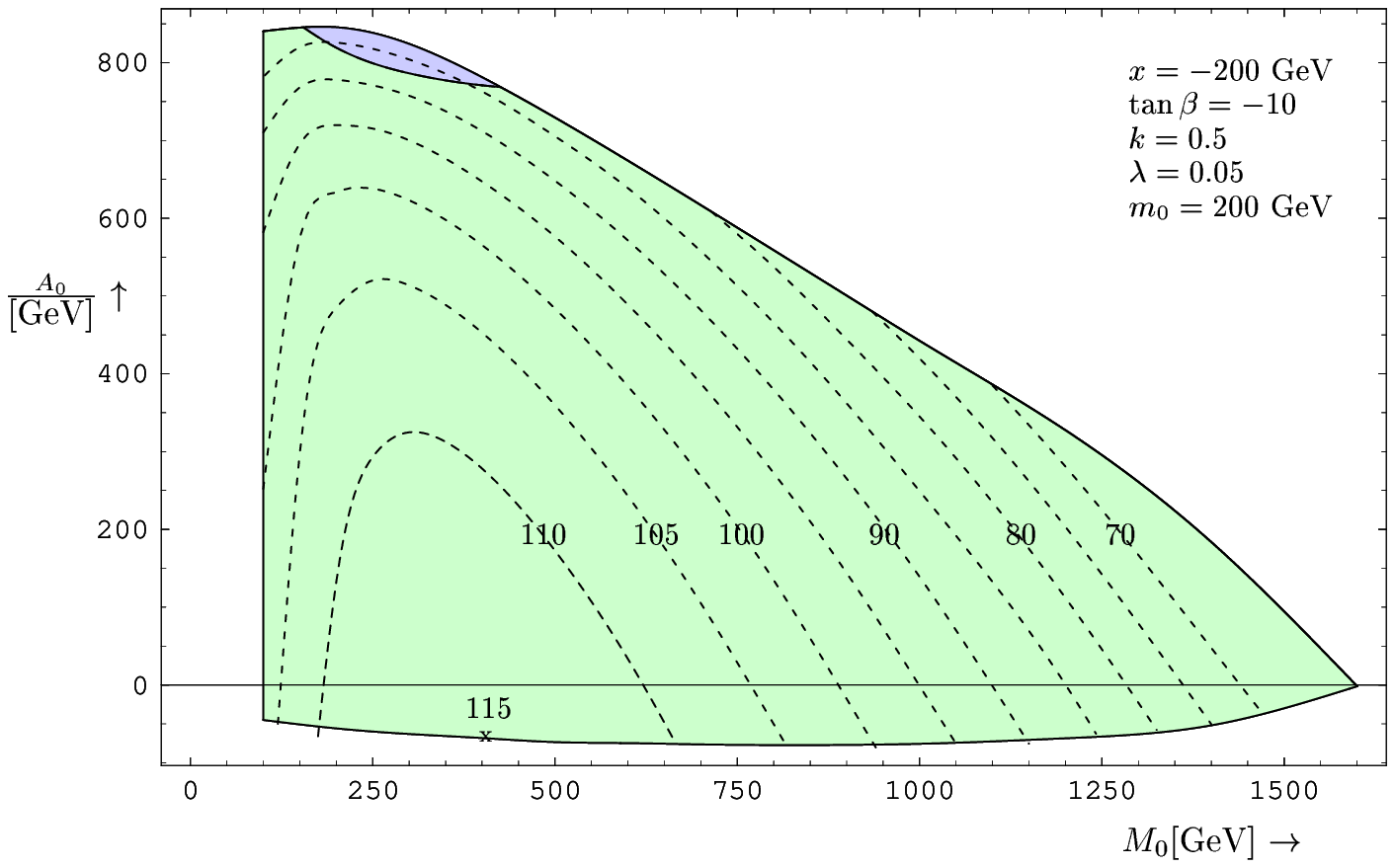}}
\put(120,136){{\footnotesize(a)}}
\put(340,136){{\footnotesize(b)}}
\end{picture}
\caption{Scan of the $M_0$-$A_0$ plane for two different
sets of $(x,\tan\beta,k)$. The colored regions represent the
phenomenologically viable range of the parameters before cuts from the
Higgs boson search are applied. The dashed
lines are curves of constant mass of the lightest CP-even Higgs
boson. In the blue (green) areas, the lightest Higgs boson
is predominantly a singlet (Higgs) state.}
\label{f_phT0}
\end{figure}

In fig.~\ref{f_phT0} we present a scan of the $M_0$-$A_0$ plane
for two different sets of $(x,\tan\beta,k)$ \cite{ich2.3}.
The parameters $M_0$
and $A_0$ are of particular interest since they determine the
values of the trilinear couplings at the weak scale 
which will play a prominent role in the discussion of the EWPT
in section 3. We have taken $m_0>100$ GeV in order avoid
light squarks and sleptons. Choosing a small value of $\lambda$
and a large value of $k$ prevents the appearance of very
light Higgs bosons.
In fig.~\ref{f_phT0}b we increased
$\tan\beta$ and the singlet vev to obtain large Higgs masses
up to 115 GeV.
The increase of $M_h$ with decreasing $A_0$
can be traced to the singlet diagonal entry in the Higgs boson
mass matrix
which is diminished by the $A_k$ contribution.
On the same lines the increasing
singlet content of the lightest CP-even Higgs state with
increasing $A_0$ can be understood. In fig.~\ref{f_phT0}b this
effect is reduced by the larger values of $|x|$ and $k$, which
render the singlet state rather heavy.

The most important constraints on $M_0$ and $A_0$ arise
from the chargino mass and from the vacuum structure.
The lower bound on $M_0$ is a consequence of eq.~(\ref{cnstr3}).
The requirement of the standard minimum being the global minimum
of the Higgs potential implies the upper and lower bounds on $A_0$
displayed in fig.~\ref{f_phT0}.
Constraints from the required absence of $A$-term induced
CCB minima are then automatically
satisfied. Note that $M_0$ is also bounded from above.
In fig.~\ref{f_phT0} we excluded parameter sets which predict
$M_h<65 $ GeV.  This leads to the upper bound on $A_0$ in  fig.~\ref{f_phT0}a.
In fig.~\ref{f_phT0}b no additional constraint arises, because of
the larger Higgs masses. On the other hand, if we would
allow for smaller Higgs masses in fig.~\ref{f_phT0}a, still an upper
bound on $A_0$ would be implied by the vacuum structure,
as is the case in fig.~\ref{f_phT0}b. Taking into account
the data from the Higgs boson search further restricts the upper
bounds on $M_0$ and $A_0$.
Roughly, the regions $M_0>600$ GeV in the $\tan\beta=-5$ set,
and $M_0>1200$ GeV in the $\tan\beta=-10$ set are additionally excluded
by the Higgs boson data \cite{opal98}. These parameter regions are
characterized by a very heavy spectrum of SUSY particles.
For that reason they are not very promising candidates for the
implementation of electroweak baryogenesis anyway, as will
be discussed in section 6.

We close this section by emphasizing some important differences
between our model and the $Z_3$-symmetric NMSSM, which
is usually considered in the literature, e.g.~in refs.~\cite{EGHRZ,ERS}.
By setting $\mu=0$
eq.~(\ref{cnstr3}) can be translated into a lower bound on the
singlet vev, $|x|\gtrsim 125$ GeV. However, the analysis of the
$Z_3$-symmetric case carried
out in ref.~\cite{ERS} shows that the actual lower bound on $|x|$
is much larger, once all phenomenological constraints are taken into
account. Typically one has a singlet vev in the multi-TeV
range and couplings $\lambda\ll1$ and $k\ll1$, with $\lambda x$ and
$kx={\cal O}(M_Z)$. In the next section we will discuss that in such
a scenario the singlet is just a spectator during the EWPT,
which proceeds more or less in the standard (i.e.~MSSM) way.
This property additionally motivates our inclusion of the $\mu$-term
in the model, while it seems questionable whether the idea of
electroweak baryogenesis can be realized in the $Z_3$-symmetric
NMSSM.  An additional peculiarity of the $Z_3$-symmetric model
is that large $A$-terms are required for successful electroweak
symmetry breaking, leading to $A_0^2>9m_0^2$. However,
this parameter range is severely constrained from the
absence of CCB minima \cite{EH_CCB99}. Some of the
special features reported in ref.~\cite{ERS} may be due
to the assumption of universal soft SUSY breaking used
in that work. It seems questionable, however,  if the
problematic bound, $|x|\gg M_Z$, can be evaded in the case of
non-universal soft terms \cite{BES95}, since the required large
values of $\lambda$ typically lead to small values of
the Higgs mass due to large Higgs singlet mixing.
The NMSSM with and without $Z_3$ symmetry also have
different properties with respect to CP-violation, as
will be discussed in section 5.

%
%
%
%
%
%
%
%

\section{Strength of the electroweak phase transition}

Order and strength of the EWPT are central questions in
electroweak baryogenesis. Only in the case of a first order phase
transition (PT) the associated departure from equilibrium is
sufficient to induce a relevant baryon number
production.
To avoid baryon number washout after the PT the even stronger
criterion $v_c/T_c \gsim1$
has to be satisfied, where $v_c$ denotes the Higgs vev at the
critical temperature $T_c$.
A first order phase transition is triggered by cubic terms in the
finite temperature effective potential \cite{dj}.
In the (MS)SM these terms
arise from 1-loop thermal corrections of bosons and therefore are small
from the very beginning.  Thus it is difficult to satisfy $v_c/T_c >1$.
In the NMSSM, on the other hand, trilinear terms enter already
the tree-level Higgs potential
due to Higgs singlet
couplings, leading to a significantly stronger EWPT
\cite{pietroni,DFM96,ich2.3}.
These contributions stem from the soft SUSY breaking
couplings $A_{\lambda}$ and $A_k$, and from the $\mu$-term
\begin{equation}  \label{strength5}
(\lambda\mu^*S+\mbox{h.c.})(|H_1^0|^2+|H_2^0|^2)+
(\lambda A_{\lambda}SH_1^0H_2^0
+\frac{k}{3}A_kS^3+\mbox{h.c.}).
\end{equation}
All these trilinear terms explicitly contain the singlet field.
Therefore, in order to induce deviations from the (MS)SM
behavior also the singlet vev must change during the phase
transition. Since at the EWPT thermal contributions to the
effective potential are of the order ${\cal O}(T^4)\sim M_Z^4$,
this requires the mass and the vev of the singlet also to be 
of the order of the electroweak scale.

At finite temperature the effective potential of the neutral Higgs
and singlet fields is modified
by the interaction with the hot plasma
\begin{equation} \label{strength1}
V_T(H_1^0,H_2^0,S)=V_{\rm tree}(H_1^0,H_2^0,S)+V^{(1)}(H_1^0,H_2^0,S)+
V_T^{(1)}(H_1^0,H_2^0,S).
\end{equation}
In the 1-loop zero temperature corrections $V^{(1)}$ we include tops, stops and 
gauge bosons, while in the 1-loop finite temperature part $V^{(1)}_T$
also Higgs bosons, neutralinos and charginos are taken into account.
We do not make a high temperature expansion, as some of the particles
can be heavy in part of the field space. Rather,   $V^{(1)}_T$ is evaluated
using a spline interpolation between the high and low temperature regions.

We stress that in the range of parameters we will
study a strong first order EWPT
is the consequence of the tree-level terms in the Higgs
potential (\ref{strength5}).
Since the stop mass will turn out to be  always larger than 200 GeV,
the thermally induced cubic terms are too small to account
for $v_c/T_c>1$ \cite{BJLS97}. The most important finite temperature
effect is the appearance of thermal effective masses
\begin{equation} \label{strength4}
      m^2\rightarrow m^2(T)=m^2(T=0)+\mbox{const}\cdot T^2,
\end{equation}
where the constant  encodes the couplings of the Higgs and singlet
fields to the particles in the plasma. It is this positive thermal contribution
to the Higgs mass that makes the symmetric phase stable at high
temperatures, and thus causes the restoration of the electroweak symmetry.
Using the complete 1-loop expression rather than the simple high
temperature approximation, is just a convenient prescription to handle
the decoupling of heavy particles. Because of the dominance of the
tree potential, we neglect
contributions to the thermal potential stemming from daisy
resummation \cite{dlhll,daisy} and 2-loop diagrams \cite{BJLS97}.

In order to determine the strength of the PT one has to
compute the critical temperature $T_c$. We define $T_c$
as being the temperature where the symmetric and the
broken minimum of the finite temperature Higgs potential
(\ref{strength1}) become degenerate.
We denote the Higgs and singlet vevs in the
broken phase by $(v_{1c},v_{2c},x_c)$.
In general, the singlet vev is different from zero even in the
symmetric phase. We refer to a PT as being ``strongly first order'',
if it avoids baryon number washout, according to the condition
$v_c/T_c>1$.
Since both Higgs fields contribute to the
sphaleron energy, $v_c$ is given by  
$v_c=\sqrt{2}\left(|v_{1c}|^2+|v_{1c}|^2\right)^{1/2}$,
where the factor of $\sqrt{2}$ is due to our choice of field
normalization.
Although CP-violation is essential for baryogenesis,
we assume CP-conservation
in this section, i.e.~all vevs, mass parameters and coupling
constants are taken real valued. Of course, we have
to verify that $V_T$ has no deeper CP-violating minimum.
Turning on phases much smaller than one would induce only
marginal changes in our results.

In this paragraph we study again the two parameter sets
already discussed in the context of fig.~\ref{f_phT0} in section 2.
We determine the critical temperature by numerical minimization of
$V_T$. After having checked
that no CP-violating minima are present we disregard
the imaginary parts of the Higgs and singlet fields.\footnote{Bubble
walls in the presence of CP-violation will be discussed in
section 5.} We
then only have to minimize $V_T$ in the real valued fields
${\rm Re}(H_1^0)$,  ${\rm Re}(H_2^0)$ and  ${\rm Re}(S)$.

\begin{figure}[t]
\begin{picture}(200,140)
\put(-74,-216){\epsfxsize12cm \epsffile{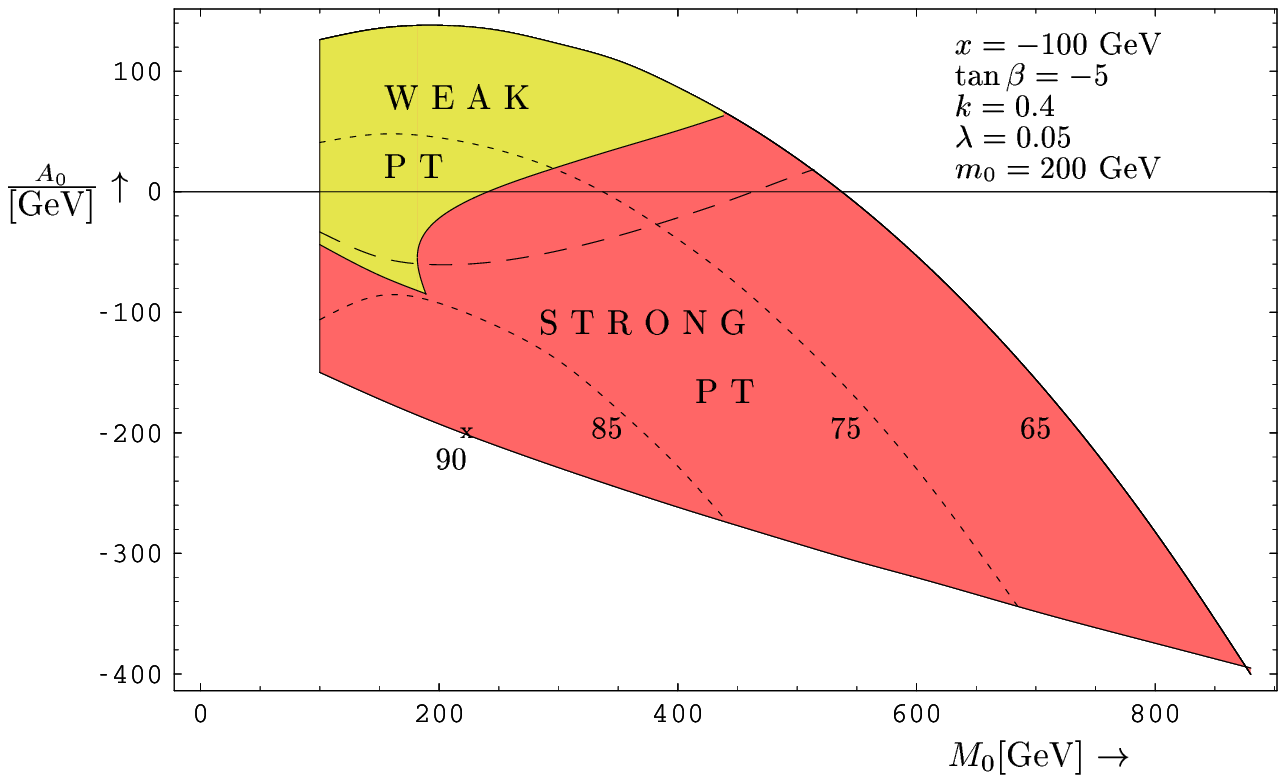}}
\put(150,-260){\epsfxsize12cm \epsffile{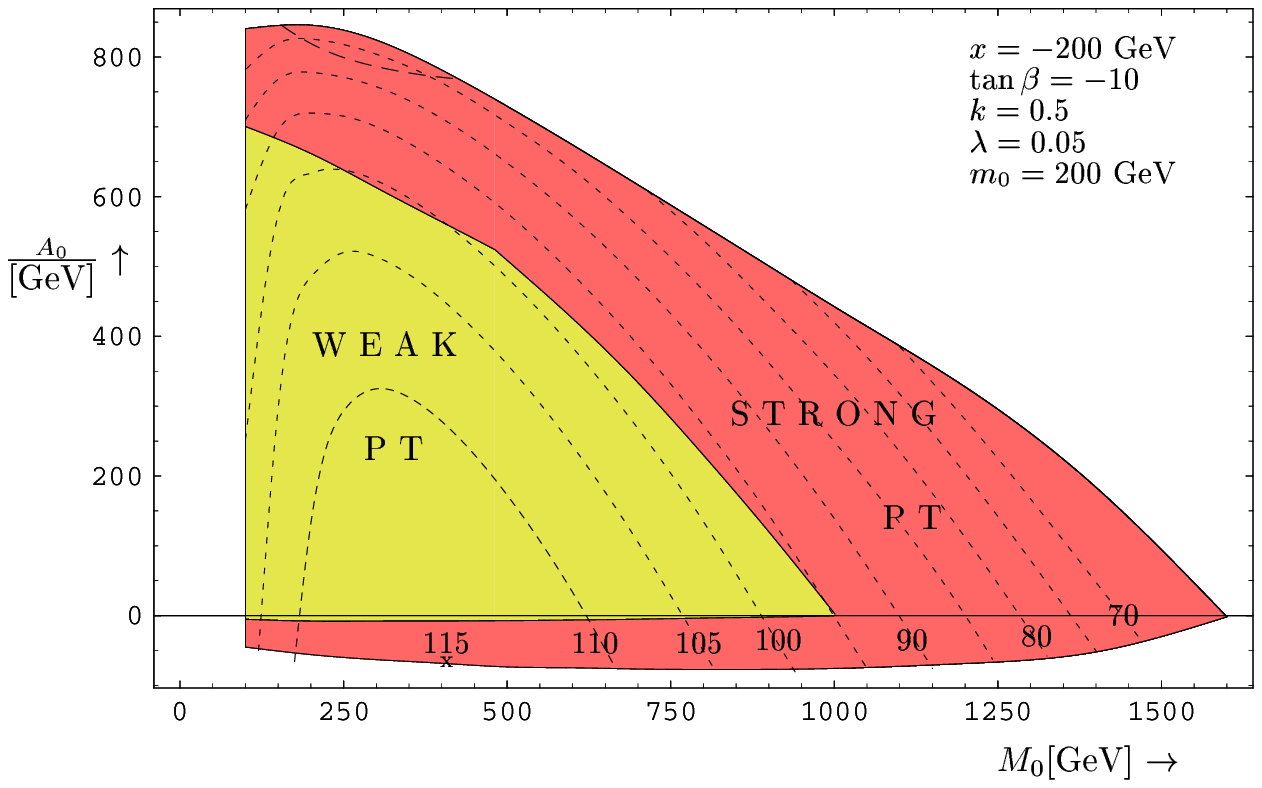}}
\put(120,131){{\footnotesize(a)}}
\put(340,131){{\footnotesize(b)}}
\end{picture}
\caption{Scan of the $M_0$-$A_0$ plane for the two different
sets of $(x,\tan\beta,k)$ which have already been considered
in fig.~\ref{f_phT0}. In the red (yellow) areas the PT is strongly
(weakly) first order, i.e.~$v_c/T_c>1$ ($v_c/T_c<1$). The dotted
lines are curves of constant mass of the lightest CP-even Higgs
boson. In the regions below (above) the dashed lines the lightest
Higgs boson is predominantly a Higgs (singlet) state.}
\label{f_AMscan}
\end{figure}

Our investigations of the strength of the PT are summarized in
fig.~\ref{f_AMscan} where the regions of strong and weak
PT in the $M_0$-$A_0$ plane are displayed for the two
parameter sets of fig.~\ref{f_phT0} \cite{ich2.3}. In case
of $\tan\beta=-5$ and
$x=-100$ GeV (fig.~\ref{f_AMscan}a) the PT is strongly
first order in most part of the phenomenologically allowed
range of parameters. However, the corresponding Higgs
masses up to 90 GeV are compatible with the experimental
Higgs mass bounds only because of the reduced Higgs
production cross section due to Higgs singlet mixing.
In the parameter set of  fig.~\ref{f_AMscan}b
we increased $|\tan\beta|$ and $|x|$ in order to obtain
larger Higgs masses. As a consequence, the region of
weakly first order PT is enlarged. However, a strong PT
occurs still for a wide range of the parameters, while Higgs
masses up to 115 GeV are consistent with $v_c/T_c>1$.

In contrast to the SM or the MSSM where the PT definitely
becomes weaker with increasing Higgs masses the situation
in the NMSSM is more complicated. Larger Higgs masses
can be related to a stronger phase transition. This may happen
for example if the broken and symmetric minima of the effective
potential are almost degenerate already at zero temperature.
The strong PT at negative $A_0$ and $M_h>100$ GeV
in fig.~\ref{f_AMscan}b results precisely from this effect.

Our results \cite{ich2.3} are in reasonable agreement
with those of
ref.~\cite{DFM96}, where the strength of the EWPT
in the general NMSSM has also been studied. The authors found that
about 50\% of their parameter sets were compatible with
$v_c/T_c>1$, and also advocated for $x\sim v$ in order to
obtain a strong phase transition. While in ref.~\cite{DFM96}
no restrictions with respect to the SUSY breaking were
made, we demonstrated that even with universal SUSY
breaking a strongly first order PT is quite natural in the
general NMSSM. Furthermore, we used updated (more restrictive)
experimental bounds to constrain the parameter space.
A further difference comes from the procedure
used in scanning the parameter space. In ref.~\cite{DFM96}
random shooting was applied, which only allowed for
statistical statements. Moreover, from the initially tested
105000 parameter sets just 2\% gave a viable zero temperature
phenomenology, leaving merely about 2000 sets
for the study of the PT. This again demonstrates the
usefulness of the systematic procedure for scanning the
parameter space, which we described in section 2.

%
%
%
%
%
%
%
%
%
%
%
%

\section{Shape of the phase boundary}
In the previous section we defined $T_c$ as being the
temperature where the symmetric and the broken minimum
of $V_T$ become degenerate. However, tunneling with
formation of bubbles of the broken phase starts at some
lower ``nucleation'' temperature, when the symmetric phase
is already meta-stable.
The bubble wall profile will crucially enter the calculation of the
baryon production during the PT which will be discussed in section 6.
Here we concentrate on CP-conserving bubble walls. The
very important case of CP-violating wall profiles will be discussed
in the next section.

At high temperatures the probability for thermal tunneling is
proportional to $e^{-S_3/T}$, where $S_3$ is the three-dimensional
action of the static field configuration $\Phi(\vec x)$ describing tunneling  
\cite{tunlinde}.
Here $\Phi$ collectively denotes the Higgs and singlet fields.
Assuming spherical symmetry the bubble configuration
(``critical bubble'') obeys the equation of motion
\begin{equation}  \label{spheric}
\frac{d^2\Phi}{dr^2}+\frac{2}{r}\frac{d\Phi}{dr}
       -\frac{\partial}{\partial \Phi}V_T(\Phi)=0.
\end{equation}
Further simplifications of eq.~(\ref{spheric}) are justified if the tunneling
occurs between two almost degenerate minima of the potential with an
energy difference $\Delta V_T$ much smaller than the height of the
potential barrier.
In such a case the radius of the bubble becomes much larger than the
thickness of the bubble wall, which thus is referred to as ``thin wall limit''. 
Neglecting therefore
the $d\Phi/dr$ term in (\ref{spheric}) we are left with
\begin{equation}  \label{dwall}
\frac{d^2\Phi}{dz^2}-\frac{\partial}{\partial \Phi}V_T(\Phi)=0.
\end{equation}
We have replaced the spatial coordinate $r$ by $z$, indicating that the
solution to (\ref{dwall}) may be viewed as a planar domain wall
with translational invariance in the directions perpendicular to the
$z$ axis.
During the period of stationary expansion the pressure
induced by the energy difference of the minima, $\Delta V_T$, is compensated
by friction \cite{moorepr,JS2000}. In the following we model this effect by taking  
the effective potential at the critical temperature.

Notice that  (\ref{dwall}) is just the classical
equation of motion of a particle moving from one maximum
of the turned around potential $-V(\Phi)$ to the other, where
is comes at rest. In this picture $z$ takes the role of the time
coordinate and $\Phi$ represents the configuration space variable.
Obviously, this a very delicate process, especially if more than
one scalar field is involved. Small changes of the initial conditions
lead to a completely different shape of the  trajectory.

For a general effective potential the bubble
wall equations have to be solved numerically. In the case of
only one scalar field the so called
``overshooting undershooting procedure'' can by applied:
The initial value $\Phi(r_0)$ is tuned such that the trajectory
approaches $\Phi=\Phi_{\rm sym}$ in the limit $r\to\infty$, which
then gives the bubble shape.  This procedure can be used for the
critical bubble as well as for the domain wall configuration.

The situation is completely different
once there are additional directions in field space. Although
in principle the shooting procedure is still applicable, in practice
the required high accuracy of the initial conditions cannot be
achieved. Thus one has to devise other numerical
methods \cite{marcos,peter,peter99,CMS99,HJLS99,peter_pr}.
They determine the bubble wall configuration as
the minimum of a functional ${\cal F}[\Phi]$  which is
built from the squared equations of motion \cite{marcos}.
Details concerning the numerical algorithms used to minimize
${\cal F}$ on a lattice are given in
refs.~\cite{marcos,peter}. It turns out that the minimization
procedure is reliable only if the starting configuration, i.e.~the
ansatz for the wall shape, differs not too much from the
actual solution. Otherwise one keeps being stuck to local minima,
which arise from discretizing the functionals  on
a lattice \cite{peter}. Thus minimization starting from
an arbitrary initial configuration is not possible up to now,
and finding an appropriate ansatz is very important.
In the case of the domain wall the kink solution
fits the exact bubble shape reasonably well, especially
if every field is allowed for having its own wall thickness
\begin{equation} \label{bub8}
\phi _i(z)=\frac{v_i}{2}\left(1-\tanh\big(\frac{z-\delta_i}{L_i}\big)\right).
\end{equation}
In general, different off-sets $\delta_i$ which shift the fields
against each other are possible as well. For the domain wall case
the mechanical analogue to energy conservation,
$E(z)=\frac{1}{2}(d\Phi/dz)^2-V(\Phi)$ = const, provides a very sensitive
criterion to check the quality of the numerically obtained solution.

In general, the numerical methods developed in
refs.~\cite{marcos,peter,peter99,HJLS99,CMS99} are inevitable to
determine the bubble shape in multi-field models. But if the
problem is effectively one dimensional some ``improved shooting method''
can still be applied.
The investigation of CP-conserving bubble wall shapes in the MSSM
\cite{marcos,peter}
revealed that for realistic Higgs masses the variation of $\tan\beta$ in
the bubble is very small, $\delta\beta\sim{\cal O}(10^{-2}-10^{-3})$.
Thus the bubble is very accurately described by taking only the combination
of the Higgs fields, $H$, which corresponds to the direction of the broken
minimum. Simple shooting along this direction provides excellent
values for the wall thickness, the surface tension or the action of the
critical bubble. Even the small variation in $\beta$ can be reliably determined by
minimizing the effective potential along the direction perpendicular to $H$.
The line which connects these minima gives a good approximation to the
trajectory in field space the actual bubble solution is corresponding to.
This can be checked by using the ``energy conservation'' criterion.
The kink ansatz turns out to be a good fit for bubble profile.

In case of the NMSSM the straight connection between the
broken and the symmetric minimum is no longer a reasonable
approximation to the true bubble wall trajectory. Ignoring
the variation of $\tan\beta$, which turns out to be small and hence can
be included afterwards, we effectively have a two field system,
$H$ and $S$. In a large part of the parameter space
(as long as $A_0$ is significantly above its lower bound shown in
fig.~\ref{f_phT0}) the effective potential is characterized by
a distinct smooth ridge which connects the symmetric and
the broken minimum. This ridge $S=F(H)$  can be
determined by minimizing the potential along the direction
perpendicular to the straight connection of the symmetric
and broken minima.\footnote{Numerically, the following
method leads to better results: Determine the saddle point
along the ridge, i.e. $\partial_H V_T=\partial_S V_T=0$. The ridge
is then defined as the trajectory of an over-damped particle
rolling from the saddle point down to the (symmetric or
broken) minimum, satisfying $d\Phi/dz+\partial_{\Phi} V_T=0$.}
Using $V_T(H,F(H))$ the set of
bubble wall equations (\ref{dwall}) reduces to only one differential
equation which can again be solved via the shooting method.
However, not every function $F$ is a possible candidate
for the actual bubble wall. Since $dH/dz$ vanishes in the symmetric
and broken phase, integrating  the bubble wall equation results in the
constraint
\begin{equation} \label{bub13}
\int_0^{H_{\rm brk}}dH\frac{\partial}{\partial H}V_T\big(H,F(H)\big)=0.
\end{equation}
In order to evade this inconsistency
we deform our ansatz by  $\tilde F(H)$=$F(H)$+$fH(H$--$v_c)$,
where the free parameter $f$ is fixed by eq.~(\ref{bub13}).
This ansatz is motivated by the mechanical analogue of the bubble
wall problem: if a ball rolls from one maximum of $-V_T$ to the
other, it moves a bit below the ridge in order to compensate for the
centrifugal force. On half way between the two maxima the
centrifugal force is most efficient and the trajectory $\tilde F$
deviates maximally from the ridge $F$. In practice, $\tilde F- F$
is small compared to the deviation of $F$ from the straight line.

Instead of expressing $S$ in terms of $\tilde F(H)$ one could just as
well eliminate
$H$ by $\tilde F^{-1}(S)$. If $\tilde F$ corresponds to the trajectory of the
actual bubble configuration the two prescriptions are equivalent.
For a general $\tilde F$ the two prescriptions will lead to different
results for the wall profile. The difference between the two
solution indicates the quality of the chosen $\tilde F$.
We emphasize that although several steps are necessary
to compute the bubble wall shape via this improved
shooting approach, the required numerics
is fairly simple. It can by carried out using computer
algebra systems like Maple or Mathematica.
The results are competitive to those obtained with the sophisticated
minimization algorithms of refs.~\cite{marcos,peter}, which
on the other hand were very helpful to check the validity
of our approximations.

\begin{figure}[t]
\begin{picture}(200,180)
\put(-140,-405){\epsfxsize17cm \epsffile{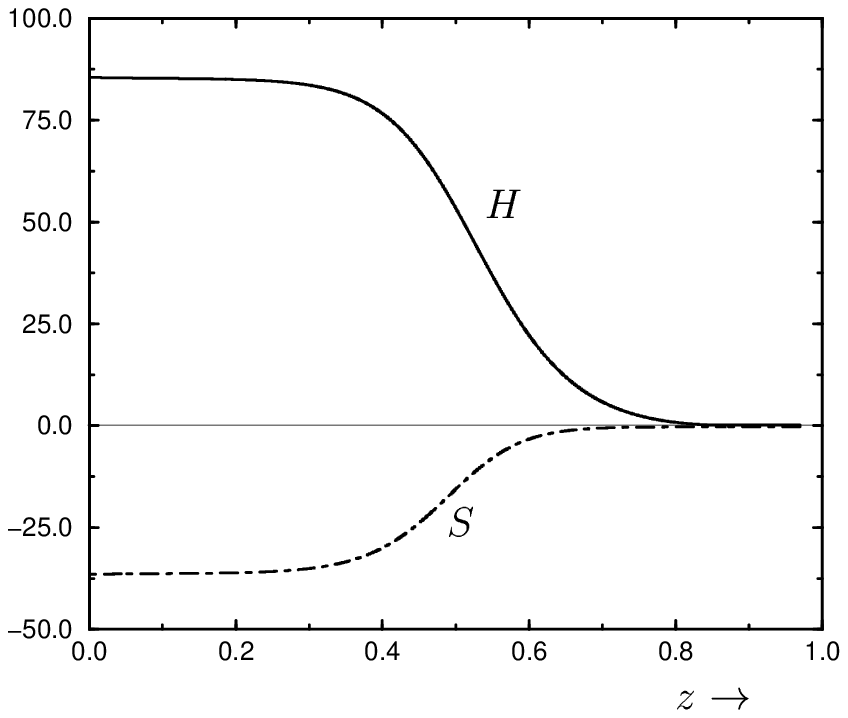}}
\put(92,-405){\epsfxsize17cm \epsffile{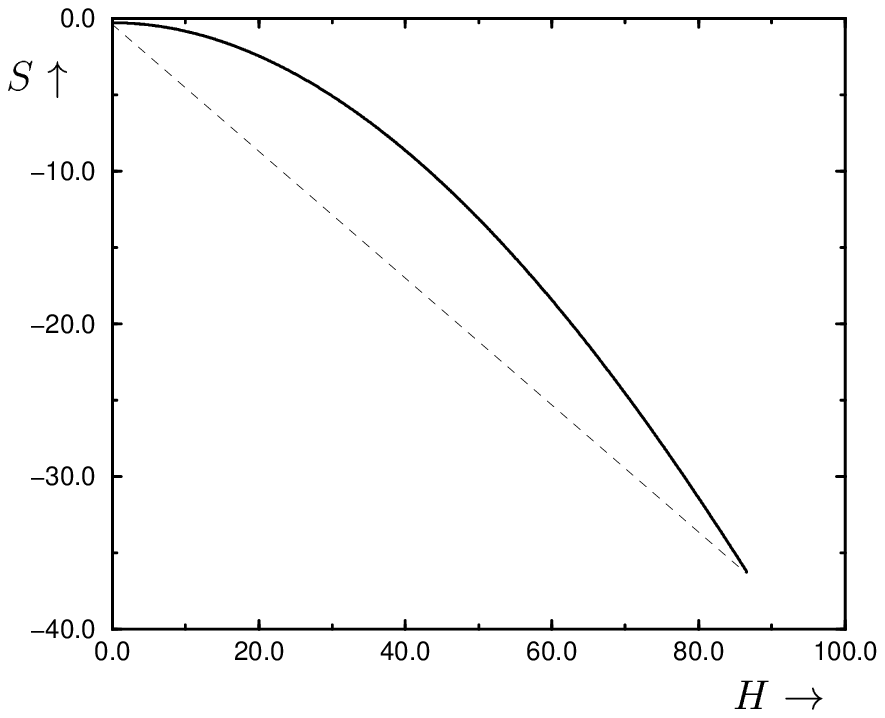}}
\put(175,153){(a)}
\put(407,153){(b)}
\end{picture}
\caption{
(a): Bubble wall profile at the critical temperature
for the parameter set of fig.~\ref{f_phT0}a with $M_0=300$ GeV and $A_0=0$.
(b): The corresponding trajectory in the $H$-$S$ plane (solid line)
and the straight connection between the symmetric and the
broken minimum (dashed line). (All units in GeV.)
}
\label{f_dwall}
\end{figure}

Let us discuss a specific example from the parameter set of fig.~\ref{f_phT0}a
with $M_0=300$ GeV and $A_0=0$. It corresponds to a
lightest Higgs mass $M_h=80$ GeV, which because of the reduced
coupling to the Z-boson is compatible with the experimental data.
We have $T_c=109$ GeV and $v_c/T_c=1.12$, i.e.~the washout of baryon
number after the phase transition is avoided.
Using our improved shooting method we find the bubble wall
profile displayed in fig.~\ref{f_dwall}a. In fig.~\ref{f_dwall}b we show
the corresponding trajectory in field space which considerable
deviates from a straight line.
Fitting the numerical solutions by the kink ansatz (\ref{bub8}) we
obtain the wall thicknesses for the Higgs and singlet fields,
$L_h=0.13$ GeV$^{-1}=14/T_c$ and $L_s=0.10$ GeV$^{-1}=11/T_c$, respectively.
For the surface tension we find $\sigma=46300$ GeV$^3$. The energy conservation 
check gives $\Delta E=6000$ GeV$^4$ which has to be compared to the barrier height 
$V_b\sim125000$ GeV$^4$. This is the same level of accuracy \cite{peter_pr}
which is accessible in the minimization approach of ref.~\cite{peter}.
On the other hand, taking the simplest ansatz for $\tilde F$,
the dashed straight line in fig.~\ref{f_dwall}b,
results in $L_h=L_s=8/T_c$
and $\sigma=82200$ GeV$^3$, which is already off by about a factor of two.

\begin{figure}[t]
\begin{picture}(200,180)
\put(-140,-405){\epsfxsize17cm \epsffile{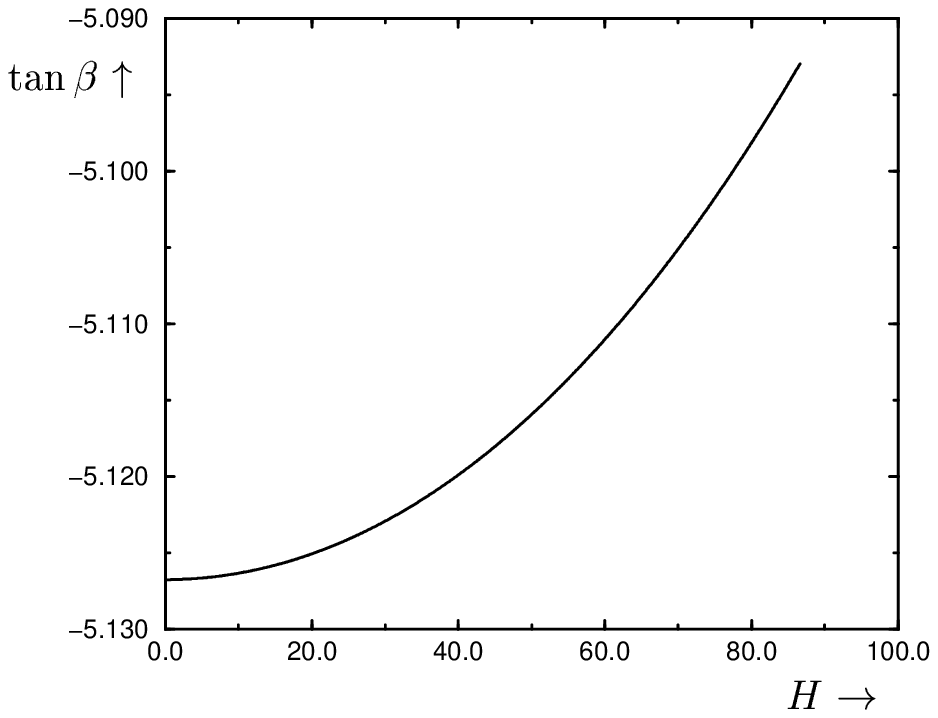}}
\put(92,-405){\epsfxsize17cm \epsffile{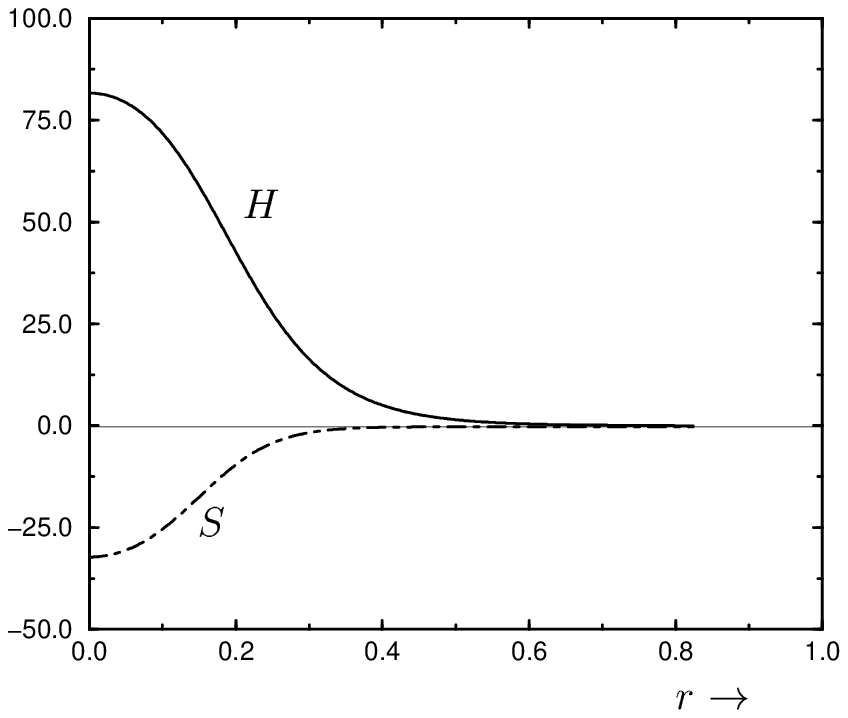}}
\put(35,153){(a)}
\put(405,153){(b)}
\end{picture}
\caption{
(a): Variation of $\tan\beta$ in the bubble wall at the critical
temperature for the parameter set of fig.~\ref{f_dwall}.
(b): Shape of the critical bubble for the same parameter
set at $T=108.5$ GeV. (Units in GeV.)
}
\label{f_dtb}
\end{figure}

Up to now we neglected the variation of $\tan\beta$ in the bubble
wall which has a strong impact on some sources for baryogenesis,
as will be discussed in section 6. In order to include this effect in our
calculation we simply minimize the effective potential in the direction
orthogonal to $H$ while keeping $H$ and $S$ fixed, implicitly
assuming that it is only a small perturbation to the trajectory $\tilde F(H)$.
In fig.~\ref{f_dtb}a we show the variation of $\tan\beta$
for the bubble wall of fig.~\ref{f_dwall}. We obtain $\delta\beta=1.2\times10^{-3}$,
i.e.~the assumption of $\delta\beta$ being a small perturbation is very
well justified. Our result is in complete agreement \cite{peter_pr} with
the one computed by using the minimization technique of ref. \cite{peter}.
Moreover, we find that the variation of $\tan\beta$ in the NMSSM and
MSSM \cite{marcos} are of the same order of magnitude.
Particularly, the singlet field provides no additional sources
for $\delta\beta$ which can be traced to its equal couplings to both
Higgs fields.

Finally, our improved shooting method can be used to
determine the shape of the critical bubble (\ref{spheric}).
The first bubbles nucleate when the action of the critical bubble
satisfies $S_3(T_n)/T_n\sim 130-140$ \cite{MSTV91}. In the computation
of the $T_n$ the small $\tan\beta$ variation in the wall can
be safely ignored.
Considering again the parameter set of fig.~\ref{f_dwall} we obtain
the bubble configuration shown in fig.~\ref{f_dtb}b, where $T=108.5$ GeV.
Since $T_c=109.2$ GeV the super-cooling amounts to 0.7 GeV,
the same order of magnitude as in the MSSM \cite{marcos}.
Notice that the wall shape is still very similar to a kink solution (\ref{bub8}), 
which is cut off at $r=0$. Calculating the corresponding action we get
$S_3/T=134$. Thus we are at the nucleation temperature.
Note that the thin wall approximation $S_3=\frac{16Pi}{3}\sigma^3/(\Delta V_T)^2$,
leading to $S_3/T=94$ is not very accurate in this case.
Here $\Delta V_T=4.04\times 10^5$ GeV$^4$ is the potential barrier.
Since  in our example bubble radius and wall thickness are of
comparable size this difference is not surprising.

We emphasize that our method to determine bubble shapes
is restricted to cases where the bubble trajectory is associated with
a smooth ridge in the effective potential,
which correspond to values of $A_0$ significantly above its lower bound.
Otherwise one is referred
to the sophisticated minimization methods of refs.~\cite{marcos,peter,CMS99}.

%
%
%
%
%
%
%
%
\section{CP-violating bubble walls}
Having established a strongly first order EWPT in the NMSSM,
successful electroweak baryogenesis still depends crucially on the
available amount of CP-violation. In models containing two Higgs
doublets, such as the MSSM or the NMSSM, complex vevs of the
Higgs (and singlet) fields provide additional sources of CP-violation.
Complex vevs can either arise
spontaneously or are induced by explicitly CP-violating couplings.
In this section we discuss their impact on bubble wall profiles in the
NMSSM. Most interestingly, we find that CP-violation may be restricted
to the phase transition itself, a phenomenon which is
called ``transitional CP-violation''. In particular, this scenario
allows for a large amount of CP-violation during the process
of baryon production, while being completely unconstrained
by experiment.

To start with, let us summarize the case of CP-violation at zero temperature.
In the MSSM the tree-level Higgs potential automatically conserves CP.
However, explicit CP-violation emerges from
possible complex phases of the soft SUSY breaking $A$-terms
and gaugino masses, and from the $\mu$-parameter \cite{wagnerpil99}.
These phases contribute to the electric dipole moments (EDMs)
of quarks and electrons
\cite{BWPW83}. From the experimental
upper limits \cite{EDM_exp} on the neutron EDM, $d_n<1.1\times 10^{-25}e$ cm,
and the electron EDM, $d_e<4.3\times 10^{-27}e$ cm, constraints on
the supersymmetric phases can be derived. Conservatively,
rather small supersymmetric phases of the order ${\cal O}(10^{-2}-10^{-3})$
are required to satisfy the experimental bounds \cite{BWPW83}. Larger
phases may only be tolerated if the first and second
generation squarks have masses in the TeV range \cite{YKNO90}, or if
accidental cancellations do occur. Recently it was realized that
these cancellations are more generic than thought previously
\cite{CP_canc}.
Explicit CP-violation occurring in the SUSY breaking
sector of the  MSSM is communicated to the Higgs sector by
radiative corrections. However, the complex phases induced in
the Higgs vevs are much too small to have any phenomenological
implications \cite{pom92}. Nevertheless, Higgs phenomenology
may be significantly changed due to mixing of the CP-even
and CP-odd Higgs states induced by complex mass parameters
and coupling constants \cite{wagnerpil99}.

The Higgs sector of the $Z_3$-symmetric NMSSM contains
one not removable phase which can be chosen to be the
phase of $\lambda k^*$. This phase is not very much constrained provided
$|\lambda|$ is significantly smaller than one \cite{mattan95}.
If we allow for $Z_3$-symmetry breaking, six possibly complex
parameters appear in the tree-level Higgs potential.\footnote{
In this section we relax the assumption of universal soft SUSY breaking
which would induce correlations between the various phases at the
weak scale.}
Two phases can be absorbed by a redefinition of the Higgs and singlet
fields. Thus we can take $\mu B+\lambda r^*$ and $kr^*$ being real valued,
without loss of generality. We parametrize the Higgs and singlet
fields according to
\begin{equation}\label{CPV1}
H_1^0=\bar h_1e^{i\theta_1}, \quad  H_2^0=\bar h_2e^{i\theta_2}, \quad
 S=\bar se^{i\theta_S},
\end{equation}
and define $\mu=\bar\mu e^{i\phi_{\mu}}$, $\lambda=\bar\lambda
e^{i\phi_{\lambda}}$, etc. Furthermore,  we introduce
\begin{equation}  \label{CPV2}
\theta=\theta_1+\theta_2, \quad \bar \theta=\theta_1-\theta_2.
\end{equation}
Because of gauge invariance the Higgs potential is independent
of the phase combination $\bar\theta$.  Using these definitions
the tree-level Higgs potential takes the form
\begin{eqnarray}  \label{CPV3}
V_{\rm tree}&=&
\big(\bar\mu^2+\bar\lambda^2\bar s^2+2\bar\lambda\bar\mu \bar s
\cos(\theta_S+\phi_{\lambda}-\phi_{\mu})\big)(\bar h_1^2+\bar h_2^2)
+\bar\lambda^2\bar h_1^2\bar h_2^2+ \bar k^2 \bar s^4
\nonumber \\
&&+2\bar\lambda\bar k\bar s^2\bar h_1\bar h_2\cos(\theta -2\theta_S
+\phi_{\lambda}-\phi_k) +\frac{g_1^2+g_2^2}{8}(\bar h_1^2-\bar h_2^2)^2
\nonumber \\
&&+2(\lambda r^*+\mu B)\bar h_1\bar h_2\cos\theta+2kr^*\bar s^2\cos(2\theta_S)
+m_{H_1}^2\bar h_1^2+m_{H_2}^2\bar h_2^2+m_S^2\bar s^2
\nonumber \\
&&+2\bar\lambda\bar A_{\lambda} \bar s \bar h_1 \bar h_2
\cos(\theta+\theta_S+\phi_{\lambda}+\phi_{A_{\lambda}})
+\frac{2}{3}\bar k\bar A_k \bar s^3
\cos(3\theta_S+\phi_k+\phi_{A_k}).
\end{eqnarray}
Note that in contrast to the MSSM the phase of the $\mu$-parameter
enters $V_{\rm tree}$. Explicit CP-violation automatically generates
complex vevs for the Higgs and singlet fields. In the general
NMSSM this is a tree-level effect which should be taken into account.
An example will be discussed below.
EDM constraints on CP-violating phases in the
general NMSSM depend on the coupling of these phases to
the (MS)SM sector.
Phases which directly enter the squark, chargino and neutralino
mass matrices, e.g.~$\phi_{\mu}$ and $\langle \theta \rangle$, have to
obey the MSSM bounds which have been discussed above. The other
phases, e.g.~$\phi_k$ and $\langle \theta_S \rangle$, are
less constrained, provided the coupling to the MSSM sector, $\lambda$,
is sufficiently small \cite{mattan95}.

Even when the Lagrangian is CP-conserving, the CP-symmetry
may be spontaneously violated by complex scalar vevs, i.e.
CP and the $SU(2)\times U(1)$ gauge symmetry are broken
together at the EWPT. In the MSSM
radiative corrections to the Higgs potential can induce a CP-violating
vacuum. However, the required Higgs boson with mass of a few GeV
is obviously ruled out by experiment \cite{pom92}.
The $Z_3$-symmetric NMSSM provides only limited improvement
on the minimal model: radiative corrections generate spontaneous
CP-violation, provided the lightest neutral Higgs mass is smaller
than about 40 GeV \cite{babubarr94}. In the general NMSSM
the situation is very different. Spontaneous CP-violation (SCPV)
is possible even at tree-level. However, Higgs spectra consistent
with the experimental Higgs mass bounds require nearly
maximal CP-violation, i.e.~$\langle \theta \rangle$, $\langle \theta_S \rangle$
of the order ${\cal O}(1)$ \cite{schotten_CP}. Phases $\theta \sim 0.1$
are only possible for Higgs masses smaller than 30 GeV.
This finding is related to the fact that CP is a discrete symmetry.
If CP is spontaneously broken there arise two degenerate
minima\footnote{For that reason explicit CP-violation
has also to be present in the model in order to avoid domain
walls when CP is broken at the EWPT. However, this
explicit phase may be much smaller than one.}
with phases $\pm \langle \theta \rangle$, which are
nearby when $\langle \theta \rangle$ becomes small. From the
SM Higgs potential it is known that the squared mass of the
Higgs boson is proportional to $\lambda' \langle h\rangle^2$,
where $\lambda'$ denotes the quartic coupling. In the case of
small SCPV we can identify $\lambda' \langle h\rangle^2\rightarrow
\lambda' v^2 \langle \theta \rangle^2$, which up to numerical
prefactors of order unity is the mass of the emerging light, 
almost CP-odd Higgs boson. Obviously, for small 
$\langle \theta \rangle$ this state becomes arbitrarily light.

Thus SCPV in the NMSSM does not appear to be very promising. Small
phases are ruled out because they require a light Higgs boson,
while large phases are tightly constrained by the EDM
experiments. We will see below that this conclusion
does not apply to SCPV which is only present during the PT.

We now derive the equations of motion for moduli and phases of the
Higgs fields.\footnote{The singlet field is most conveniently split into
real and imaginary part which can be treated along the lines discussed
in the previous section.}
Using the definitions (\ref{CPV1}) and  (\ref{CPV2}) we can
rewrite the kinetic terms for the Higgs bosons in the Lagrangian
according to
\begin{equation} \label{thlag}
{\cal L}=\partial_{\mu}\bar h_1\partial^{\mu} \bar h_1+\partial_{\mu}\bar
h_2\partial^{\mu} \bar h_2
+\frac{\bar h_1^2+\bar h_2^2}{4}(\partial_{\mu}\theta\partial^{\mu}\theta+
\partial_{\mu}\bar\theta\partial^{\mu}\bar\theta)
+\frac{\bar h_1^2-\bar h_2^2}{2}\partial_{\mu}\theta\partial^{\mu}\bar\theta+\dots.
\end{equation}
Since $\partial_{\bar\theta}V_T=0$ the Euler-Lagrange equation
for $\bar\theta$ implies
\begin{equation} \label{thconst}
 (\bar h_1^2+\bar h_2^2) \partial^{\mu}\bar\theta+ (\bar h_1^2-\bar h_2^2)
\partial^{\mu}\theta=c_{\theta}^{\mu}={\rm const}.
\end{equation}
In general, the  integration constant $c_{\theta}^{\mu}$
cannot be set to zero without introducing a pure gauge field.
However, a non-vanishing $c_{\theta}^{\mu}$  would enhance the
energy of a bubble configuration, so we
will set $c_{\theta}^{\mu}=0$ in the following. After elimination
of $\partial_{\mu}\bar\theta$ by help of (\ref{thconst}) the equations of
motion for $\bar h_1$, $\bar h_2$ and $\theta$ take the form
\begin{eqnarray}
\label{h1eq}
2\frac{d^2}{dz^2}\bar h_1+\frac{2\bar h_1\bar h_2^4}{(\bar h_1^2+\bar h_2^2)^2}
\frac{d^2}{dz^2}\theta-\frac{\partial}{\partial \bar h_1}V_T=0 \\
\nonumber \\
2\frac{d^2}{dz^2}\bar h_2+\frac{2\bar h_2\bar h_1^4}{(\bar h_1^2+\bar h_2^2)^2}
\frac{d^2}{dz^2}\theta-\frac{\partial}{\partial \bar h_2}V_T=0 \label{h2eq} \\
\nonumber \\
\label{thetaeq}
\frac{d}{dz}\left[\frac{2\bar h_1^2 \bar h_2^2}{\bar h_1^2+\bar  
h_2^2}\frac{d}{dz}\theta\right]
-\frac{\partial}{\partial \theta}V_T=0
\end{eqnarray}
where we restricted ourselves to the case of the static domain
wall perpendicular to the $z$-direction (\ref{dwall}). The equations
for the critical bubble follow along the same lines. In
eqs.~(\ref{h1eq}) - (\ref{thetaeq}) the phase $\theta$ is dynamical
only if both moduli, $\bar h_1$ and $\bar h_2$, are different
from zero. This demonstrates that in the bubble wall
$\theta$ cannot be assigned to one of the Higgs fields.

In the previous section we noted that $\tan\beta$ varies only
slightly in the bubble wall. Neglecting this small perturbation
the equations (\ref{h1eq}) - (\ref{thetaeq}) can be further reduced
and we obtain for the variation of the phases of the Higgs fields
$H_1^0$ and $H_2^0$
\begin{equation}  \label{CPbub5}
\delta \theta_1=\sin^2(\beta_T) \delta \theta, \quad
\delta \theta_2=\cos^2(\beta_T) \delta \theta.
\end{equation}
Here $\tan\beta_T$ represents the ratio of the Higgs field
moduli in the broken minimum at the critical temperature.
In the limit of large $\tan(\beta_T)$ the variation of $\theta$
is almost completely absorbed by $\theta_1$, while $\theta_2$
remains more or less constant. For example, in case of
$\tan\beta_T=5$ we obtain $\delta \theta_2=0.038 \delta \theta$, which
is small even for $\delta \theta ={\cal O}(1)$. In section 6 we will
discuss that the variations of complex phases fuel baryon production,
rather than the complex phases themselves. As a consequence, baryon
number generation resulting from a varying phase in the top quark
mass, $m_t=y_t\sin (\beta_T)\bar h e^{i\theta_2}$, will turn out to be
highly suppressed in the limit of large $\tan(\beta_T)$. Since in the region
of the parameter space considered in this work the Higgs vev
ratio changes only by a few percent when $T$ is raised from zero to
the critical temperature, this is the case in the regime of large $\tan\beta$.    

We now discuss the implications of the general equations of motion
derived above. Let us start with explicit CP-violation.
In the MSSM the complex phases which are induced in
the Higgs vevs by CP-violating couplings are completely
negligible for realistic Higgs masses. In ref.~\cite{HJLS99}
we found this behavior confirmed at finite temperature:
$\theta \lesssim {\cal O}(10^{-3})$, even for explicit phases
of the order of one. The variation of $\theta$ in the
bubble is of the same order of magnitude. We conclude
that in the MSSM with explicit
CP-violation, the Higgs vevs can be taken real to very
good approximation.

In the NMSSM the situation is different, because
explicit CP-violation is possible in the tree-level Higgs
potential (\ref{CPV3}).
In general, the behavior of the complex valued Higgs and singlet
fields in the bubble wall can only be reliably determined
by using the minimization techniques of ref.~\cite{peter}.
In order to make this possible, the equations of motion
which describe s=Re$(S)$, c=Im$(S)$,
$\bar h_1$, $\bar h_2$ and $\theta$ have to be included
in the ${\cal F}$. In ref.~\cite{HJLS99} this
approach was applied to compute CP-violating bubble
wall profiles in the MSSM and NMSSM. It turned out that
also the phase $\theta$ is rather accurately described
by a kink-ansatz (\ref{bub8}). Moreover, it was verified
that in the calculation of the wall shape small CP-violating phases
$\theta \lesssim {\cal O}(10^{-1})$
can indeed be treated as perturbations of a CP-conserving
solution. In this case the bubble wall profile can be conveniently
determined in two steps: first, one computes the CP-conserving
profile of the fields $\bar h_1$, $\bar h_2$ and $s$,
where the CP-violating fields $\theta$
and $c$ are set to zero. In a second step, the CP-violating
constituents of the bubble wall may be computed by
simply minimizing the effective potential $V_T(\bar h_1, \bar h_2,\theta, s,c)$
with respect to $\theta$ and $c$, while keeping $\bar h_1$,
$\bar h_2$ and $s$ fixed. The variation of the Higgs
vev ratio can be neglected in this calculation.
We emphasize that this prescription works only in case
of small CP-violation in the Higgs and singlet fields.

If $V_T(\bar h_1, \bar h_2,\theta$$=$$0, s,c$$=$$0)$ has a smooth ridge,
the profile of the CP-conserving fields can be obtained from the
improved shooting method discussed in section 4.  Hence the
complete CP-violating bubble profile can be computed without
minimizing the functional ${\cal F}$.

\begin{figure}[t]
\begin{picture}(200,180)
\put(-140,-405){\epsfxsize17cm \epsffile{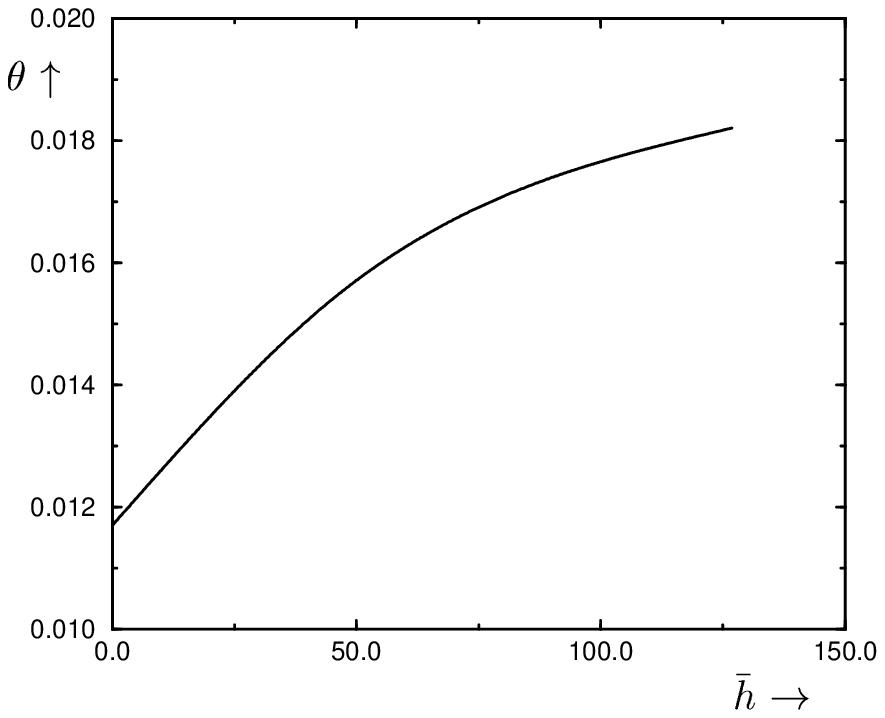}}
\put(92,-405){\epsfxsize17cm \epsffile{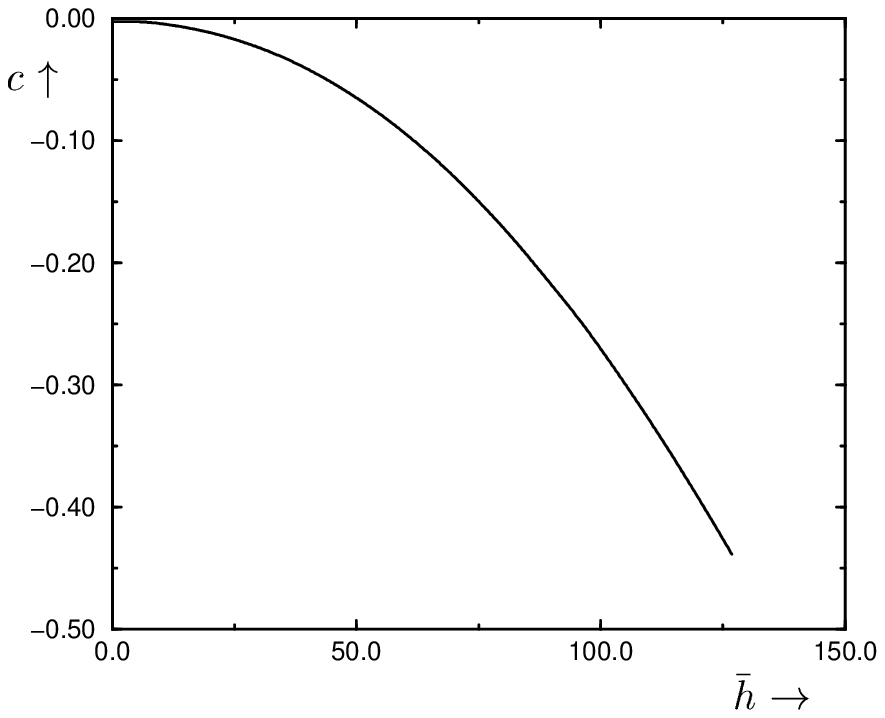}}
\put(35,153){(a)}
\put(405,153){(b)}
\end{picture}
\caption{
(a): Variation of the CP-violating fields $\theta$ (a)
and $c$ (b) in the bubble wall as a function of $\bar h=\sqrt{\bar h_1^2+\bar  
h_2^2}$  at the critical
temperature for $\phi_{\mu}=0.1$, $A_0=-100$ GeV and
$M_0=125$ GeV. The remaining parameters are chosen
as in fig.~\ref{f_phT0}a. (Units in GeV.)
}
\label{f_ECPV}
\end{figure}

In fig.~\ref{f_ECPV} we display an example of a bubble wall in presence
of explicit CP-violation using
$\phi_{\mu}=0.1$, $A_0=-100$ GeV and $M_0=125$ GeV, the remaining
parameters are chosen as in fig.~\ref{f_phT0}a. The lightest Higgs boson
has a mass $M_h= 86$ GeV.
We find $T_c=101$ GeV and $v_c/T_c=1.77$, so the PT is strongly first order. 
For the wall thickness we obtain $L_w=5/T$. 
The CP-even singlet field evolves from $s=115$ GeV in the symmetric phase
to  $s=-79$ GeV in the Higgs phase. From fig.~\ref{f_ECPV}a we take
$\theta=0.0182$ in the broken phase, and $\delta\theta=0.0065$.
In the MSSM, $\phi_{\mu}\sim 1$ would be required to induce
an effect of the same magnitude. Also the CP-odd component of the
singlet field, $c$, varies in the bubble wall, as is shown in fig.~\ref{f_ECPV}b.
In the symmetric phase we have $c=0$, since  $\phi_{\mu}$ couples
to the singlet only in case of non-vanishing $\bar h_{1,2}$.

We now turn to the case of spontaneous CP-violation. Although this
scenario is disfavored at zero temperature, the  Higgs and singlet fields
may still acquire large complex phases at high temperatures,
or even only during the EWPT. Since it will turn out in section 6
that explicit CP-violation can account for the observed baryon asymmetry
only in special regions of the NMSSM parameter space, spontaneous
CP-violation at finite temperature becomes an interesting alternative
scenario.

\begin{figure}[t]
\begin{picture}(200,170)
\put(120,205){\epsfxsize6cm \epsffile{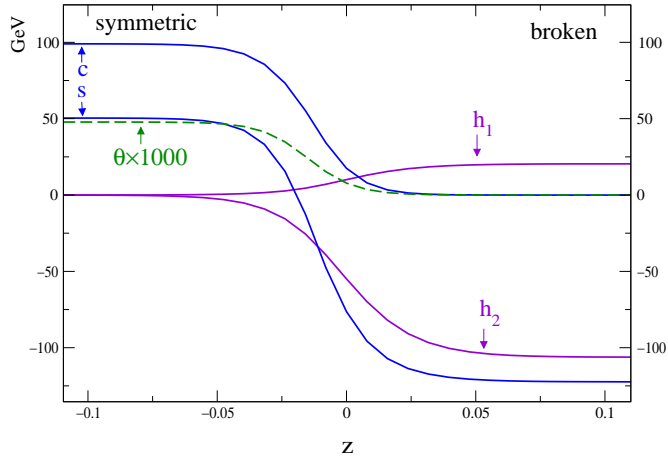}}
\end{picture}
\caption{Transitionally CP-violating bubble wall profile from the
parameter set of table \ref{t_SCPV}. The position variable $z$ is
given in units of GeV$^{-1}$.
}
\label{f_SCPV}
\end{figure}

In the $Z_3$-symmetric NMSSM spontaneous CP-violation at finite
temperature has already been investigated some years ago in
ref.~\cite{CPR94}, where an effective potential truncated at the
renormalizable level was used. It was found that spontaneous
CP-violation cannot occur in the
finite temperature broken phase  while having a
viable temperature zero phenomenology, since thermal
corrections to the effective potential have the tendency to restore
symmetries. The authors suggested that spontaneous
CP-violation may take place during the EWPT
(``transitional CP-violation''), i.e.~the Higgs and
singlet fields may acquire complex phases in the bubble wall,
although no definite example was given.  However, the truncated
effective potential used in their analysis proved to give misleading
results in case of the MSSM \cite{HJLS99}, as will be discussed below.

\begin{table}[b] \centering
\begin{tabular}{|c|c|c|c|c|c|c||c|c|c|} \hline
$x$ & $\tan\beta$ & $\lambda$ & $k$ & $M_0$ & $A_0$ & $m_0$ &
$A_{\lambda}$ & $A_k$ & $m_{S}^2$  \\ \hline
-150 & -5 & 0.05 & 0.4 & 100 & -100 & 200 & $-150$ & $50$& $-2000$  \\ \hline
\end{tabular}
\caption{Parameter set used in fig.~\ref{f_SCPV}
(all dimensionful parameters in GeV).}
\label{t_SCPV}
\end{table}

In our analysis of finite temperature spontaneous CP-violation
in the general NMSSM we use the full 1-loop effective potential
(\ref{strength1}) without any truncations. It turns out that
spontaneous CP-violation does not occur for the universal pattern
of SUSY breaking discussed in section 2, which induces too
large masses for the CP-odd Higgs bosons.
However, spontaneous
CP-violation can take place if universality is violated in the
singlet sector. In this case $A_{\lambda}$, $A_k$ and $m^2_S$
are free parameters.\footnote{It
is even sufficient to violate universality only via $m_S^2$.}
Fig.~\ref{f_SCPV}, which is taken from ref.~\cite{HJLS99}, shows an example
of a transitionally CP-violating bubble wall for the parameters given in
table \ref{t_SCPV}, which was computed
using the minimization algorithm of ref.~\cite{peter}. More precisely,
the scenario corresponds to a phase transition between an electroweak symmetric
high temperature phase, where CP is violated by a complex singlet vev,
and an electroweak broken, CP-conserving low temperature phase,
i.e.~$(s,c)_{\rm sym}=(50{\rm GeV},99{\rm GeV})\rightarrow
(s,c)_{\rm brk}=(-122{\rm GeV},0)$. The critical temperature
is found to be 101 GeV, and we have a strong first order PT
with $v_c/T_c=1.6$.
The mass of the lightest CP-even Higgs boson is 84 GeV
which because of its reduced coupling to the Z-boson is still
compatible with the experimental data.
There is also a rather light CP-odd
state with a mass of 105 GeV, which is almost a pure singlet. The bubble
wall is found to be rather thin with $L_w\sim 3/T$. 
In our example spontaneous CP-violation is triggered in the
singlet sector. The small coupling $\lambda$   communicates
it to the Higgs fields. As a consequence, the almost maximal
phase of the singlet field, $\theta_s$, induces only a small
phase $\theta\sim1/20\sim\lambda\theta_s$ in the Higgs sector.
Up to now we only identified one particular region in the NMSSM
parameter space where transitional CP-violation occurs. A more systematic
study would be very desirable. Especially it would be interesting to
find an example where transitional CP-violation originates
from the Higgs fields rather than the singlet field. In that case
$\theta$ is expected to be of the order one.

Let us finally comment on transitional CP-violation in the MSSM,
where there also have been claims in the literature that transitional
may occur \cite{MSSM_SCPV}. In ref.~\cite{jap99} even an example
of a transitionally CP-violating parameter set has been presented.
All these studies were carried out using an effective potential that
has been truncated at the renormalizable level, i.e.~the logarithmic
terms, etc., were neglected.
In ref.~\cite{HJLS99} we performed a systematic search for transitional
CP-violation in the MSSM using the full 1-loop thermal effective potential
without finding any viable parameter set. In particular, the example of transitional
CP-violation presented in ref.~\cite{jap99} could not be confirmed.
This discrepancy appears to be due to the different approximations
which have been made concerning the effective potential. Our results
indicate that the truncated effective potential is not appropriate to
discuss transitional CP-violation.

%
%
%
%
%
%
%
%
%
%
\section{Baryon Asymmetry in the Semi-Classical Limit}
\subsection{Generalities}
Between the initial nucleation and the completion of the
phase transition expanding bubbles convert the symmetric phase
into the broken phase. The early universe is filled with a hot
plasma of particles, most of them have rather different masses
and mixings inside and outside the bubble. Thus the bubble wall
behaves like a potential on which the particles scatter. At the
bubble surfaces the plasma is thrown out of equilibrium by the
motion of the phase boundary. The higher pressure inside
the bubble tends to accelerate the wall, while the interaction
between the wall and the particles in the plasma
dissipates energy and slows the wall down. Finally, a stationary
situation is reached, where the different forces balance and the wall
propagates with constant velocity $v_w$. In general, the wall
velocity depends on the shape of the effective potential and on
the composition of the plasma.

In the symmetric phase baryon number violation occurs frequently
due to hot sphaleron processes.
A baryon asymmetry is produced, if the departure from equilibrium
at the phase boundary biases the baryon number violating processes
in a CP-violating fashion.  The most efficient mechanisms of baryon
number generation rely on transport.
If CP is violated in the interaction between the bubble wall and
the particles in the plasma, different population densities for particles
and anti-particles are induced in front of the wall. The difference
between the particle and anti-particle populations is then transported
into the symmetric phase, where it biases rapid baryon number violation.
The total amount of baryon number which gets produced during
the phase transition depends crucially on the shape and motion of
the bubble wall.

Different methods have been suggested in the literature
in order to describe the effects of CP-violating interactions between
particles in the plasma and the propagating bubble wall,
which finally generate CP-violating source terms that enter
the diffusion equations for particle transport.
Depending on the properties of the bubble wall,  in the first
place  its velocity $v_w$ and thickness $L_w$, different
approximations may be applied.
The most rigorous approach is based on the closed time-path (CTP)
formulation of non-equilibrium quantum field theory \cite{ctp}. It leads to
a set of quantum Boltzmann equations \cite{kadbaym} which describe the
temporal evolution of particle densities including
particle number changing interactions and CP-violating
source terms \cite{riotto_ctp}. However, the approximations and results
are still controversely discussed. There have also been
efforts to use the CTP formalism only in calculating the
CP-violating source terms which in turn are inserted into
classical Boltzmann equations \cite{cqrvw}.

If the thickness of the bubble wall $L_w$ is smaller than the
mean free path $l$ of the particle under consideration (``thin wall''),
one may neglect the influence of the plasma during
the scattering of the particle off the  wall. The interaction
of a fermion or boson with the wall can then approximately be
described by using a ``free'' Dirac or Klein-Gordon equation,
respectively. CP-violation is encoded in different reflection
and transmission coefficients for particles and anti-particles
\cite{refcoeff,FKOTT,JPT1,AOS97,DKW98}. In ref.~\cite{HN}
also the effects of thermal scattering were
taken into account, which due to decoherence have a
negative impact on the generation of a CP-violating
observable \cite{GHOOQ94}.
Baryon production is fueled by a net flux of charge into
the symmetric phase (``charge transport mechanism'' \cite{refcoeff}),
which is induced by the reflection asymmetry. This current is
subsequently inserted into a set of classical Boltzmann equations
which determines the evolution of the particle distributions.
Recently, reflection and transmission probabilities have also been calculated
by using standard quantum field theory methods \cite{RV99}.
There, thermal scattering, i.e.~damping, was included via an imaginary part 
of the self-energy.

As the bubble wall becomes thicker, $L_w\sim l$, interactions with
the plasma must inevitably be taken into account, when a particle
encounters the propagating wall. In general, this requires
non-equilibrium quantum field theory methods \cite{riotto_ctp}.
However, if $L_w \gg 1/T$, most particles have inverse momenta
$1/p \ll L_w$ and may therefore be treated  semi-classically
\cite{TZ91,CKN_WKB91,JPT2,CJK}.
Using the WKB approximation one can derive dispersion relations
which in the presence of CP-violation are different for particles and anti-particles.
The dispersion relations then enter classical Boltzmann equations,
leading to a unified description of CP-violating source terms,
particle scattering and transport. Recently, it was
clarified how the semi-classical description arises from the
quantum Boltzmann equations in the limit of a
slowly varying background field (thick wall regime) \cite{JKP}.
Of course, even in the thin wall limit the high-momentum particles
may be described semi-classically. However, in that case they only give 
a sub-leading contribution to the CP-violating source, which
in this case is dominated by the low-momentum particles.
Even in case of thick walls there may be important
contributions from low-momentum modes neglected
in the semi-classical treatment, which
according to ref.~\cite{riotto_ctp} significantly
enhance the produced amount of baryon number.
There is also a very recent calculation \cite{moore2000} showing
that gauge fields in the hot phase lower significantly the
wall velocity, such also leading to a more efficient baryon
production.
In ref.~\cite{moorepr}
the semi-classical approximation to particle dynamics has been
applied to calculate the bubble wall velocity in the SM. In
ref.~\cite{JS2000} it was shown that the stops in the MSSM
considerably lower the wall velocity.

Notice that different particle species with different (gauge) interaction
have different mean free paths. Thus whether one is in the
thin or thick wall regime depends on the particle species
under consideration. In the (N)MSSM especially particles with
strong interactions (quarks and squarks) require a thick wall treatment.

There is some disagreement among the groups that have
estimated the baryon asymmetry generated during the EWPT
in the MSSM.  In refs.~\cite{riotto_ctp,cqrvw,HN}
the CP-violating source terms are proportional to the
variation of $\tan\beta$ in the wall, which according to
the discussion of section 4 causes a suppression of
at least ${\cal O}(10^{-2})$. This dependence is due to
taking into account only the leading order of the Higgs
insertion expansion, which has been used to
calculate source terms. At higher orders in the expansion
there arise contributions, which escape the $\delta\beta$
suppression, as was recently shown in ref.~\cite{RV99}.
However, these corrections turn out very small and
are competitive to the leading order contributions only
in case of $\delta\beta<{\cal O}(10^{-3})$.
In the recent paper \cite{76} it was argued convincingly
that the $(\tan \beta)'$ source cancels and that one has to consider
closely a source term symmetrical in the two Higgs fields. 

In this section we generalize the method introduced in ref.~\cite{CJK}
to calculate the baryon asymmetry in the NMSSM, which
has not been considered so far. Different from ref.~\cite{CJK} our 
formulas also cover the case of CP-violating bubble walls.
We obtain a non-vanishing baryon asymmetry even in case
of constant $\tan\beta$, which is due to the variation of
the singlet field or the presence of CP-violating bubble walls.

\subsection{WKB approximation and dispersion relations}
The idea of deriving CP-violating source terms using a
semi-classical approximation was developed in context of the
two Higgs doublet model \cite{JPT2} and
afterwards applied to the charginos in the MSSM \cite{CJK}.
In the following we review the derivation of
semi-classical dispersion relations and generalize this
method to account for the effects of CP-violating bubble walls,
which potentially are present in the NMSSM.
These additional sources of CP-violation will turn out to be
very helpful in order to generate a baryon to entropy ratio
in the observed range: $\eta_B=n_B/s\sim2-7\times10^{-11}$ \cite{eta}.
We also give the dispersion
relations in the bosonic case. In the next section these
dispersion relations enter the classical Boltzmann equations
that describe particle transport induced by the propagating
bubble wall.

\subsubsection{The fermionic case}
We start the derivation of fermionic dispersion relations in presence
of CP-violation with a simple example: A single (Dirac-) fermion
$\Psi_D^T=(\eta_{\alpha},\bar\chi^{\dot\alpha})$
that couples only to one of the two Higgs doublets,
which we denote by $H$.
The most prominent realizations are the quarks and leptons
of the (N)MSSM in the presence of a CP-violating bubble wall.
Due to its coupling to the Higgs, $y$, the fermion obtains a mass
proportional to the Higgs vev, $M=y H$. During the passage of the
bubble wall the fermion mass becomes space-time dependent.
Assuming that the bubble has grown to macroscopic size
and reached its final velocity, we can neglect the curvature of
the wall and boost to the rest frame of the bubble. Then the fermion
mass only depends on one position coordinate, which we
denote by $z$, i.e.~$M=M(z)$. As a consequence,
the energy of the particle, $E$,
and its momentum perpendicular to the $z$-direction, $p_{\perp}$,
are constants of motion. Writing $M=me^{i\theta}$, possible
CP-violation is encoded in a non-vanishing phase $\theta$.\footnote{Here
$\theta$ denotes the phase of the Higgs boson the fermion
is coupling to. It should not be mixed up with the common phase
of both Higgs fields defined in eq.~(\ref{CPV2}) which is denoted
by the same symbol.}
Any constant $\theta$ can be absorbed by a redefinition of the
fermion field. When couplings to additional fields are neglected
only a varying phase contributes to CP-violation.

In the presence of a complex mass term the fermion field is 
described by the free Dirac equation
\begin{eqnarray}  \label{fwkb1a}
(i\gamma^{\mu}\partial_{\mu}-P_LM-P_RM^*)\Psi_D\equiv
  \left(\begin{array}{cc}-M &
                      i\sigma^{\mu} \partial_{\mu} \\  i\bar\sigma^{\mu}
                      \partial_{\mu}  & -M^*\end{array}\right)
                {\eta \choose \bar\chi}=0.
\end{eqnarray}
Notice that at this level all interactions between the particle and the
plasma are neglected. The scattering effects will be accounted for
in the next section when Boltzmann equations are written down
that describe the local phase space distributions.
We are working in the chiral representation of the $\gamma$-matrices.
Exploiting conservation of energy and boosting to
the Lorentz frame  where $p_{\perp}=0$ we can take the ansatz
$\Psi_D=e^{-iEt}\xi(z)$ and are left with a one dimensional problem.
The interaction between the fermion and the wall conserves the
$z$-component of the spin, $S_z$. Thus eq.~(\ref{fwkb1a}) splits
into two equations \cite{JPT1}
\begin{equation} \label{fwkb2}
i\partial_z\xi_{\pm}=\pm Q(z)\xi_{\pm}, \qquad Q(z)=\left(\begin{array}{cc} E&
-m(z)e^{-i\theta(z)} \\ m(z)e^{i\theta(z)} &-E
\end{array}\right)
\end{equation}
where $\xi_+=(\xi_1,\xi_3)$ and $\xi_-=(\xi_2,\xi_4)$ are
the $S_z=\pm\frac{1}{2}$ components of $\xi$.
To solve eq.~(\ref{fwkb2}) one brings
the $z$-dependent matrix $Q(z)=D(z)Q_D(z)D(z)^{-1}$
into a diagonal form,
where  \cite{CJK}
\begin{equation} \label{fwkb2a}
Q_D=\left(\begin{array}{cc} \sqrt{E^2-m^2}&0\\ 0 & -\sqrt{E^2-m^2}
\end{array}\right),
D=\left(\begin{array}{cc} \cosh X&e^{-i\theta}\sinh X\\
e^{i\theta}\sinh X &\cosh X \end{array}\right)
\end{equation}
and $\tanh 2X=m/E$. In the  local helicity basis,
$\tilde\xi_{\pm}=D^{-1}\xi_{\pm}$, the Dirac equation (\ref{fwkb2})
takes the form
\begin{equation} \label{fwkb3}
i\hbar\partial_z\tilde\xi_{\pm}=(\pm Q_D-D^{-1}i\hbar\partial_zD)
\tilde\xi_{\pm}
\end{equation}
which still is an exact equation. In general, the
correction term $D^{-1}i\hbar\partial_zD$
caused by the  position dependent field
redefinition is not of diagonal form. The two components
of $\tilde\xi_{\pm}$ are still coupled. However, the off-diagonal
part is proportional to $\partial_z D\sim D/L_w$. Typical
momenta of the particles in the plasma are of the order of the temperature
$T$, which is much larger than $1/L_w$ for the bubbles under
consideration. We therefore expand eq.~(\ref{fwkb3}) in powers
of $\partial_z$ or more precisely in powers of $\hbar$
(WKB approximation) that we already reintroduced for that reason.

To order $(\hbar)^0$ we can neglect the $D^{-1}\hbar i\partial_z D$
contribution. Thus the two components of $\tilde\xi_{\pm}$
decouple in eq.~(\ref{fwkb3}).  Inserting the WKB ansatz for the fermion field
\begin{equation} \label{wkb0}
\tilde\xi_{\pm}^{(1)}={1 \choose 0} e^{-\frac{i}{\hbar}\int^zp_z(z')dz'}, \qquad
\tilde\xi_{\pm}^{(2)}={0 \choose 1} e^{-\frac{i}{\hbar}\int^zp_z(z')dz'}
\end{equation}
into (\ref{fwkb3}), we obtain the dispersion relations $p_z(E)$
\begin{eqnarray} \label{fwkb0a}
&&\tilde\xi_+^{(1)}, \tilde\xi_-^{(2)}:
\qquad p_z=\sqrt{E^2-m^2}, \nonumber\\
&&\tilde\xi_+^{(2)}, \tilde\xi_-^{(1)}: \qquad p_z=-\sqrt{E^2-m^2}.
\end{eqnarray}
The momenta $p_z$ are the eigenvalues of the matrix entering the
RHS of eq.~(\ref{fwkb3}).
The eigenfunctions may more transparently be labeled by the
chirality states they correspond to in the limit $m\rightarrow 0$,
e.g.~$\tilde\xi_+^{(1)}\sim \xi_1\sim\eta_1$, etc.
The dispersion relations may then be combined
to $p_z={\rm sgn}(p_z)\sqrt{E^2-m^2}$ which holds for left-handed
particles $\eta$ and right-handed particles $\bar\chi$.
Obviously, in the classical limit the $\theta$-dependence completely
disappears, demonstrating that CP-violation is indeed a
quantum-mechanical phenomenon.

To solve the Dirac equation (\ref{fwkb3}) to order $\hbar$  we have
to take into account the $D^{-1}\hbar i\partial_z D$ term
which reintroduces a coupling between the two components
of $\tilde\xi_{\pm}$. The dispersion relations $p_z(E)$ are obtained
from the eigenvalues of the matrix $\pm Q_D-D^{-1}i\hbar\partial_zD$.
Since to order $\hbar$ the off-diagonal terms do not
contribute, we are left with \cite{CJK}
\begin{eqnarray} \label{fwkb5}
L\mbox{ }(\eta): &&p_z={\rm sgn}(p_z)\sqrt{E^2-m^2}-
\hbar\theta'\sinh^2X, \nonumber\\
R\mbox{ }(\bar\chi): &&p_z={\rm sgn}(p_z)\sqrt{E^2-m^2}+
\hbar\theta'\sinh^2X,
\end{eqnarray}
where $\theta'=\partial_z \theta$  and
\begin{equation}\label{6.39A}
\sinh^2X=\frac{E-\sqrt{E^2-m^2}}{2\sqrt{E^2-m^2}}.
\end{equation}
Again the states are labeled by their asymptotic chirality properties.
Notice that the variation of $m$, which is encoded in $\partial_z X$, drops in the 
dispersion relations.
The  CP-violating
part of the dispersion relation, $\Delta p_z=\hbar\theta'\sinh^2X$,
is proportional to the {\em derivative} of the phase $\theta$.
Thus only a varying phase contributes to CP-violation
in the semi-classical limit.
Furthermore, CP-violation is proportional to
$\sinh^2 X$, which guarantees
that its effect is turned off in the limit $m\rightarrow 0$,
where $\theta$ is no longer well defined. Because of the different
dispersion relations, left- and right-handed particles
feel a different (semi-classical) force in their interaction
with the wall. For the anti-particles, $\bar \eta$
($\bar L$) and $\chi$ ($\bar R$), the CP-violating part comes
with the opposite sign. Besides the force term there is a second
manifestation of CP-violation: The phase $\theta$ enters also
the transformation matrix to the helicity basis $D$.
Since the interaction eigenstates are different from the
helicity states, particles and anti-particles interact
in a different way with the surrounding plasma.
This generates a CP-violating source term which drives
``spontaneous'' baryogenesis \cite{CKN_spont}.

As pointed out recently in ref.~\cite{76}, one should better use the  
kinetic momentum $p_{\rm kin}=m v_{\rm group}=m\frac{\partial E}{\partial p}$
instead of the canonical momentum $p$ (which we used up to now) in
the quasi-classical limit of particles in the Boltzmann transport
equations. This is quite in the spirit of the correspondence principle
of basic quantum mechanics. Calculating the (inverse) group
velocity from (\ref{fwkb5}), and using (\ref{6.39A}),
the kinetic moment beyond the zeroth order contains (order
$\hbar$) correction terms 
\begin{equation}
\Delta p_{\rm kin}=\pm\frac{\hbar\theta'm^2}{2E\sqrt{E^2-m^2}}.
\end{equation}
In the following we will need the dispersion relation for
energy in terms of momentum which to order $\hbar$ is obtained as
\begin{eqnarray} \label{fwkb6}
E_{\pm}\equiv E_0\pm\Delta E=\sqrt{p_{\rm kin}^2+m^2}\pm{\rm sign}(p_z)\theta'
\frac{m^2}{2(p^2_{\rm kin}+m^2)}
\end{eqnarray}
$E_+$  is the energy of left-handed particles and
anti-particles, whereas $E_-$ corresponds to right-handed particles
and anti-particles.\footnote{Notice that in ref.~\cite{CJK}
the phase $\theta$ has been defined with the opposite sign.}
In the derivation of eq.~(\ref{fwkb6}) we
transformed to a general Lorentz frame with non-zero momentum
parallel to the wall.
When boosting to the plasma frame the dispersion
relation (\ref{fwkb6}) is preserved to linear order in $v_w$ and $\theta'$.

In our derivation of the dispersion relations we followed basically
ref.~\cite{CJK}. CP-violation arises because of a position
dependent phase in the transformation  from the interaction
(i.e.~chirality) states to the local mass (i.e.~helicity) eigenstates.
In ref.~\cite{JPT2} a slightly different approach was used to
obtain the dispersion relation of a fermion (e.g.~top quark)
in the presence of a CP-violating Higgs field background, $H$,
in context of the 2HD model:
The complex phase in the fermion mass
was removed by a {\em gauge} transformation, which
induced a gauge field in the kinetic term of the fermion.
The dispersion relation
obtained with this technique is completely analogous to our
result (\ref{fwkb5}). However, it is not clear how to generalize
this method to cover the case where several species mix,
as occurs with the gauginos and Higgsinos in the (N)MSSM,
or where the NMSSM singlet field background contributes
to CP-violation. In both situations the CP-violating phase
cannot be removed by a gauge transformation. On the other hand,
the method described above is still applicable \cite{CJK}.

We are now in the position to address the problem of
mixing Dirac fermions $\Psi_{DI}$, where the ``flavor''
index $I=1,...,N$. In order to solve the corresponding Dirac equation
we diagonalize the matrix
\begin{equation} \label{fwkb7}
Q(z)=\left(\begin{array}{cc} {\bf 1}E & -{\bf M}^{\dagger}(z)\\
{\bf M}(z) & -{\bf 1}E \end{array}\right)
\end{equation}
which is a straight forward generalization of (\ref{fwkb2}). Here
${\bf 1}$ denotes the unity matrix in flavor space and ${\bf M}$
is a position dependent complex mass matrix for the fermions,
without further specifications. As a first step we write
${\bf M}$ and ${\bf M}^{\dagger}$ in terms of two unitary matrices
${\bf U}$ and ${\bf V}$, and a diagonal matrix
${\bf M}_D={\bf m}e^{i \boldsymbol{\theta}}$
\begin{equation} \label{fwkb8}
{\bf M}={\bf V}{\bf M}_D{\bf U}^{\dagger}, \qquad
{\bf M^{\dagger}}={\bf U}{\bf M}_D^{\dagger}{\bf V}^{\dagger}.
\end{equation}
${\bf U}$ and ${\bf V}$ may be obtained by diagonalizing
the hermitian matrices ${\bf M}^{\dagger}{\bf M}$
and ${\bf MM}^{\dagger}$, respectively.
The transformation to the helicity basis takes the form
\begin{equation}   \label{fwkb9}
\boldsymbol{\tilde \xi}_{\pm}=
\left(\begin{array}{cc} \cosh{\bf X} & -e^{-i\boldsymbol{\theta}}\sinh{\bf X}\\
-e^{i\boldsymbol{\theta}}\sinh{\bf X} & \cosh{\bf X} \end{array}\right)
\left(\begin{array}{cc} {\bf U}^{\dagger} & 0\\
0 & {\bf V}^{\dagger} \end{array}\right)\boldsymbol{\xi}_{\pm}
\equiv {\bf T}^{-1}\boldsymbol{\xi}_{\pm}
\end{equation}
where ${\bf X}$ is a diagonal matrix in flavor space, obeying
$\tanh(2{\bf X})={\bf m}/E$.
Like in the single fermion case
the CP-violating part of the dispersion relation is encoded in the
diagonal elements of ${\bf T}^{-1}i\hbar\partial_z{\bf T}$. We find
\begin{eqnarray}   \label{fwkb10}
\tilde \xi_{\pm}^{(1)}:\mbox{ }
p_z&=&\pm\sqrt{E^2-m^2_I}-\theta_I'
\sinh^2 X_I-\cosh^2 X_I[{\bf U}^{\dagger}i\partial_z{\bf U}]_I
+\sinh^2 X_I[{\bf V}^{\dagger}i\partial_z{\bf V}]_I, \qquad \nonumber\\
\tilde \xi_{\pm}^{(2)}:\mbox{ }
p_z&=&\mp\sqrt{E^2-m^2_I}+\theta_I'
\sinh^2 X_I-\cosh^2 X_I[{\bf V}^{\dagger}i\partial_z{\bf V}]_I
+\sinh^2 X_I[{\bf U}^{\dagger}i\partial_z{\bf U}]_I. \qquad
\end{eqnarray}
Here the subscript $I$ on the RHS denotes the $Ith$ diagonal element
of the corresponding matrix, i.e.~$m_I=m_{II}$, and we dropped
the factor $\hbar$. In addition to the $\theta'$ contribution,
which we already encountered in the case of a single Dirac
fermion, the dispersion relations receive corrections due
to the position dependent rotations in flavor space, ${\bf U}$ and
${\bf V}$. 

In eqs.~(\ref{fwkb9}) and (\ref{fwkb10}) we allowed for complex
phases in the diagonal matrix ${\bf M}_D$.
Since ${\bf U}$ and ${\bf V}$ contain $2N^2-2$ real parameters
it would be sufficient to take real values in ${\bf M}_D$ in order to
reproduce the $2N^2$ real parameters of ${\bf M}$.
However, we have to omit  the $N-1$ rotations in ${\bf U}$ and ${\bf V}$
which belong to the Abelian subgroup of $SU(N)$. These are not
related to CP-violation, but rather correspond to
artificial redefinitions of the interaction (chirality) states, as can be
deduced from the case vanishing fermion mixing.
Thus allowing for complex values for the masses in ${\bf M}_D$ we
end up with a one by one correspondence between the $N^2$ parameters
in ${\bf M}$ and the ones contained in ${\bf U}$, ${\bf V}$ and ${\bf M}_D$.

Having resolved the ambiguity in the definition of the helicity
basis, it is possible to determine the matrices
${\bf U}$, ${\bf V}$ and ${\bf M}_D$ numerically.
The dispersion relations are then easily obtained by evaluating
the expressions (\ref{fwkb10}). Here we will concentrate
on the charged Winos and Higgsinos in the (N)MSSM. Since they mix
via the $2 \times 2$ chargino mass matrix
\begin{equation} \label{char_mass}
{\cal L}= \cdots+
(i\tilde W^-,\tilde h_1^-)\left(\begin{array}{cc}
M_2&g_2(H_2^0)^* \\ g_2(H_1^0)^* &\mu+\lambda S
\end{array}\right) {i\tilde W^+ \choose \tilde h_2^+}
\end{equation}
analytic formulas can be obtained.
The charged Winos and Higgsinos
can be combined to a Dirac spinor $\Psi_D=(\Psi_L,\Psi_R)^T$, where
$\Psi_L=(\tilde W^+,\tilde h_2^+)^T$
and $\Psi_R=(\overline{\tilde W^-},\overline{\tilde h_1^-})^T$.
We parametrize the SU(2) matrices ${\bf U}$ and ${\bf V}$ by
\begin{equation}  \label{fwkb11}
{\bf U}=
\left(\begin{array}{cc} \cos a & e^{-i\gamma}\sin a \\
  -e^{i\gamma}\sin a & \cos a
\end{array}\right), \qquad
{\bf V}=
\left(\begin{array}{cc} \cos b & e^{-i\delta}\sin b \\
  -e^{i\delta}\sin b & \cos b   \end{array}\right).
\end{equation}
According to the discussion above we dismissed phases
multiplying the cos-terms.
The diagonal elements of ${\bf U}^{\dagger}i\partial_z{\bf U}$ which
enter the dispersion relations  (\ref{fwkb10}) read
\begin{equation}\label{fwkb12}
[{\bf U}^{\dagger}i\partial_z{\bf U}]_1=-
[{\bf U}^{\dagger}i\partial_z{\bf U}]_2=-\gamma'\sin^2 a.
\end{equation}
We observe that only the derivative of the complex phase $\gamma$
contributes, whereas the derivative of $a$ drops.
Similar relations hold for ${\bf V}^{\dagger}i\partial_z{\bf V}$.
Inserting these expressions into eq. (\ref{fwkb10})
we obtain the dispersion relations for left- and right-handed
particles and anti-particles
\begin{eqnarray} \label{fwkb13}
L_I:\mbox{ }
p_z=\mbox{sgn}(p_z)\sqrt{E^2-m^2_I}-(\theta_I'+\delta'\sin^2 b)
\sinh^2 X_I+\gamma'\sin^2 a\cosh^2 X_I, \mbox{ }\nonumber \\
\bar L_I:\mbox{ }
p_z=\mbox{sgn}(p_z)\sqrt{E^2-m^2_I}+(\theta_I'+\delta'\sin^2 b)
\sinh^2 X_I-\gamma'\sin^2 a\cosh^2 X_I, \mbox{ }\nonumber \\
R_I:\mbox{ }
p_z=\mbox{sgn}(p_z)\sqrt{E^2-m^2_I}+(\theta_I'-\gamma'\sin^2 a)
\sinh^2 X_I+\delta'\sin^2 b\cosh^2 X_I, \mbox{ }\nonumber \\
\bar R_I:\mbox{ }
p_z=\mbox{sgn}(p_z)\sqrt{E^2-m^2_I}-(\theta_I'-\gamma'\sin^2 a)
\sinh^2 X_I-\delta'\sin^2 b\cosh^2 X_I. \mbox{ }
\end{eqnarray}
In the symmetric phase $L_2$ and $\bar R_2$
evolve to the left-handed Higgsinos states $\tilde h_2^+$ and
$\tilde h_1^-$, respectively.
The flavor transformations
${\bf U}$ and  ${\bf V}$ are related to the
parameters of the chargino mass matrix (\ref{char_mass}):
\begin{eqnarray} \label{fwkb14}
&&\sin^2a=2|A|^2/\Lambda(\Lambda+\Delta)\quad {\rm with}
\nonumber\\
&&A=g_2((M_2H_2^0)^*+(\mu+\lambda S) H^0_1)\nonumber\\
&&\Delta=|M_2|^2-|\mu+\lambda S|^2+g^2_2(|H^0_1|^2-|H_2^0|^2)\nonumber\\
&&\Lambda=(\Delta^2+4|A|^2)^{1/2}
\end{eqnarray}
and $\gamma={\rm arg}A$. This gives
\[\gamma'\sin^2a=2 {\rm Im}(A^*A')/\Lambda(\Lambda+\Delta)\]
and there are similar relations for $\sin^2b, \delta$
and $\delta'\sin^2b$ exchanging $a$ and $b$, $H^0_1$ and $H^0_2$,
$\gamma$ and $-\delta$.
The mass eigenvalues read (in non-symmetric notation)
\begin{eqnarray} \label{fwkb15}
[M_D]_{11}&=&M_2\frac{\cos a}{\cos b}-g_2(H_2^0)^*
\frac{\sin a}{\cos b}e^{-i\gamma},
\nonumber\\ { }
[M_D]_{22}&=&(\mu+\lambda S)\frac{\cos a}{\cos b}+g_2(H_1^0)^*
\frac{\sin a}{\cos b}e^{i\gamma}. 
\end{eqnarray}
For  $\lambda=0$ these expressions agree with the results of
ref.~\cite{76}. In this case one is left with the MSSM, and
the phases
$\gamma$ and $\delta$ only vary due to a change in the
Higgs vev ratio $\tan\beta$ or because of transitional CP-violation in
the bubble wall.  The first contribution is highly suppressed, since
the variation of $\beta$ is at most $\sim 10^{-2}$ for
realistic Higgs masses \cite{marcos,peter}, while transitional CP-violation
most probably does not occur at all in the MSSM \cite{HJLS99}.
On the other hand, the contribution to the
chargino dispersion relations stemming from the variation of
the complex phases in ${\bf M}_D$ requires only explicit
CP-violating phases in $\mu$ or $M_2$. Eq.~(\ref{fwkb15})
demonstrates that even though the phases in the two terms
entering $[M_D]_{11,22}$ are position independent, their contribution to the
resulting phase varies due to the change in the (real) Higgs vevs \cite{CJK}.
According to the discussion in section 5,
the corrections due to complex Higgs vevs induced by the
explicitly CP-violating phases can also be safely neglected in the MSSM
\cite{HJLS99}.

\begin{figure}[t]
\begin{picture}(200,180)
\put(0,5){\epsfxsize7cm \epsffile{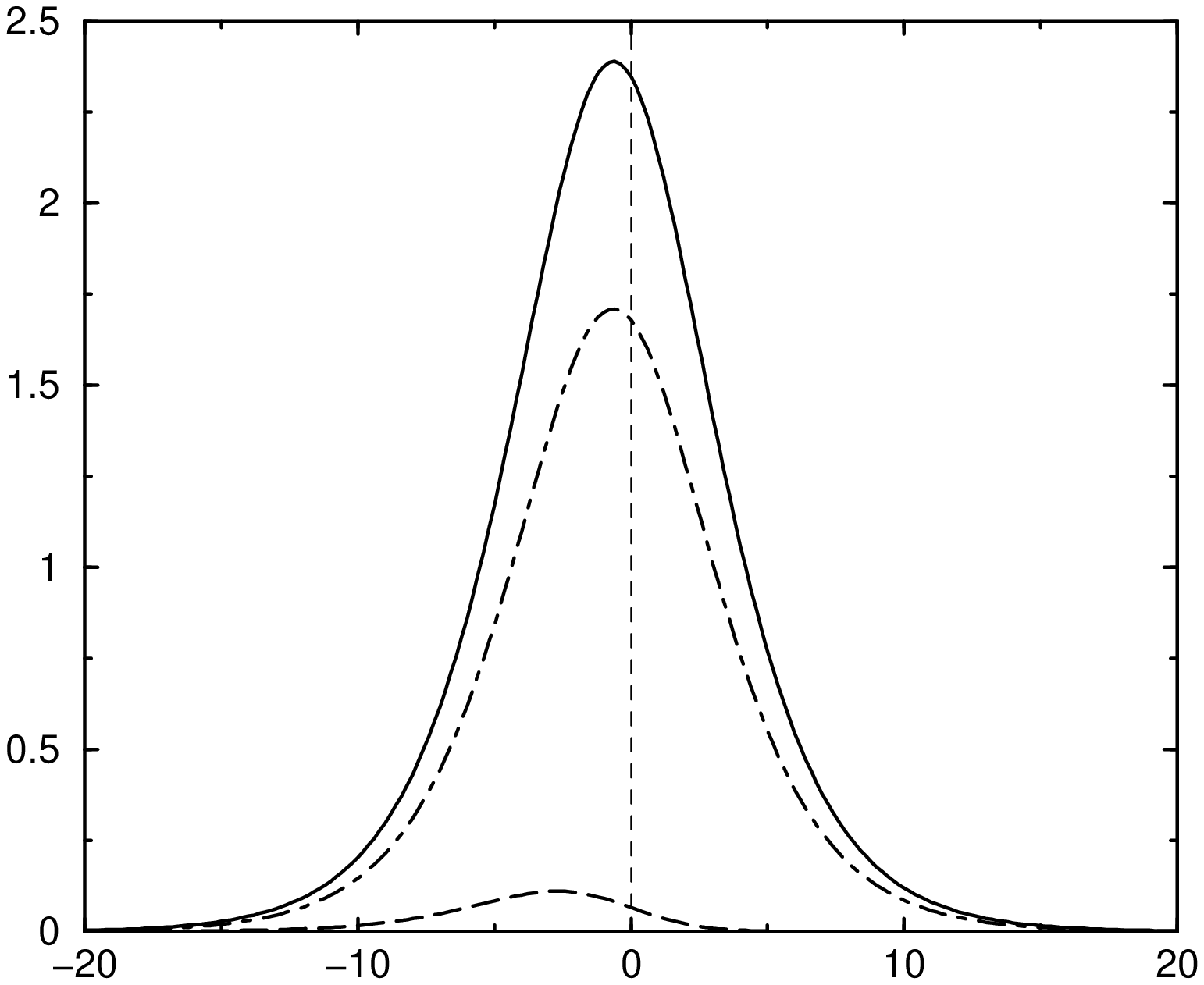}} 
\put(250,5){\epsfxsize7cm \epsffile{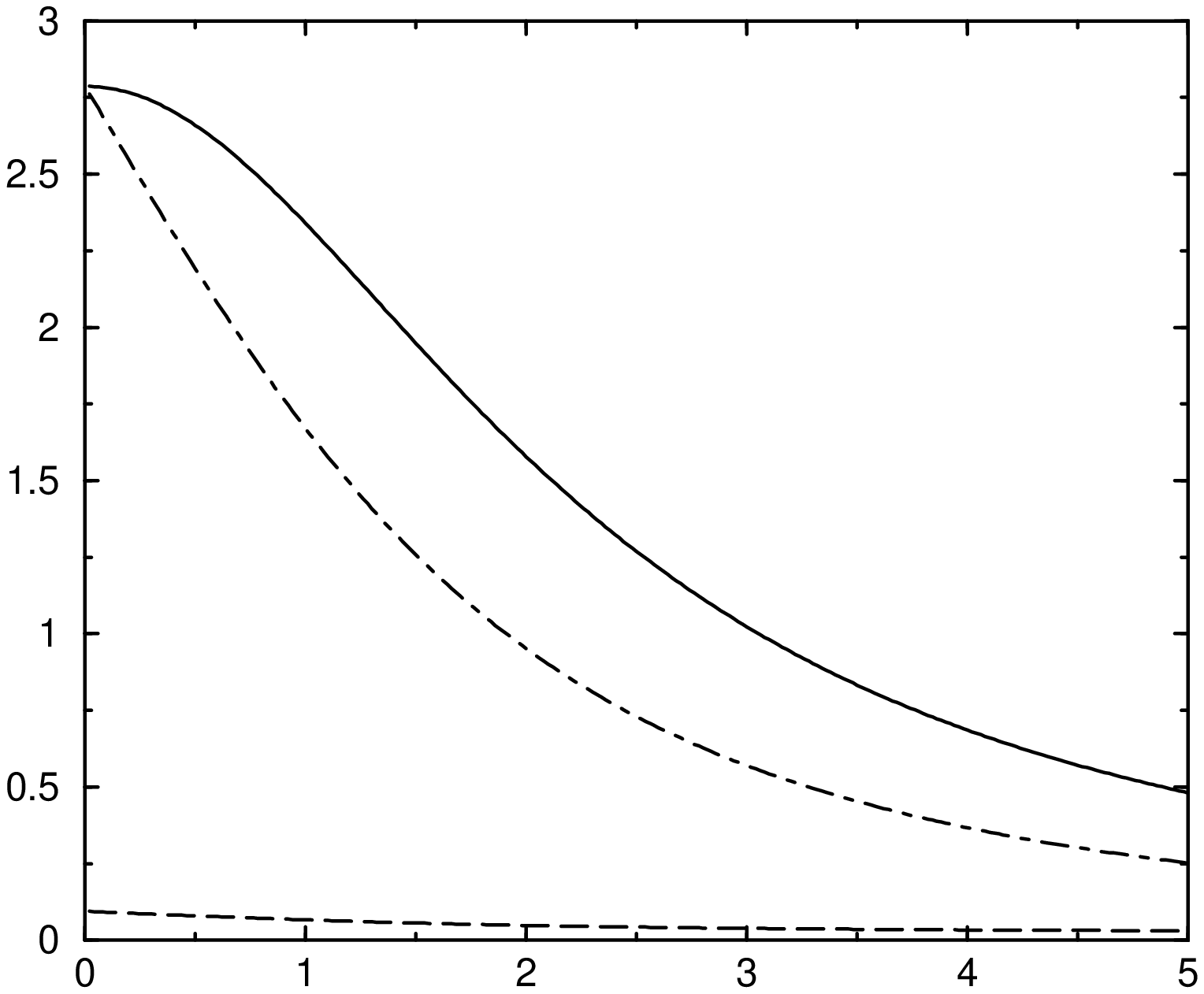}} 
\put(175,150){(a)}
\put(425,150){(b)}
\put(150,-5){$zT_c\rightarrow$}
\put(400,-5){$p/T_c\rightarrow$}
\put(225,100){$\uparrow$}
\put(200,85){$\frac{\Delta E}{T_c}\times 10^4$}
\end{picture}
\caption{
CP-violating contributions to the chargino dispersion relation
for the explicitly CP-violating parameter set of fig.~\ref{f_ECPV}, 
(a) as a function of $z$ for $p=T_c$
and (b) as a function of $p$ for $z=0$.
The dashed-dotted, dashed and solid lines
represent the helicity and flavor contributions (\ref{fwkb13})
and the result using the kinetic momentum (\ref{6.42}), respectively.
}
\label{f_dischar}
\end{figure}

In the NMSSM CP-violation enters the dispersion relations
in several new ways. As already discussed in section 5, spontaneous
CP-violation in the bubble wall occurs for specific values
of the SUSY parameters, which leads to a variation
of $\gamma$, $\delta$ and $\theta_{1,2}$. If this effect is present, it
dominates the CP-violating part of the dispersion relations.
Even in the absence of transitional CP-violation and with constant
$\tan\beta$ there are contributions to $\gamma'$ and $\delta'$:
The complex phases induced in the vevs by the explicitly CP-violating
phases may considerably change in the bubble wall according to section 5.
Also the phase of the effective $\mu$-term, $\mu+\lambda S$,
is position dependent because of the variation of the singlet vev.
However, this effect is suppressed when the coupling $\lambda$
is small.

Again we calculate the (inverse) group velocity, now from (\ref{fwkb13}).
$E$-independent terms drop out, i.e.
$\cosh^2X_I$ can be substituted by $\sinh^2X_I$. The kinetic
momenta beyond the zeroth order (\ref{fwkb5}) then contain (order
$\hbar$) correction terms $\pm(\theta_I'+\delta'\sin^2b-\gamma'
\sin^2a)m^2/2E(E^2-m^2)^{1/2}$. The dispersion relation
can be inverted, and to leading order in the derivatives
the CP-violating part of the dispersion relation for the eigenstate
$L_2$ corresponding to $\tilde h^+_2$ in the symmetric phase is
\begin{equation}\label{6.42}
\Delta E=-{\rm sign}(p_z)(\theta_2'+\delta'\sin^2b-\gamma'\sin^2a)  
\frac{m^2_2}{2(p^2_{\rm kin}+m_2^2)}\end{equation}
where $m^2_2=|[M_D]_{22}|^2$ from eq. (\ref{fwkb15}).
For $p_{\rm kin}\gg m$ this is $2 |p_z|/p$ times 
the result one would obtain with canonical
momentum after the substitution of $\cosh^2$ by $\sinh^2$ in (\ref{fwkb13})
(see fig.~\ref{f_dischar}b).
Note that $\Delta E$ is now totally symmetric under the exchange of
$H_1$ and $H_2$. This destroys the most prominent source term
$\sim H_1H_2'-H_1'H_2$ of older work.

In fig.~\ref{f_dischar} we compare the CP-violating contributions to
the chargino dispersion relations stemming from the
local helicity and flavor transformations (\ref{fwkb13}) with 
the result from using kinetic variables (\ref{6.42}). We assume 
that $p_{\perp}=0$, and  approximate
the bubble wall profile by a kink-ansatz with a common
wall thickness for all fields, i.e.~we take
the straight connection between the symmetric and the
broken minimum in field space. In this approximation $\tan\beta$ is automatically
constant along the wall. We study the
case of explicit CP-violation induced by $\phi_{\mu}=0.1$, which
has already been considered in fig.~\ref{f_ECPV} in section 5. It is
characterized by $L_w=5/T_c$ and $T_c=101$ GeV. Fig.~\ref{f_dischar}a
shows the position dependence of $\Delta E$ for $p=p_z=T_c$. In  
fig.~\ref{f_dischar}b we present $\Delta E$ as a function of momentum
for $z=0$. One observes that $\Delta E$ 
is dominated by the helicity contribution proportional to $\theta_2'$ (dashed-dotted line),
while the flavor contribution (dashed line) only gives a small correction.
The result from the kinetic approach is larger by a factor between 1 and 2.
For the transitionally CP-violating parameter set of fig.~\ref{f_SCPV}
the results are similar. The amplitude of $\Delta E$, however,
is by a factor of about 7 larger, due to the larger amount of CP-violation
and the thinner bubble wall ($L_w=3/T$). In any case, $\Delta E$ provides
only a small correction to dispersion relation.

We close this paragraph by briefly considering the case
of Majorana fermions. Majorana spinors have the
special property of being invariant under charge conjugation,
i.e.~particles and anti-particles are contained in the
same four-component spinor.
The mass matrix of Majorana particles is symmetric,
and the entries are complex numbers, in general.
In the NMSSM,  the neutralinos are Majorana
particles with a $5\times 5$  mass matrix.
From the symmetry property of the mass matrix we
conclude ${\bf M}{\bf M}^{\dagger}={\bf M M}^*=({\bf M}^*{\bf M})^*$.
The flavor transformations therefore obey  ${\bf V}={\bf U}^*$.
Inserting this result into the general dispersions relations
(\ref{fwkb10}) (or into its two-dimensional version (\ref{fwkb13}))
we find that ``left-handed'' and ``right-handed'' states acquire
exactly the opposite CP-violating contribution, in contrast
to the general Dirac case. This is not surprising, since in case
of Majorana fermions the two helicity states describe particles
and anti-particles, respectively.

%
%
%
%
%
%
%
%
\subsubsection{The bosonic case}
In supersymmetric models the scalar superpartners of the
top quarks may give important contributions to the CP-violating
source that fuels baryon production. The stops efficiently interact with
the bubble wall via the large top Yukawa coupling. They contain
many degrees of freedom, and due to renormalization group flow
the right-handed stop is probably lighter than any other squark.
In the following we consider the interaction between the squarks and
the bubble wall in the semi-classical limit. Our approximation is valid
if the thickness of the bubble wall is much larger than the typical
wavelength of particles in the plasma which is of order $1/T$. As in
the fermionic case, we derive dispersion relations in which the
CP-violating part is generated by varying phases in the
scalar mass matrix.

Consider $N$ complex scalar fields ${\bf A}=(A_1,...A_N)^T$ with
mass term ${\bf A}^{\dagger}{\bf M^2A}$.
The entries of the hermitian mass matrix ${\bf M^2}$ may be complex.
Due to the interaction with the wall, ${\bf M^2}$ depends on the
position variable $z$.
We represent ${\bf M^2}$ by a unitary matrix ${\bf U}$ and a real
diagonal matrix  ${\bf M^2_D}=\mbox{diag}(M^2_{D1},...,M^2_{DN})$
\begin{equation} \label{wkbb1}
{\bf M^2}={\bf UM^2_DU}^{\dagger}.
\end{equation}
Like in the fermionic case we omit the $N-1$ rotations in ${\bf U}$
which belong to the Abelian subgroup of SU(N). These transformations
artificially redefine the interaction eigenstates. Their phases
are not related to CP-violation, as can be deduced from the case
of vanishing mixing. We then have a one to one correspondence
between the $N^2$ real parameters in ${\bf M^2}$, and in ${\bf U}$
and ${\bf M^2_D}$.

From the Klein-Gordon equation for the
scalar fields $(\hbar^2\partial_{\mu}\partial^{\mu}+{\bf M^2}){\bf A}=0$
we obtain the equation of motion for the
local mass eigenstates ${\bf \tilde A}={\bf U}^{\dagger}{\bf A}$
\begin{equation} \label{wkbb2}
\big[-\hbar^2(\partial_z^2+2{\bf U}^{\dagger}\partial_z{\bf U}\partial_z
+{\bf U}^{\dagger}\partial_z^2{\bf U})+{\bf M^2_D}-E^2\big]{\bf \tilde A}(z)=0.
\end{equation}
To derive this equation we implemented energy conservation in the
wall frame by the ansatz ${\bf \tilde A}(t,z)={\bf \tilde A}(z)e^{-iEt}$
and boosted to the Lorentz frame where there is no momentum
parallel to the wall, i.e.~$p_{\perp}=0$.

We solve eq.~(\ref{wkbb2}) by the WKB method.
To order ${\cal O}(\hbar^0)$ we can neglect the ${\bf U}^{\dagger}\partial_z{\bf U}$
and ${\bf U}^{\dagger}\partial_z^2{\bf U}$ contributions and
find the wave functions
${\bf \tilde A}^{(I)}(z)={\bf e}_I\exp(-i\int^zp_z(z')dz')$, where ${\bf e}_I$
is the $I$th unit vector in flavor space. Furthermore,  we obtain the dispersion
relation $p_z^2=E^2-M_{DI}^2$. As we already found in the fermionic
case, CP-violation vanishes in the classical limit.

We now include the order  ${\cal O}(\hbar^1)$ corrections. We still can
neglect the ${\bf U}^{\dagger}\partial_z^2{\bf U}$ term, while
the ${\bf U}^{\dagger}\partial_z{\bf U}$ contribution becomes relevant now.
Its off-diagonal entries reintroduce a coupling between the $N$
components of eq.~(\ref{wkbb2}).
In the dispersion relation
only the diagonal part of ${\bf U}^{\dagger}\partial_z{\bf U}$
enters. We obtain
\begin{equation} \label{wkbb4}
{\bf \tilde A}^{(I)}: \mbox{ } E^2=M_{DI}^2+p_z^2
+2p_z[U^{\dagger}i\partial_zU]_I,
\end{equation}
where $[U^{\dagger}i\partial_zU]_I$ denotes the $I$th
diagonal entry of the corresponding matrix.
To transform the dispersion relations to a general Lorentz frame with
non-vanishing $p_{\perp}$ one has to replace $E^2$
by $E^2-p_{\perp}^2$.

Let us evaluate the general expression (\ref{wkbb4}) for the
case $N=2$. Using the representation of ${\bf U}$ from
eq.~(\ref{fwkb11}), we obtain
\begin{equation} \label{wkbb5}
{\bf \tilde A}^{(1,2)}: \mbox{ } E^2=M_{D_{1,2}}^2+p_z^2
-2p_z\gamma'\sin^2 a,
\end{equation}
Again, CP-violation arises due to the varying phase $\gamma$
in the transformation to the local mass eigenstates. A variation
in $a$ gives no contribution to the dispersion relations. For the
anti-particles, ${\bf \tilde A}^{\dagger (1,2)}$, the CP-violating
part in eq.~(\ref{wkbb5}) enters with the opposite sign. If we
apply these expressions to the top squarks of the (N)MSSM with mass
matrix
\begin{eqnarray} \label{stop_mass}
{\cal L}&=&\cdots-
(\tilde t^*,\tilde t^c)\left(\begin{array}{cc}
m^2_{LL}&m^2_{LR} \\ (m^2_{LR})^* &m^2_{RR}
\end{array}\right) {\tilde t \choose (\tilde t^c)^*} \nonumber \\ \nonumber \\
m^2_{LL}&=&m^2_{Q_3}+y_t^2|H_2^0|^2+\left(\frac{g_2^2}{4}-
 \frac{g_1^2}{12}\right)(|H_1^0|^2-|H_2^0|^2)
\nonumber \\
m^2_{RR}&=&m^2_{U_3}+y_t^2|H_2^0|^2+\frac{g_1^2}{3}(|H_1^0|^2-|H_2^0|^2)
\nonumber\\
m^2_{LR}&=&y_t(\mu+\lambda S)H_1^0+y_tA_t^*(H_2^0)^*,
\end{eqnarray}
the parameters of ${\bf U}$ are given similarly to (\ref{fwkb14}) by
$\sin^2a=2|A|^2/\Lambda(\Lambda+\Delta),\ \gamma={\rm arg}~ A$
now with $A=y_t((\mu+\lambda S) H^0_1+A^*_tH^{0*}_2)$.
Solving (\ref{wkbb5}) for $p$, one obtains
$p(E)=(E^2-M^2_{D_{1,2}})^{1/2}+\gamma'\sin^2a$ to first
order in $\gamma'$. The group velocity $\left(\frac{\partial p}{\partial  
E}\right)^{-1}$
is independent of $\gamma'$, thus stops do not contribute to CP-violation
in this order in the kinetic approach. 

%
%
%
%
%
%
%
%
\subsection{Diffusion equations}
\subsubsection{The fluid approximation}
In this section we study the coupled differential equations that
describe particle interactions and transport during the phase
transition. We treat the plasma as consisting of quasi-classical particles
with definite canonical position and momentum. This is an approximation
to quantum Boltzmann equations which have to be discussed in principle.
This picture is justified for thick walls ($p \gg 1/L_w$) if it predicts
a sizable effect, not dominated by non-leading terms in the derivative
expansion \cite{JKP}.
The dynamics of the particles is then governed by the dispersion
relations $E(\vec{x},\vec{p})$ derived in the previous section.

The information about the particle distributions is encoded
in the phase space densities $f_i(\vec{x},\vec{p},t)$. Their
temporal evolution follows from the Boltzmann equation
\begin{equation} \label{diffeq1}
d_tf_i=(\partial_t+\dot{\vec{x}}\cdot\partial_{\vec{x}}+
\dot{\vec{p}}\cdot\partial_{\vec{p}})f_i={\cal C}_i[f].
\end{equation}
The time derivatives of
position and momentum obey the Hamilton equations
$\dot{\vec{x}}=\partial_{\vec{p}}E(\vec{x},\vec{p})$ and
$\dot{\vec{p}}=-\partial_{\vec{x}}E(\vec{x},\vec{p})$.
The Boltzmann equation
can in principle be solved numerically. However, to make it
analytically tractable we use the fluid-type truncation \cite{JPT2}
\begin{equation} \label{diffeq3}
f_i(\vec{x},\vec{p},t)=\frac{1}{e^{\beta(E_i-v_ip_z-\mu_i)}\pm 1}
\end{equation}
for the phase space densities of fermions (+) and bosons (--) in the
rest frame of the plasma.
Here $v_i$ and $\mu_i$ denote the velocity perturbations and
chemical potentials for each fluid, respectively.  We also split
$E_i$ into a dominant part $E_{0i}=\sqrt{p^2+m_i^2}$ and a perturbation
$\Delta E_i\sim\partial_z\theta$ which is related to CP-violation.
The chemical potentials are the central quantities
that finally will determine the baryon asymmetry. The velocity
perturbation on the other hand,
is only introduced to allow the particles to
move in response to the force, giving rise to chemical potential
perturbations.
The fluid truncation is valid as long as perturbations beyond
the ansatz (\ref{diffeq3}) are attenuated faster than chemical
potential perturbations. As discussed in  ref.~\cite{JPT2} this
requires $v_w<\frac{L_w}{3D}$, where the diffusion constant $D$
will be introduced below.

We are looking for a ``stationary'' solution of the Boltzmann equation,
because at late times the wall moves with constant velocity $v_w$. This means that
any explicit time dependence enters in the combination $\bar z\equiv z-v_wt$.
Inserting the fluid ansatz into the Boltzmann equation (\ref{diffeq1}) we obtain 
to linear order in the perturbations $\Delta E_i$, $\mu_i$ and $v_i$
\begin{equation}  \label{diffeq4}
f'_{\pm}\Big\{-v_w(\Delta E_i'-v_i'p_z-\mu_i')-\frac{p_z}{E_{0i}}
(v_i'p_z+\mu_i')+\frac{(m_i^2)'}{2E_{0i}}v_i \Big\}={\cal C}_i[f]
\end{equation}
where $f_{\pm}'=df_{\pm}/dE_0=-\beta e^{\beta E_0}/
(e^{\beta E_0}\pm 1)^2$ denotes the derivative of the
unperturbed Fermi-Dirac or Bose-Einstein distribution. The remaining
primes in eq.~(\ref{diffeq4}) denote $\partial_{\bar z}$. More precisely,
eq.~(\ref{diffeq4}) is the difference of the corresponding equations
for particles and anti-particles. The parameters $\Delta E_i$, $\mu_i$
and $v_i$ therefore represent the differences of these quantities for
particles and anti-particles. For that reason the term $f'_{\pm}v_wE_{0i}'$,
which provides the main contribution to friction in the calculation of the
wall velocity \cite{moorepr}, cancels out. In order to obtain differential  
equations for
the perturbations we average over momentum, weighting eq.~(\ref{diffeq4})
by 1 and $p_z$, respectively, and receive
\begin{eqnarray}  \label{diffeq5}
-\langle \frac{p_z^2}{E_{0i}}\rangle v_i'+\kappa_iv_w\mu_i'
&=&\langle {\cal C}_i\rangle,  \\
\label{diffeq6}
v_w\langle p_z^2\rangle v_i'-\langle \frac{p_z^2}{E_{0i}}\rangle \mu_i'
-v_w\langle p_z\Delta E_{0i}'\rangle &=& \langle p_z{\cal C}_i\rangle.
\end{eqnarray}
The average is defined according to
\begin{equation}  \label{diffeq7}
\langle \cdot \rangle \equiv \frac{\int d^3pf_{\pm}'(\cdot)}{\int d^3pf'_+(m=0)}
\equiv \kappa_i\frac{\int d^3pf_{\pm}'(\cdot)}{\int d^3pf'_{\pm}}.
\end{equation}
The statistical factor $\kappa$ is 1 for massless fermions, 2 for
massless bosons and exponentially small for particles much
heavier than $T$.
In the derivation of eqs.~(\ref{diffeq5}) and
(\ref{diffeq6}) one uses that the energy perturbations $\Delta E_i$
are odd in $p_z$, according to the results of the previous section.
Furthermore, we neglected the contribution of the last term in
the curly brackets of eq.~(\ref{diffeq4}) which turns out to be
numerically less important than the
$\langle \frac{p_z^2}{E_{0i}}\rangle v_i'$ term taken into account
in eq.~(\ref{diffeq5}).

To linear order in the perturbations the collision terms on the
RHS of the Boltzmann equation take the form \cite{CJK}
\begin{equation}  \label{diffeq10}
\langle {\cal C}_i \rangle = \sum_p \Gamma_p^d \sum_j \mu_j,\quad
\langle {p_z\cal C}_i \rangle = v_i\bar p_z^2\sum_p \Gamma_p^e.
\end{equation}
where $\Gamma_p$ denotes the rate of the process $p$. In the sum
over chemical potentials, incoming particles enter with positive sign,
outgoing particles with negative sign.
We distinguish between
elastic interactions with rates $\Gamma^e$ and decays (inelastic
interactions) with rates $\Gamma^d$. Since elastic scattering
conserves the number of particles these processes do not contribute
to $\langle {\cal C}_i \rangle $. In the hot electroweak plasma
the elastic processes dominated by gauge boson exchange reactions
are much more  efficient than the particle
decays coming from Yukawa interactions, sphalerons, etc.
It is therefore justified to neglect the contribution of inelastic processes
to $\langle {p_z\cal C}_i \rangle$.

In the collision integral a delta function $\delta (\sum p_i)$  usually
represents the conservation of energy and momentum
in the interactions. However, in the plasma frame this is no longer
true, because of the moving bubble wall. The Boltzmann
equation is formulated in terms of mass eigenstates which
change in space and time due to the varying Higgs (and singlet) vevs.
As a result, additional CP-violating contributions to the
collision term arise, which are related to what has been
dubbed ``spontaneous'' baryogenesis \cite{CKN_spont}.
The final form of the collision term is therefore given by \cite{CJK}
\begin{equation}  \label{diffeq11}
\langle {\cal C}_i\rangle =\sum_p\Gamma_p^d\left(\sum_j\mu_j+
\Delta E_{{\rm sp},p}\right),
\end{equation}
where $\Delta E_{{\rm sp},p}$ denotes the bubble wall induced
deviation from energy conservation in the corresponding process.

Finally, we reduce the two coupled transport equations
(\ref{diffeq5}) and (\ref{diffeq6}) to a single one by differentiating
(\ref{diffeq6}) and eliminating $v_i$ in favor of $\mu_i$.
We approximate the thermal averages according to
$\langle p_z^2/E_0 \rangle\sim\kappa_i\langle p_z^2/E_0 \rangle_0$,
$\langle p_z^2 \rangle \sim \kappa_i\langle p_z^2 \rangle _0$, where the  
subscript ``0''
denotes averaging with the massless, unperturbed Fermi-Dirac distribution.
Defining the diffusion constant $D_i=\kappa_i\langle p_z^2/E_0\rangle_0^2/
(\bar p_z^2\Gamma_i^e)$ we find the diffusion equation \cite{CJK}
\begin{eqnarray} \label{diffeq12}
-\kappa_i(D_i\mu_i''+v_w\mu_i')+\sum_p\Gamma_p^d\sum_j\mu_j=S_i,
\nonumber\\
S_i=\frac{D_iv_w}{\langle p_z^2/E_0 \rangle_0}
\langle p_z\Delta E_i'\rangle'-\sum_p\Gamma_p^d
\langle\Delta E_{{\rm sp},p}\rangle.
\end{eqnarray}
In order to obtain eqs.~(\ref{diffeq12}) we neglected derivatives of
the CP-conserving thermal averages and rates,
and left aside ratios of inelastic to elastic scatterings.
We also ignored terms beyond leading
order in the wall velocity, an approximation that certainly breaks down if the wall 
velocity approaches the speed of sound in the plasma,
$v_s=1/\sqrt{3}\sim 0.58$.
If the wall moves faster than $v_s$, perturbations cannot propagate
in the region in front of the wall any more.
Notice that the first contribution to the CP-violating source term $S_i$ in
(\ref{diffeq12}), which
is due to the semi-classical force, is proportional to the diffusion
constant. This is simply because particles must move in order
to build up perturbations.
The second, ``spontaneous'' source term is
independent of transport properties of the corresponding particles
and therefore dominates in the limit of inefficient transport.
In $S_i$ the thermal averages over the CP-violating energy
perturbations $\Delta E$ are performed using
the massive distribution functions in order to account
for Boltzmann suppression of heavy particles.

%
%
%
%
%
%
%
%
\subsubsection{Diffusion equations for supersymmetric models}
In the previous paragraph we derived a complicated network
of diffusion equations (\ref{diffeq12})  that couples all
particle species in the hot plasma. In principle, after specification
of decay rates, diffusion constants and CP-violating sources, it is
possible to solve the transport equations numerically. However,
analytic progress can be made by using conservation laws and
neglecting interactions that are slow compared to the
relevant time scale. An interaction with rate $\Gamma$
can be neglected if the typical interaction time is large compared
to the average time a particle spends diffusing in front of the wall
before being caught by the wall, which is equivalent to  \cite{JPT2}
\begin{equation}\label{diffeq15}
\frac{D}{v_w^2}\ll\Gamma^{-1}.
\end{equation}

The electroweak (``weak'') sphaleron interaction with rate $\Gamma_{ws}$ is
slow in precisely the sense of (\ref{diffeq15}) (unless the wall
velocity is particularly small, i.e.~$v_w\lsim 0.01$).
In the following we will therefore assume baryon and
lepton number conservation and include the weak sphalerons
only at the end of the calculation. The neglect of the weak
sphalerons allows us to completely forget about leptons in our
transport equations and compute only the quark and Higgs
densities.

The processes we do take into account are the supergauge
interactions, the strong sphaleron interactions,
and those described by the Lagrangian
\begin{eqnarray}\label{diffeq17}
{\cal L_{\rm int}}&=&y_t t^cq_3H_2+y_t \tilde t^cq_3\tilde h_2 +
y_t t^c\tilde q_3\tilde h_2 - y_t\mu \tilde t^{c*}\tilde q_3^*H_1
+y_tA_t\tilde t^c\tilde q_3H_2  +\mbox{h.c.}
\end{eqnarray}
Via terms of type $\lambda \tilde s \tilde h_1H_2$,
the singlino, $\tilde s$, is coupled to the quark-Higgs system.
In the case of transitional CP-violation triggered in the singlet sector
(see fig.~\ref{f_SCPV}), the singlino receives the largest CP-violating 
source term of all particles, even though its impact on baryogenesis is 
suppressed by the small coupling $\lambda$.
For the sake of simplicity we ignore this interesting contribution
in the following. We assume the supergauge
interactions to be in equilibrium. The chemical potential of any particle
is then equal to that of its superpartner, with exception of the singlet field. 
It is convenient to define the chemical potentials
$\mu_{U}=(\mu_{u^c}+\mu_{\tilde u^c})/2$,
$ \mu_{Q_1}=(\mu_u+\mu_d+\mu_{\tilde u}+\mu_{\tilde d})/4$,
$\mu_{H_1}=(\mu_{H_1^0}+\mu_{H_1^-}+\mu_{\tilde h_1^0}+
\mu_{\tilde h_1^-})/4$, etc.
In this notation the interaction terms take the form\footnote{In contrast
to refs. \cite{CJK,HN} we count all left-handed particles and the corresponding
superpartners with positive signs.}
\begin{eqnarray}
(\Gamma_y+\Gamma_{yA})(\mu_{H_2}+\mu_{Q_3}+\mu_T), \quad
\Gamma_{y\mu}(\mu_{H_1}-\mu_{Q_3}-\mu_T), \quad~~~~~~~~~~~~
\nonumber\\
\Gamma_{ss}(2\mu_{Q_3}+2\mu_{Q_2}+2\mu_{Q_1}+
\mu_T+\mu_B+\mu_C+\mu_S+\mu_U+\mu_D), ~~~~~~~~~~~~
\nonumber\\
 \label{diffeq19}
\Gamma_{hf}(\mu_{H_1}+\mu_{H_2}), \quad
\Gamma_m(\mu_{Q_3}+\mu_T), \quad \Gamma_{H_1}\mu_{H_1}, \quad
\Gamma_{H_2}\mu_{H_2}.~~~~~~~~~~~~~~
\end{eqnarray}
The rates in the first line are related to the interactions (\ref{diffeq17}).
$\Gamma_{ss}$ denotes the strong sphaleron rate. $\Gamma_{hf}$
is due to Higgsino helicity flips induced by the $\mu \tilde h_1\tilde h_2$
term. $\Gamma_{H_{1,2}}$
and $\Gamma_m$ correspond to Higgs and axial top number
violating processes, present only in the phase boundary and the
broken phase.

Because of the small Yukawa couplings of the first and second
family quarks, these particles are in very good approximation
only produced by strong sphalerons.
Hence their number densities are algebraically constrained.
If the system is near thermal equilibrium,
number densities and chemical potentials are related by
\begin{equation}\label{diffeq21}
n_i=\frac{1}{6}k_i\mu_iT^2
\end{equation}
where $k_i$ is the appropriate sum over statistical factors
$\kappa$ introduced in (\ref{diffeq7}),
e.g.~
$k_{Q_1}=N_c(\kappa_u+\kappa_d+\kappa_{\tilde u}+\kappa_{\tilde d})$,
$k_U= N_c(\kappa_{u^c}+\kappa_{\tilde u^c})$,
$k_{H_1}=(\kappa_{H_1^0}+\kappa_{H_1^-}+\kappa_{\tilde h_1^0}+
\kappa_{\tilde h_1^-})$, etc.
$N_c=3$ denotes the number of colors. In the massless limit
used in  ref.~\cite{CJK} one obtains
$k_{Q_{1,2,3}}=18$,  $k_U=k_D=...=k_T=9$, $k_{H_{1,2}}=6$.
Using baryon number conservation the strong sphaleron rate reads
\begin{eqnarray} \label{diffeq24}
\Gamma_{ss}(2\mu_{Q_3}+\dots+\mu_D)=
\Gamma_{ss}\left[\left(2+9\frac{k_{Q_3}}{k_B}\right)\mu_{Q_3}+
\left(1-9\frac{k_T}{k_B}\right)\mu_T\right]
\end{eqnarray}
To arrive at this expression we assumed that all the squark
partners of the light quarks are degenerate in mass. Assuming
equilibrium for the strong sphalerons we obtain
\begin{equation}\label{mssm1}
\mu_T=\frac{2k_B+9k_{Q_3}}{9k_T-k_B}\mu_{Q_3}.
\end{equation}
The validity of this assumption will be discussed below.

We are now able to write down the reduced set of diffusion
equations for the relevant particle species $Q_3$, $H_1$ and $H_2$
\begin{eqnarray}\label{diffeq25}
-A{\cal D}_{q}\mu_{Q_3}+(\Gamma_y+\Gamma_{yA})[\mu_{H_2}+B\mu_{Q_3}]-
\Gamma_{y\mu}[\mu_{H_1}-B\mu_{Q_3}]+B\Gamma_m\mu_{Q_3}=S_{Q_3}~~~
\nonumber \\[.2cm]
-k_{H_1}{\cal D}_h\mu_{H_1}+\Gamma_{y\mu}[\mu_{H_1}-B\mu_{Q_3}]
+\Gamma_{hf}[\mu_{H_1}+\mu_{H_2}]+\Gamma_{H_1}\mu_{H_1}=S_{H_1}~~~
\nonumber \\[.2cm]
-k_{H_2}{\cal D}_h\mu_{H_2}+(\Gamma_y+\Gamma_{yA})[\mu_{H_2}+B\mu_{Q_3}]
+\Gamma_{hf}[\mu_{H_1}+\mu_{H_2}]+\Gamma_{H_2}\mu_{H_2}=S_{H_2}~~~
\end{eqnarray}
where
\begin{eqnarray}\label{mssm2}
A&=&\frac{9k_Tk_{Q_3}+9k_Bk_{Q_3}+4k_Bk_T}{9k_T-k_B}
\nonumber \\
B&=&\frac{k_B+9k_T+9k_{Q_3}}{9k_T-k_B}
\end{eqnarray}
and ${\cal D}_i\equiv D_i\frac{d^2}{d\bar z^2}+v_w\frac{d}{d\bar z}$.
These equations result from summing the diffusion equations of
particles which belong to the same color and SU(2) multiplets.
We have taken a common diffusion constant, $D_q$, for the (s)quarks, as well 
as one for the two Higgs doublets, $D_h$.
Notice that the effects of hypercharge
screening have been neglected, which can be shown to affect the created
baryon asymmetry at most by a factor of order one \cite{CK95}.

We keep the rates related to the top Yukawa interactions finite.  If
these interaction are in equilibrium, the resulting diffusion
equations are sourced only by the combination $S_{H_1}-S_{H_2}$,
because of the constraint $\mu_{H_1}+\mu_{H_2}=0$.  As a result the
dominant contribution to the chargino source terms cancels, because the 
corresponding terms for $\tilde h_1^-$ and
$\tilde h_2^+$ are exactly of the same size.  This would not be
true for the $\gamma',\delta'$ contribution if we used the
dispersion relations for canonical momenta.

In the MSSM the full diffusion equations (\ref{diffeq25}) have already been
studied in ref.~\cite{76} and later in ref. ~\cite{HJS01}. Applying the closed 
time path formalism to calculate the source terms,
the diffusion equations (\ref{diffeq25}) were analyzed recently in 
ref.~\cite{CMQSW01}.

%
%
%
%
%
%
%
%
%
%
%
%
%

\subsection{The baryon asymmetry}
\subsubsection{Solution of the diffusion equations}
In the previous section we derived differential equations for the
chemical potentials of the various particle species contained in
the hot electroweak plasma. Baryon number violation has been
neglected throughout this calculation. However, what we set out
to compute was the total baryon asymmetry created during the phase
transition. So before solving the network of diffusion equations, we turn
to baryon number generation by weak sphaleron processes,
which is fueled by  the chemical potential of left-handed quarks, 
$\mu_{{\cal B}_L}\equiv\mu_{Q_1}+\mu_{Q_2}+\mu_{Q_3}$.
Using baryon number conservation and eq. (\ref{mssm1}) we find
\begin{equation}\label{mubl}
\mu_{{\cal B}_L}=
\left[1-\frac{k_{Q_3}+2k_T}{9k_T-k_B}\left(\frac{2k_B}{k_{Q_1}}+
\frac{2k_B}{k_{Q_2}}\right)\right] \mu_{Q_3}\equiv C\mu_{Q_3}.
\end{equation}
The evolution of the baryon number density $n_{\cal B}$ is governed by
\begin{equation} \label{bar3}
-{\cal D}_qn_{\cal B}+3\Theta(\bar z)\Gamma_{ws}(T^2\mu_{{\cal B}_L}-an_{\cal B})=0,
\end{equation}
where we have assumed identical diffusion constants for all quarks
and squarks, and neglected contributions of leptons.
The position dependence of the weak sphaleron rate is modeled
by a step function $\Theta (\bar z)$: anomalous baryon number
violation is unsuppressed in the symmetric phase $(\bar z>0)$ and
suddenly switched off in the broken phase $(\bar z<0)$.
The second term in Eq.~(\ref{bar3}) describes damping of the baryon
asymmetry by weak sphalerons in the symmetric phase. The parameter $a$
depends on the degrees of freedom present in the hot plasma.  Taking
only the right-handed stop to be light gives $a=48/7$ \cite{76}.

From Eq.~\ref{bar3} one can easily obtain the baryon to entropy ratio
in the broken phase
\begin{equation}   \label{bar5}
\eta_B\equiv \frac{n_{\cal B}}{s}=\frac{135 \Gamma_{ws}}{2\pi^2g_*v_wT}
\int_0^{\infty}d\bar z \mu_{{\cal B}_L}(\bar z)e^{-\nu\bar z}
\end{equation}
where we have taken the entropy density $s=(2\pi^2g_*/45)T^3$ and
$\nu=3a\Gamma_{ws}/(2v_w)$ \cite{76}.
$g_*\sim 126$ is the effective number of degrees of freedom at the
phase transition temperature. Eq.~(\ref{bar5}) shows that the
integral over the left-handed quark number, $n_L\propto\mu_{{\cal B}_L}$
in the symmetric phase determines the final baryon asymmetry.

We now return to eqs.~(\ref{diffeq25}) 
in order to compute $\mu_{{\cal B}_L}$. These linear second order
differential equations can be solved by finding the
appropriate Green's function. We keep the discussion general
and consider the following set of $N$ coupled diffusion equations
\begin{equation}  \label{bar6}
\left(\begin{array}{ccc}
-k_{11}{\cal D}_{11} +\Gamma_{11} & \cdots & -k_{1N}{\cal D}_{1N} +\Gamma_{1N} \\ 
\vdots & \ddots & \vdots \\
-k_{N1}{\cal D}_{N1} +\Gamma_{N1} & \cdots & -k_{NN}{\cal D}_{NN} +\Gamma_{NN}
\end{array}\right)
\left(\begin{array}{c} \mu_1 \\ \vdots \\ \mu_N \end{array}\right)=
\left(\begin{array}{c} S_1 \\ \vdots \\ S_N \end{array}\right)
\end{equation}
where ${\cal D}_{ab}=D_{ab}\frac{d^2}{d\bar z^2}+v_w\frac{d}{d\bar z}$.
The corresponding boundary conditions
read $\mu_a(|\bar z|\rightarrow\infty)=0$.
The matrix valued Green's  function $G_{ab}$  is defined by
$\sum_{c=1}^N(-k_{ac}{\cal D}_{ac} +\Gamma_{ac})G_{cb}(\bar z)=
\delta_{ab}\delta(\bar z)$.
In the transport equations (\ref{diffeq25})
also position dependent rates are present, e.g.~$\Gamma_m$.
They typically vanish in the symmetric phase and become maximal in the
broken phase. In order to keep the problem analytically tractable
we  simply model the position dependence of these rates by
step functions, i.e.~$\Gamma_{ab}(\bar z)=\Gamma_{+ab}\Theta(\bar z)+
\Gamma_{-ab}\Theta(-\bar z)$.

The general structure of the Green's function then reads
\begin{equation}\label{bar14_n}
G_{ab}(\bar z)=\Theta(\bar z)\sum_{i=1}^N c_{i+ab}e^{-\lambda_{i+}\bar z}+
                     \Theta(-\bar z)\sum_{i=1}^N c_{i-ab}e^{-\lambda_{i-}\bar z}.  
\end{equation}
The constants $\lambda_{i\pm}$ can be computed
from $\det[k_{ab}(-D_{ab}\lambda_{\pm}^2+v_w\lambda_{\pm})+\Gamma_{\pm ab}]=0$. 
Of course, this procedure supplies
us with $4N$ solutions. However, $2N$ of them
($\lambda_{i+}<0$ for $\bar z >0$ and $\lambda_{i-}>0$  for $\bar z >0$)
correspond to exponentially growing solutions of eq.~(\ref{bar6})
and have to be discarded. The coefficients $c_{i\pm ab}$ can then
be computed from the $2N^3$ dimensional set of linear equations
\begin{eqnarray} \label{bar15a}
\sum_{b=1}^N
\big[-k_{ab}D_{ab}\lambda_{i+}^2+k_{ab}v_w\lambda_{i+}+
\Gamma_{+ab}\big]c_{i+bc}&=&0
\\  \label{bar15b}
\sum_{b=1}^N
\big[-k_{ab}D_{ab}\lambda_{i-}^2+k_{ab}v_w\lambda_{i-}+
\Gamma_{-ab}\big]c_{i-bc}&=&0
\\ \label{bar15c}
\sum_{i=1}^N \big[c_{i+ab}-c_{i-ab}\big]&=&0
\\ \label{bar15d}
\sum_{i,b=1}^N k_{ab}D_{ab}\big[\lambda_{i+}c_{i+bc}-
\lambda_{i-}c_{i-bc}\big]&=&\delta_{ac}
\end{eqnarray}
where $\Gamma_{\pm ab}$ denote the rates for positive and
negative $\bar z$. Eqs.~(\ref{bar15a}) and (\ref{bar15b})
result from solving the homogeneous  version of eq.~(\ref{bar6})
in the range of positive and  negative $\bar z$, respectively.
Only $2N^2(N-1)$ of them are independent.
The continuity of the Green's function at $\bar z=0$ is guaranteed
by eq.~(\ref{bar15c}). Finally, eq.~(\ref{bar15d}) is obtained by
integrating the definition equation for the Green's function
on an infinitesimally small interval around $\bar z=0$.
Diffusion constants $D_{ab}$ and statistical
factors $k_{ab}$ differing in  broken and symmetric
phase can be treated along the same lines.

Once the Green's function is known one can easily compute the
chemical potentials. Applying the general
formulas to the diffusion equations (\ref{diffeq25}) we
identify $\mu_1\equiv \mu_{Q_3}$, $\mu_2\equiv\mu_{H_1}$, 
$\mu_3\equiv\mu_{H_2}$, $S_1\equiv S_{Q_3}$, $S_2\equiv S_{H_1}$ 
and $S_3\equiv S_{H_2}$.
Using eqs.~(\ref{bar5}) and (\ref{mubl}) we obtain for the
baryon to entropy ratio
\begin{equation}   \label{bar16}
\eta_B=\frac{135 \Gamma_{ws}}{2\pi^2g_*v_wT}C
\int_0^{\infty}d\bar z e^{-\nu \bar z} \int_{-\infty}^{\infty} d\bar z'
G_{1a}(\bar z-\bar z')S_a(\bar z').
\end{equation}
Since the Green's function consists only of exponentials,
the $\bar z$-integration can be performed analytically.
The evaluation of this expressions is performed in the next paragraph.
Also the various approximations we used in its derivation will be discussed.
%
%
%
%
%
%
%
%
\subsubsection{Numerical evaluation and discussion}
Before starting to calculate a numerical value
for the emerging baryon asymmetry we discuss the validity
of the assumptions and approximations made in the
derivation of eq.~(\ref{bar16}). This requires the specification
of the various parameters that enter the diffusion equations.
For the diffusion constants of quark and Higgs fields we take \cite{CMQSW01}
\begin{equation} \label{num1}
D_q=\frac{6}{T}, \quad D_h=\frac{110}{T}. 
\end{equation}
We use the rates \cite{CMQSW01}
\begin{eqnarray} \label{num2}
\Gamma_y+\Gamma_{yA}=0.015T, \quad \Gamma_{y\mu}=0, \quad  \Gamma_m=0.05T\Theta(-\bar z)
\nonumber\\
 \Gamma_{hf}=0.016T, \quad \Gamma_{H_1}=\Gamma_{H_2}=0.0036T\Theta(-\bar z).
\end{eqnarray}
$\Gamma_{y\mu}$ is strongly suppressed because it involves
heavy left-handed stop states (\ref{diffeq17}). $\Gamma_m$,
$\Gamma_{H_1}$ and $\Gamma_{H_2}$ are present only broken phase.
We model their position dependence by a step function.
The weak and strong sphaleron rates are given by \cite{moore_ss97}
\begin{eqnarray}  \label{num3}
\Gamma_{ws} &\approx& 6 \alpha_w^4T \approx 2.2\times 10^{-5}T
\nonumber\\
\Gamma_{ss} &\approx& 1500\Gamma_{ws}\approx 0.033 T.
\end{eqnarray}
It has been shown recently that parametrically
$\Gamma_{ws}=C\ln(1/g^2)\alpha_w^5$ \cite{dietrich}, but lattice
measurements of the rate are consistent with $C\sim1/\alpha_w$.
The thickness of the bubble wall varies considerably in the NMSSM,
$1/T \lesssim L_w \lesssim 20/T$, i.e.~the wall may become much thinner
than in case of the MSSM, where $20/T \lesssim L_w \lesssim 30/T$ has
been found \cite{marcos}. Calculations of the wall velocity
in the SM lead to $0.36<v_w<0.44$ \cite{moorepr}.
Gauge fields in the hot plasma diminish this result \cite{moore2000}.
In supersymmetric models there arise additional friction terms from
the SUSY particles, in the first place from a light top squark.
In the MSSM, this can bring down the wall velocity to $v_w\sim 10^{-2}$  
\cite{JS2000}.
In the following we treat $v_w$ as a free parameter
and examine its impact on the emerging baryon asymmetry.

Let us now summarize the approximations leading to eq.~(\ref{bar16})
for the baryon to entropy ratio:

$\bullet$ Assumption 1: $L_w>\frac{1}{T}$ so that most particles in the
plasma are indeed accurately described by the WKB approximation in
their interaction with the bubble wall. For typical wall thicknesses in the
NMSSM, $5/T \lesssim L_w \lesssim 15/T$, this assumption is very well
justified, although for the very thinnest walls $L_w \sim 1/T$
the WKB approximation becomes questionable.

$\bullet$ Assumption 2:  $v_w<\frac{L_w}{3D}$, the thermalization
condition that guarantees the applicability of the fluid ansatz. In
case of (s)quarks this condition is satisfied for $v<0.4$ even for
rather thin walls with $L_w\sim5/T$. Higgs particles thermalize
much slower, as can be deduced from their large diffusion constant.
Even for the largest wall thicknesses, $L_w\sim20/T$,
rather small velocities, $v_w<0.1$, are required. 

$\bullet$ Assumption 3: $v_w<\frac{1}{\sqrt{3}}$. We work to linear order
in the wall velocity, which is only justified if the wall moves slower
than the speed of sound in the plasma. Otherwise, diffusion in the
region in front of the wall, giving rise to ``non-local'' baryogenesis,
is no longer possible. However, in case of $v_w>\frac{1}{\sqrt{3}}$
the fluid approximation would break down anyway, i.e.~from
assumption 3 no new constraints result.

$\bullet$ Assumption 4: $\Gamma_{ws}<v_w^2/D$, so that the
back-reaction of the baryon number violating processes can
be neglected in the diffusion equations (\ref{diffeq25}). According
to eqs.~(\ref{num1}) and (\ref{num3}) this requires $v_w>0.01$ (quarks)
and $v_w>0.04$ (Higgs particles).

$\bullet$ Assumption 5: $\Gamma_{ss}>v_w^2/D$ in order to put
the strong sphaleron interaction to equilibrium. In this case the
strongest constraint, $v_w<0.45$, result from the quarks, which
is easily satisfied.

The calculation of the final baryon asymmetry still requires
the specification of the source terms (\ref{diffeq12}) which
enter eq.~(\ref{bar16}).
We concentrate on the source terms induced by the semi-classical
force and neglect those related to spontaneous baryogenesis.
We include the source terms of the top quark (\ref{fwkb6})
and the charged Higgsinos (\ref{fwkb13}). 
For sake of simplicity we neglected the neutral Higgsinos, which
should affect the  final baryon asymmetry at most by a factor of order one.
Also the singlino is disregarded.

In order to evaluate the source terms (\ref{diffeq12}) the bubble
wall profile is required. We approximate our numerical solutions
by a kink-ansatz with a common wall thickness for all fields
present in the bubble. However, we allow  for a variation in $\tan\beta$ in
the bubble wall by taking 
$|H_2^0(z)|=|H_1^0(z)|(\tan\beta_0+{\rm const}\cdot|H_1^0(z)|^2|)$,
in the spirit of fig.~\ref{f_dtb}a. The relevant source terms
are obtained from combining eqs.~(\ref{fwkb6}),
(\ref{fwkb13}) and  (\ref{diffeq12})
\begin{eqnarray} \label{num5}
S_t&=&\frac{N_cD_qv_w}{\langle p_z^2/E \rangle_0}(\langle
A\rangle \theta_t''
+\langle B \rangle
(m_t^2)'\theta_t')'
\\[.3cm] \label{num7}
S_{\tilde h_1}&=&\frac{D_hv_w}{\langle p_z^2/E \rangle_0}(\langle A
\rangle \Big(\theta_{\tilde h}'-\gamma_{\tilde h}'\sin^2(a_{\tilde h})+
\delta_{\tilde h}'\sin^2(b_{\tilde h})\Big)' +
\nonumber \\[.3cm]
&&~~~~~~~
+\left.\langle B \rangle
(m_{\tilde h}^2)'\Big(\theta_{\tilde h}'-\gamma_{\tilde h}'\sin^2(a_{\tilde h})+
\delta_{\tilde h}'\sin^2(b_{\tilde h})\Big) \right).'  
\\[.3cm] \label{num7a}
S_{\tilde h_2}&=&S_{\tilde h_1}
\end{eqnarray}
with
\begin{equation}\label{6.81a}
\langle A\rangle = \left\langle\frac{|p_z||m|^2}{2E^2}\right\rangle_+
\quad \langle B\rangle = \left\langle\frac{|p_z|(E^2-|m|^2)}{2E^4}
\right\rangle_+
\end{equation}
Here $N_c=3$ is the number of colors and $E^2=\vec{p}^2+m_i^2$,
$E^2_z=p_z^2+m_i^2$ and $p_{\perp}^2=\vec{p}^2-p_z^2$.
With exception of $\langle p_z^2/E \rangle_0$ all thermal averages
in eqs.~(\ref{num5}) - (\ref{num7a}) are performed with  massive
distribution functions for fermions  (\ref{diffeq7}),
which ensures the decoupling of heavy particles.
Taking into account only
the source terms of eqs.~(\ref{num5}) - (\ref{num7}), we have
$S_{Q_3}=S_t$,  $S_{H_1}=S_{\tilde h_1}$ and 
$S_{H_2}=S_{\tilde h_2}$.

\begin{table}[b] \centering
\begin{tabular}{|c|c|c||c|c|c|c||c|c|c|c|c|} \hline
 & $M_0$ & $A_0$ & $M_{\tilde t_1}$ & $M_{\tilde t_2}$ & $M_{\tilde b_2}$ &
$M_{\tilde u}$ & $k_{Q_3}$ & $k_T$ & $k_B$ & $k_{Q_1}$ & $C$\\ \hline
A  & & & 0& 0& 0& 0& 18 & 9 & 9 & 18 & 0\\ \hline
B  & & & $\infty$ & 0&$\infty$ &$\infty$ & 6 & 9 & 3 & 6 & 0.385\\ \hline
C  &125 &-100 & 418& 238& 334& 375& 7.26 & 4.56 & 3.60 & 7.20 & 0.125\\ \hline
D  &100 &-100 & 371& 201& 217& 325& 8.43 & 4.98 & 3.87 & 7.74 & 0.102\\ \hline
\end{tabular}
\caption{Squark spectra and corresponding statistical factors
used in the discussion of the baryon asymmetry. For the sets
C and D we used $\lambda=0.05$, $k=0.4$, $m_0=200$ GeV
and $\tan\beta=-5$. Furthermore, we took $x=-100$ $(-150)$ GeV
in case C (D). The sets A and B do not follow the universal pattern of
SUSY breaking considered in section 2. (Units in GeV.)}
\label{t_squark}
\end{table}

In the approach with canonical momenta (which we had in the first
version of this paper), there was a source $S_{\tilde t}$ for the stop and 
$S_{\tilde h_1}\not=S_{\tilde h_2}$. The latter leads to an important
contribution to the $H_1-H_2$-combination, the well-known term
proportional $(\tan\beta)'$ and to an additional term for the singlet
field. These are absent now. The brackets $\langle A\rangle$ and
$\langle B\rangle$ in the approach with canonical momenta are
\begin{equation}
\label{6.81b}
\langle A^{\rm can}\rangle=\left\langle\frac{|p_z|E_z-p_z^2}{2E}\right\rangle
_+\ ,\ \langle B^{\rm can}\rangle=\left\langle\frac{p_\bot^2|p_z|+p_z^2E_z}{4E^3e_z}\right\rangle
_+\end{equation}
For $p\gg m$ they differ by a factor $\frac{1}{2}\frac{|p|}{|p_z|}$
from the ones for kinetic momenta (\ref{6.81a}).

The generated baryon asymmetry depends on the squark spectrum
because of the potential suppression due to strong sphalerons. 
In the following we consider four different squark spectra which
are listed in table \ref{t_squark}.
Case A corresponds to the massless
limit, where strong sphaleron suppression is most efficient. In case
B all squarks are assumed to be heavy, with exception of the right-handed
stop, which is taken massless. The spectra C and D are obtained
from the universal pattern of SUSY breaking discussed in section 2.
The first and second generation squarks are almost degenerate in
mass with $\tilde b_2$, thus $2k_B=k_{Q_1}=k_{Q_2}=2k_U=\dots$
Case C corresponds to the example of explicit CP-violation discussed
in context of fig.~\ref{f_ECPV}, while case D is the squark
spectrum of the  transitionally CP-violating example of fig.~\ref{f_SCPV}.

According to eq.~(\ref{bar16}) the produced baryon asymmetry is
proportional to the parameter $C$, which is related to the chemical
potential of left-handed quarks (\ref{mubl}) and also given in 
table \ref{t_squark}. It turns out that the results for different spectra
can indeed be obtained by rescaling with the relevant $C$ parameters,
i.e.~the indirect impact of the statistical factors in the diffusion equations
(\ref{diffeq25}) is small.
For the massless case A, and more general for the case of degenerate
squarks, $C$ vanishes. Baryon production is completely suppressed
by rapid strong sphalerons transitions, which is a well known 
result \cite{giush}. The cancellation disappears for non-degenerate
squarks. Even for the realistic spectra C and D, resulting from universal
SUSY breaking, there is only a mild suppression by a factor of 3--4 
relative to the idealized spectrum B (usually assumed in context of 
the MSSM).

\begin{figure}[t]
\begin{picture}(200,180)
\put(140,5){\epsfxsize7cm \epsffile{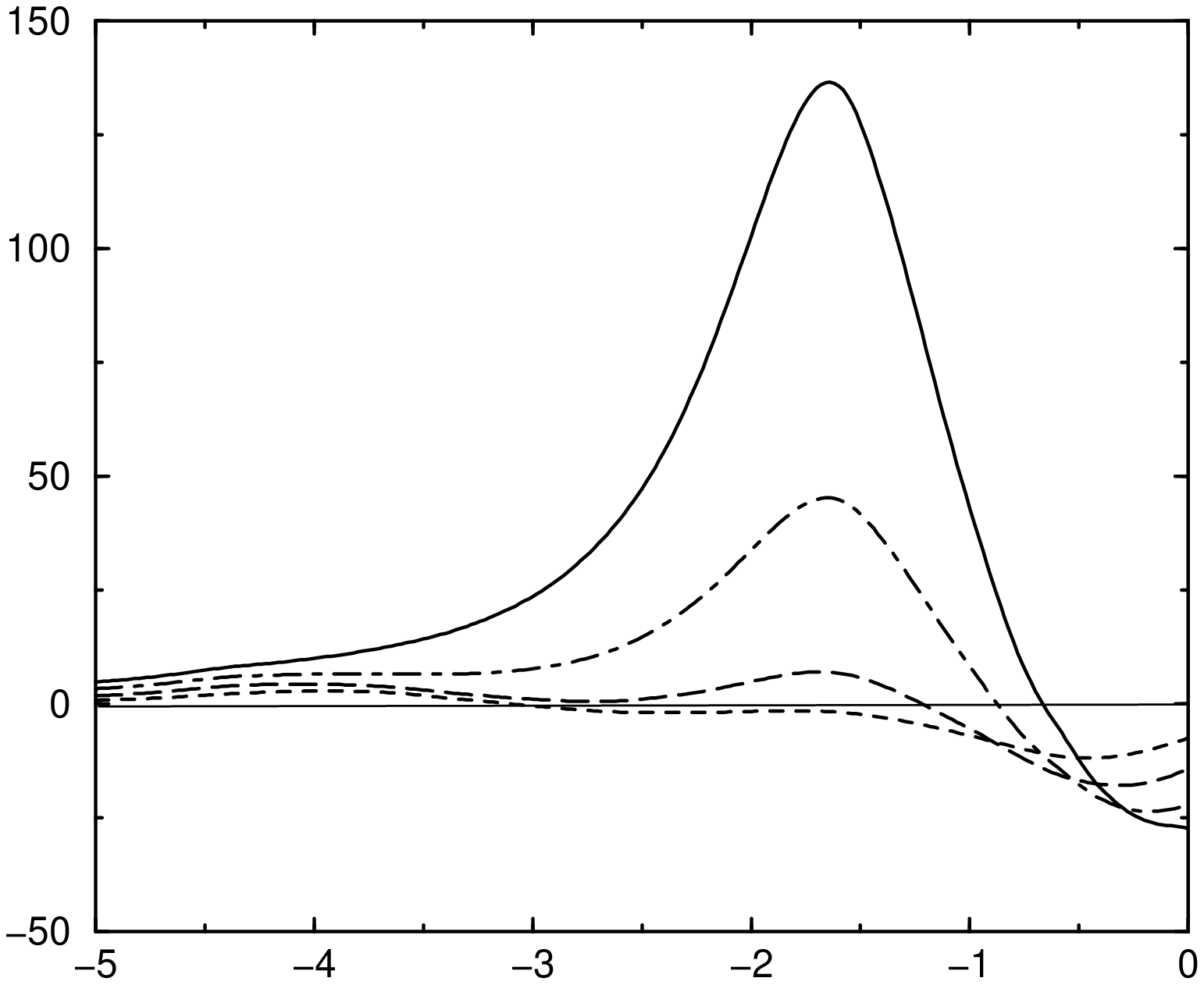}}
\put(355,0){${\rm log}_{10}(v_w)\rightarrow$}
\put(115,85){$\frac{\eta^*}{\delta\theta_t}\uparrow$}
\end{picture}
\caption{Top quark contribution to the baryon asymmetry
for the spectrum D in units of $2\times10^{-11}$ normalized by $\delta\theta_t$
as a function of the wall velocity for different values of the
wall thickness $L_w=20/T,10/T,5/T,3/T$ (from below).
}
\label{f_top}
\end{figure}

To begin with let us consider the source term of the top quarks (\ref{num5}),
which in the first place depends on the variation of the phase of the
top quark mass along the wall, $S_t\propto\delta\theta_t$, and the wall
thickness $L_w$.
From eqs.~(\ref{CPV1}) and (\ref{CPbub5})
we obtain $\delta\theta_t=\delta\theta_2=\cos^2(\beta)\delta\theta$,
where $\theta_2$ denotes the phase of $H_2^0$ and $\theta$ is the (gauge
invariant) sum of the phases of the two Higgs fields, that entered the
bubble equations in section 5. As a consequence, $S_t$ is rather
sensitive to the Higgs vev ratio. Baryon production from the top
quark source is particularly efficient for small values of $\tan(\beta_T)$.
In fig.~\ref{f_top} we present the baryon asymmetry induced by $S_t$ as
a function of the wall thickness and velocity, where we used $v_c/T_c=1.60$,
$T_c=101$ GeV and $y_t=1.015$ (corresponding to $\tan\beta=5$).
 
Our results\footnote{In numerical evaluations we still used the
brackets $\langle A^{\rm can}\rangle, \langle B^{\rm can}\rangle$ of eq.
(\ref{6.81b}) instead of $\langle A\rangle, \langle B\rangle$ of eqs. (\ref{6.81a}) in order not to be forced to completely
repeat the whole analysis. This introduces a deviation of order (1), in the
case $|p|\gg m$ a factor $\frac{1}{2}$ instead of $\frac{|p_z|}{|p|}$ in
the brackets $\langle A\rangle$ and $\langle B\rangle$. This is well within
the accuracy of the present investigation.}
 would be slightly enhanced by larger values of $v_c/T_c$.
We measure the baryon asymmetry in units
of $2\times10^{-11}$, which is the lower observational bound \cite{eta}
\begin{equation} \label{num8}
\eta^*=\frac{\eta}{2\times10^{-11}}.
\end{equation}
In fig.~\ref{f_top} we display the dependence
of the baryon asymmetry  on the wall velocity for different values
of the wall thickness $L_w=20/T,10/T,5/T,3/T$.
We used the squark spectrum D in the evaluation. Results for
other spectra can be easily obtained by rescaling with the relevant
$C$ parameters of table~\ref{t_squark}. 
For $L_w\lsim10/T$ baryon production is most efficient for $v_w\sim0.02-0.03$.
Interesting enough, this is precisely the range of wall velocities
found in recent calculations in the MSSM \cite{moore2000,JS2000}.
$\eta$ becomes small for large values of $v_w$ because transport in 
front of the wall gets more and more inefficient. For small wall
velocities deviation from equilibrium becomes too small to generate
a relevant baryon asymmetry. We observe an approximate $1/L_w^2$
dependence in the generated baryon number.
This behavior is expected from the explicit expression for
the source term (\ref{num5}), which contains three derivatives
with respect to $\bar z$. Since the calculation of $\eta$ (\ref{bar16})
requires one (numerical) integration of the source over $\bar z'$, approximating
$\partial_{\bar z}\sim1/L_w$ we obtain $\eta\propto1/L_w^2$.
As a result baryon production from the top quark source is
most efficient in the case of slow ($v_w\sim 10^{-2}$) and thin walls, and for
non-degenerate squark spectra. It turns out that the 
Higgs source term behaves in a similar way. In fig.~\ref{f_top}
we assumed $\delta\beta=10^{-3}$ in the bubble wall which 
is a typical value according to section 4. However, even large 
values, $\delta\beta=10^{-2}$, change the result only by a few percent.
From fig.~\ref{f_top} we approximately obtain for the maximal value
of the baryon asymmetry
\begin{equation} \label{num9}
\eta^*_{\rm max}\sim47\left(\frac{5}{L_wT}\right)^2\delta\theta_t.
\end{equation}
However,  for medium and large Higgs vev ratios
even small values of $\delta\theta_t \sim{\cal O}(10^{-2})$
are difficult to achieve, since $\delta\theta_t=\cos^2(\beta)\delta\theta$.
In the case of $\tan\beta=5$, for instance,  we obtain $\delta\theta_t=\delta\theta/26$, which
leads to a significant suppression. In the example of transitional CP-violation
(fig.~\ref{f_SCPV}) we have $\tan\beta=5$, $L_w=3$ and $\delta\theta\sim1/20$ 
leading to $\eta^*_{\rm max}\sim1/4$. In case of explicit CP-violation, the top quark 
contribution to the baryon asymmetry is much smaller even, and can be safely
neglected.

We now come to the charged Higgsino contribution to the baryon
asymmetry fueled by the source terms (\ref{num7}) and (\ref{num7a}) for
the examples of explicit and transitional CP-violation already
discussed in the context of figs.~\ref{f_ECPV} and \ref{f_SCPV}.
Again, we present our results$^{14}$ as a function of the wall velocity for
different values of the wall thickness  $L_w=20/T,10/T,5/T,3/T$.
Concerning the dependence of the generated baryon asymmetry
on the wall velocity and the wall thickness we find a  similar 
behavior as in the case of the top quark source: $\eta$ becomes large
for thin walls and $v_w\sim10^{-2}$.

\begin{figure}[t]
\begin{picture}(200,185)
\put(0,5){\epsfxsize7cm \epsffile{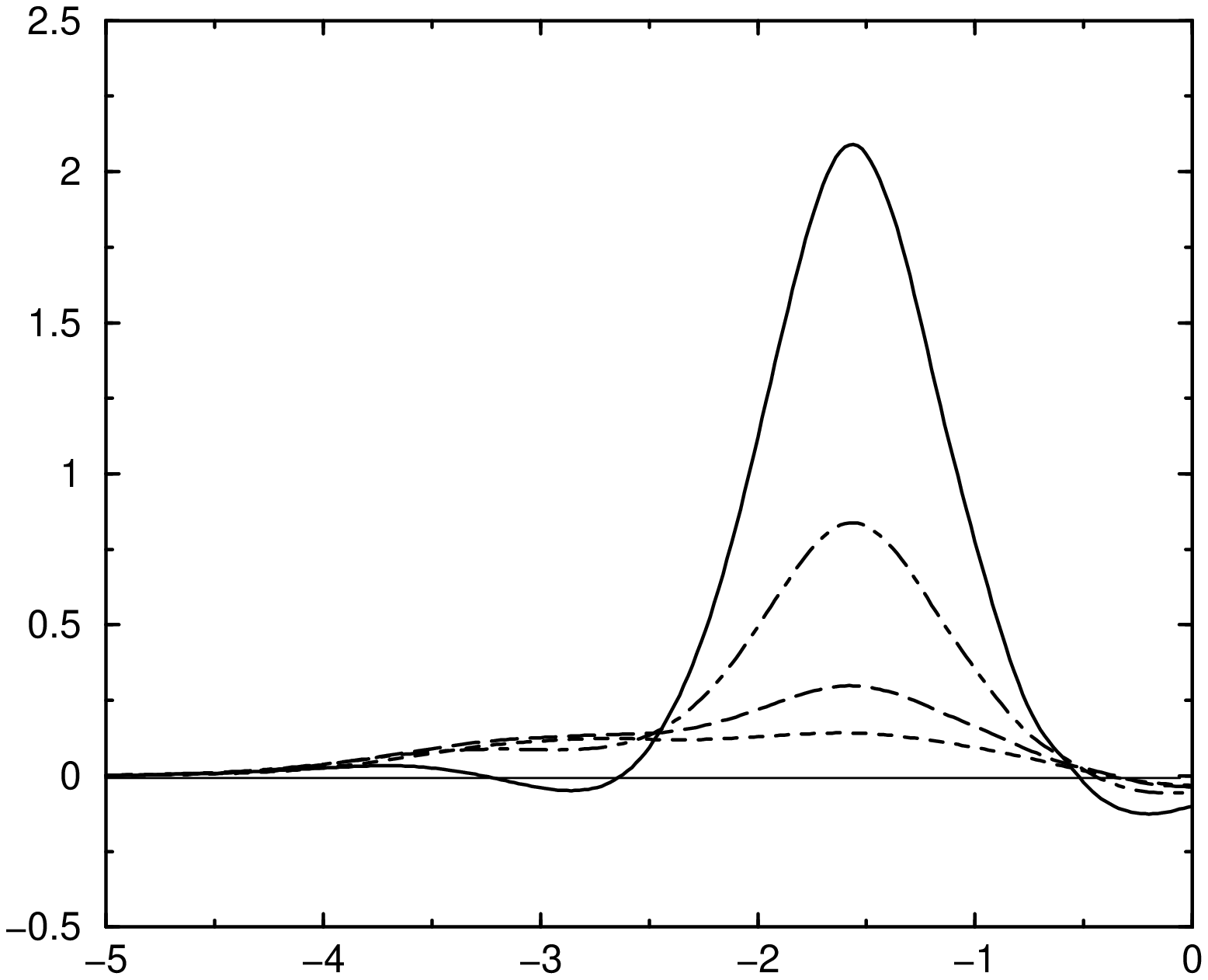}}
\put(240,5){\epsfxsize6.8cm \epsffile{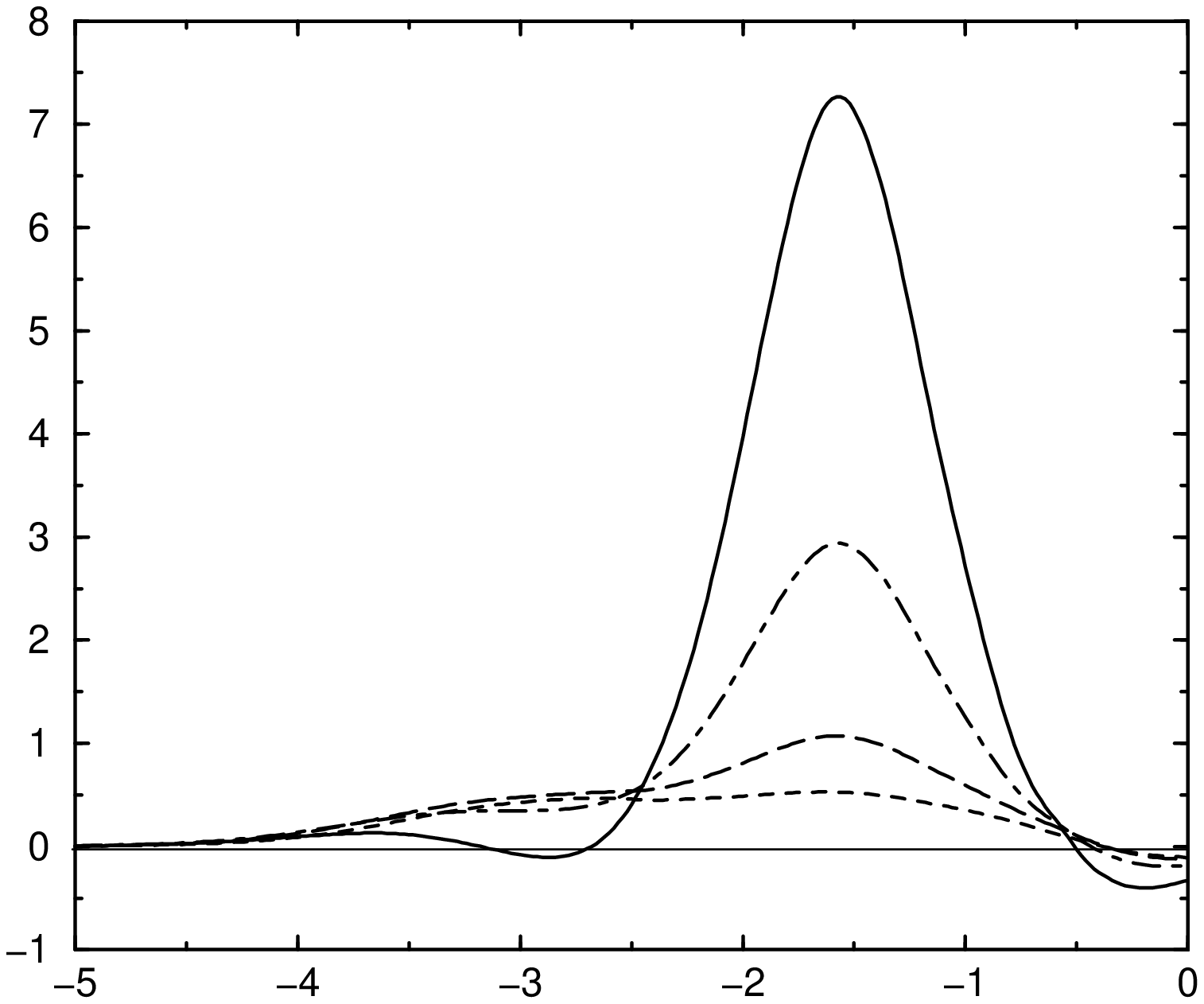}}
\put(175,145){(a)}
\put(405,145){(b)}
\put(220,85){$\eta^*$}
\put(220,100){$\uparrow$}
\put(135,-9){${\rm log}_{10}(v_w)\rightarrow$}
\put(375,-9){${\rm log}_{10}(v_w)\rightarrow$}
\end{picture}
\caption{The chargino contribution to the baryon asymmetry
in units of $2\times10^{-11}$ as a function of the wall
velocity  for different values of the
wall thickness $L_w=20/T,10/T,5/T,3/T$ (from below).
We use the squark spectrum C and the example of explicit CP-violation 
considered in the context of fig.~\ref{f_ECPV}.
(b): The same quantity for the transitionally CP-violating bubble
wall of fig.~\ref{f_SCPV} and the squark spectrum D.
}
\label{f_char_hel}
\end{figure}

The baryon asymmetry generated from the 
chargino dispersion relation is shown in fig.~\ref{f_char_hel}.
If the top Yukawa interactions were in equilibrium, as was assumed in
elder work on the subject, this contribution is completely erased,
because of the equal source terms for both Higgsinos (\ref{num7},\ref{num7a}).
Even with finite top Yukawa rates no baryon asymmetry is generated
from this source, if both Higgs fields have equal rates in the diffusion
equations (\ref{diffeq25}), i.e.~if $\Gamma_y+\Gamma_{yA}=\Gamma_{y\mu}$ 
and $\Gamma_{H_1}=\Gamma_{H_2}$. In our scenario an asymmetry in the
top Yukawa rates is inevitably induced by the heavy left-handed stop,
leading to a strong suppression of $\Gamma_{y\mu}$ (\ref{num2}).
In the evaluations of fig.~\ref{f_char_hel} we used $\delta\beta=10^{-3}$.
If we increase the change in the Higgs vev ratio by a factor of 10, the
results only change by about 15 percent.

From fig.~\ref{f_char_hel}a for the case of explicit CP-violation 
we read of
\begin{equation}
 \eta^*_{\rm max}\sim 0.8 \left(\frac{5}{L_wT}\right)^2\sin(\phi_{\mu}).
\end{equation} 
Thus CP-violating phases of order $10^{-1}$ are required to
account for the observed baryon asymmetry in this scenario.
Phases of this size are compatible with the EDM experiments only
if the first and second generation squarks are heavy or in the case of
accidental cancellations. Our findings for baryogenesis from
explicit CP-violation in the NMSSM are very similar to the results 
in the MSSM \cite{HJS01}. Still, baryogenesis in the NMSSM is 
slightly more effective because of the thinner bubble walls.

The scenario of transitional CP-violation is a very attractive
alternative, being not in conflict with the EDM experiments. 
As shown in fig.~\ref{f_char_hel}a, it can generate
the baryon asymmetry for a considerable range of wall velocities,
especially for $L_w=3/T$. It even allows for a certain baryon 
``over-production'' to compensate for a partial washout of
the generated baryon asymmetry due to the existence of
bubbles with opposite CP-properties. The introduction of small
explicit phases lifts the energetic degeneracy of the two
sort of bubbles, leading to a more rapid nucleation of bubbles
of the true vacuum and a net baryon production.

\begin{figure}[t]
\begin{picture}(200,165)
\put(15,15){\epsfxsize6.5cm \epsffile{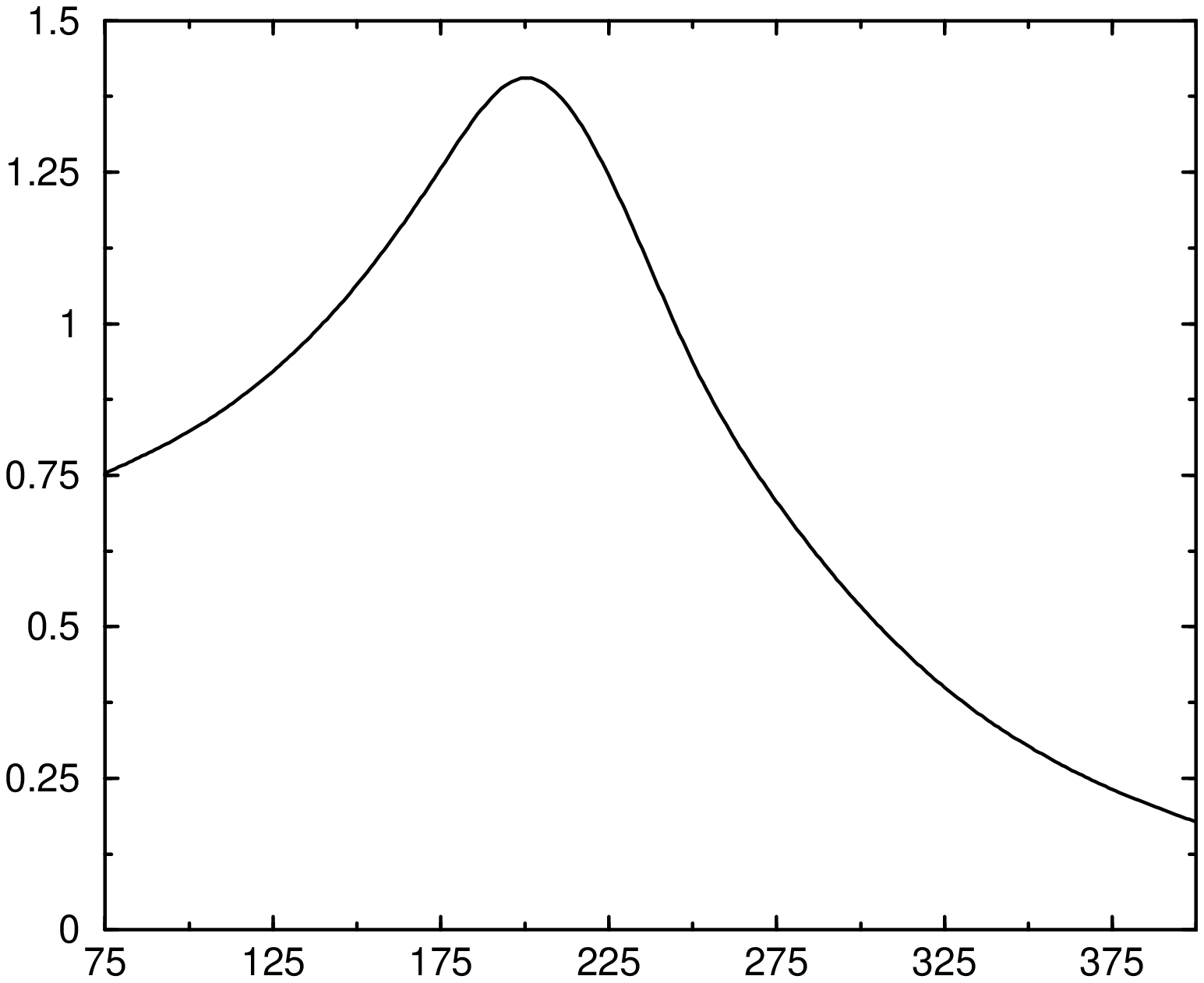}}
\put(240,15){\epsfxsize6.7cm \epsffile{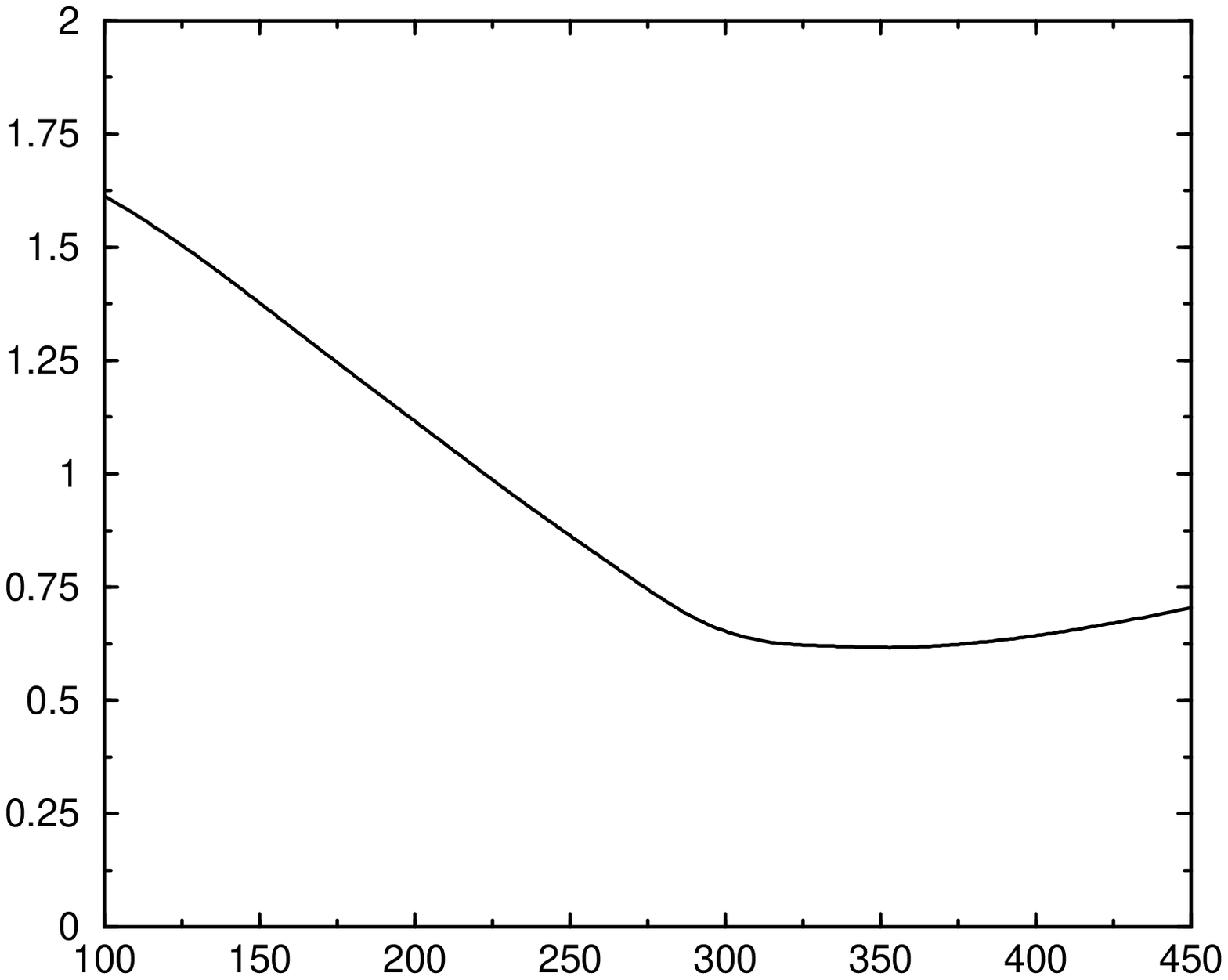}}
\put(175,145){(a)}
\put(400,145){(b)}
\put(225,85){$\eta^*$}
\put(225,100){$\uparrow$}
\put(135,-2){$M_2~[\rm GeV]\rightarrow$}
\put(355,-2){$m_{U_3}~[\rm GeV]\rightarrow$}
\end{picture}
\caption{Chargino contribution to the baryon asymmetry
in units of $2\times10^{-11}$ as a function of the gaugino
mass parameter $M_2$ (a) and as a function of the soft scalar mass 
parameter $m_{U_3}$. In both cases we otherwise used the parameter 
set of explicit CP-violation (fig.~\ref{f_ECPV}), $\phi_{\mu}=0.1$, 
$L_w=5/T$, $v_w=0.03$ and $\delta\beta=10^{-3}$.
}
\label{f_13}
\end{figure}

Up to now we centered the discussion of
baryon production around two particular sets of NMSSM parameters.
However, the results may considerably change if different regions
in the parameter space are considered. Of course, baryon production becomes
inefficient if the particles which supply the CP-violating source terms,
in the first place the Higgsinos, are heavy and decouple from the plasma.
The thermal averages, which enter the sources (\ref{num5}) - (\ref{num7a}),
already decrease by a factor of 10 if $m/T$ is raised from one to five.
In case of $m/T=10$ the suppression is of order ${\cal O}(10^{-3})$
and successful baryogenesis becomes extremely difficult. If universal
SUSY breaking is assumed this occurs in the regime $M_0 \gtrsim 700$ GeV.

The chargino source term is rather sensitive to the relative size
of the Higgsino mass parameter, $\mu$, and the mass of the
SU(2)-gauginos, $M_2$. In fig.~\ref{f_13}a we present the dependence
of the generated baryon asymmetry on $M_2$, while the remaining
parameters as well as the bubble wall profile are taken from the
example of explicit CP-violation considered in fig.~\ref{f_ECPV},
i.e.~the effect of a change in $M_2$ on the bubble wall is ignored.
Moreover, we use the squark spectrum C, $\phi_{\mu}=0.1$ $L_w=5/T$ 
and $v_w=0.03$.
We find a resonance structure for $\eta$, with the peak located
at $M_2\sim |\mu|=226$ GeV.  Fig.~\ref{f_char_hel}a corresponds
to $M_2=103$ GeV.
In fig.~\ref{f_13}b we present the dependence of the baryon
asymmetry on the soft breaking scalar mass parameter $m_{U_3}$.
Even though this parameter has no direct impact on the chargino
source terms (\ref{num7}), (\ref{num7a}) it alters the statistical factors 
$k_{Q_3}$ and $k_T$ due to varying stop masses. Baryogenesis
becomes less efficient if the two stop stated are more or less
degenerate. In the case under consideration this happens for
$m_{U_3}\sim m_{Q_3}=329$ GeV. In fig.~\ref{f_char_hel}a we
used $m_{U_3}=256$ GeV.

\section{Conclusions}
In our computation
we used the semi-classical approximation to describe the interaction
of the particles in the plasma and the propagating bubble wall which
in context of the MSSM has been introduced in ref.~\cite{CJK}. Our
work differs in various aspects from that analysis. We included
contributions from top quarks which have previously
been neglected and took into account also
position-dependent mixing in the chargino mass matrix (``flavor
contribution'').  In ref.~\cite{CJK} different results were obtained due
to an error in the transport equations, which prevented the cancellation
between the dominating helicity parts in the $\tilde h_1^-$ and $\tilde h_2^+$ 
source terms. With dispersion relations for canonical 
momenta, the charged
Higgsinos provide in the MSSM  CP-violating source terms which
are proportional to the variation of $\tan\beta$ in the bubble wall. This result would agree
well with refs.~\cite{HN,cqrvw,riotto_ctp} where different methods
where used to determine the CP-violating source terms. Recently
$\delta\beta$ independent contributions to the source terms in the
MSSM have been calculated \cite{RV99}. However, these arise from
higher orders in a Higgs insertion expansion and should not be included
in our semi-classical approach.
In the NMSSM non-vanishing source terms for stops and charginos
(and for the top quark) are generated even for constant $\tan\beta$ because
of the changing singlet field, and due to CP-violating bubble walls.
These arise either from explicit CP-violation, which in case of the
NMSSM is possible already in the tree-level Higgs potential, or from
transitional CP-violation.

Using kinetic momenta in the dispersion relations as suggested convincingly
in ref. ~\cite{CJK} $\tilde h_1$ and $\tilde h_2$ sources are exactly
equal and $\tan \beta'$ effects cancel. Thus one has to consider a 
source symmetric in the Higgses. This contribution can be also sizeable,
if the Yukawa interactions are not in equilibrium ~\cite{CJK} and if the
left-handed stop is heavy ~\cite{CMQSW01,HJS01}. This statement
ist unchanged in our singlet model. The WKB results and their
difference to other work should be further discussed
using quantum transport equations.

In our work we investigated the impact of different squark spectra
on the emerging baryon asymmetry, while in ref.~\cite{CJK} only
the massless approximation was considered. As a consequence,
we obtained a non-vanishing baryon asymmetry, while assuming
the strong sphalerons to be in equilibrium. Baryogenesis becomes
most efficient for non-degenerate squarks. The generated baryon 
asymmetry is roughly proportional to $1/L_w^2$. Since in the NMSSM
the bubble walls can be considerably thinner than in the MSSM, more
baryons can be produced. The baryon asymmetry strongly depends
on the velocity of the bubble wall, which in the NMSSM has not been 
computed so far. Interesting enough, we find that the maximal baryon 
number is generated for $v_w\sim 10^{-2}$. This is a typical value
for the wall velocity in the MSSM.

Finally, we studied the dependence of the chargino 
source terms on $\mu,M_2,m_{Q_3}^2$ and $m_{U_3}^2$.
We found an enhancement if $\mu$ and $M_2$ become degenerate.
This resonance behavior has also been reported in
refs.~\cite{cqrvw,riotto_ctp,CJK,RV99}. If $m_{Q_3}^2\sim m_{U_3}^2$,
baryon production is partially suppressed by strong sphalerons.
Concluding, the  NMSSM can account for the observed
baryon asymmetry, especially if $\mu\sim M_2$ and if the squark
spectrum is rather non-degenerate, or if transitional CP-violation
occurs. Both scenarios point to a non-universal pattern of SUSY
breaking.
%
%
%
%
%
%
%
%
%
%

\section*{Acknowledgment}
We thank P.~John
for useful discussions.
This work was supported in part by the
TMR network {\it Finite Temperature Phase Transitions in Particle
Physics}, EU contract no. ERBFMRXCT97-0122.

\end{document}